\documentclass[times,twocolumn]{aastex62}
\usepackage{amsmath}
\usepackage{multirow}
\usepackage{cases}
\usepackage{graphicx}
\usepackage{subfigure}
\usepackage{natbib}
\usepackage{color}
\usepackage{tabularx}
\usepackage{bm}
\usepackage{threeparttable}
\usepackage{gensymb}

\newcommand{\thetae}{\theta_{\rm E}}

\newcommand{\pie}{\pi_{\rm E}}

\newcommand{\te}{t_{\rm E}}
\newcommand{\eventa}{KMT-2017-BLG-0849}
\newcommand{\eventb}{KMT-2017-BLG-1057}
\newcommand{\eventc}{OGLE-2017-BLG-0364}
\newcommand{\eventd}{KMT-2017-BLG-2331}
\newcommand{\evente}{KMT-2017-BLG-0958}
\newcommand{\an}{\theta_{*}}

\DeclareGraphicsExtensions{.pdf,.png,.jpg}

\shorttitle{}
\shortauthors{Gui et al.}

\begin{document}
\title{{\large Systematic KMTNet Planetary Anomaly Search. XII. Complete Sample of 2017 Subprime Field Planets}}

\correspondingauthor{Weicheng Zang}
\email{3130102785@zju.edu.cn}

\author[0009-0007-9365-9806]{Yuqian Gui}
\affiliation{Department of Astronomy, Tsinghua University, Beijing 100084, China}

\author[0000-0001-6000-3463]{Weicheng Zang}
\affiliation{Center for Astrophysics $|$ Harvard \& Smithsonian, 60 Garden St.,Cambridge, MA 02138, USA}
\affiliation{Department of Astronomy, Tsinghua University, Beijing 100084, China}

\author[0009-0004-1650-3494]{Ruocheng Zhai}
\affiliation{Department of Astronomy, Tsinghua University, Beijing 100084, China}

\author[0000-0001-9823-2907]{Yoon-Hyun Ryu} 
\affiliation{Korea Astronomy and Space Science Institute, Daejon 34055, Republic of Korea}

\author[0000-0001-5207-5619]{Andrzej Udalski}
\affiliation{Astronomical Observatory, University of Warsaw, Al. Ujazdowskie 4, 00-478 Warszawa, Poland}

\author[0000-0003-0626-8465]{Hongjing Yang}
\affiliation{Department of Astronomy, Tsinghua University, Beijing 100084, China}

\author[0000-0002-2641-9964]{Cheongho Han}
\affiliation{Department of Physics, Chungbuk National University, Cheongju 28644, Republic of Korea}

\author[0000-0001-8317-2788]{Shude Mao}
\affiliation{Department of Astronomy, Tsinghua University, Beijing 100084, China}
\affiliation{National Astronomical Observatories, Chinese Academy of Sciences, Beijing 100101, China}

\collaboration{(Leading Authors)}

\author[0000-0003-3316-4012]{Michael D. Albrow}
\affiliation{University of Canterbury, School of Physical and Chemical Sciences, Private Bag 4800, Christchurch 8020, New Zealand}

\author[0000-0001-6285-4528]{Sun-Ju Chung}
\affiliation{Korea Astronomy and Space Science Institute, Daejeon 34055, Republic of Korea}

\author{Andrew Gould} 
\affiliation{Max-Planck-Institute for Astronomy, K\"onigstuhl 17, 69117 Heidelberg, Germany}
\affiliation{Department of Astronomy, Ohio State University, 140 W. 18th Ave., Columbus, OH 43210, USA}

\author[0000-0002-9241-4117]{Kyu-Ha Hwang}
\affiliation{Korea Astronomy and Space Science Institute, Daejeon 34055, Republic of Korea}

\author[0000-0002-0314-6000]{Youn Kil Jung}
\affiliation{Korea Astronomy and Space Science Institute, Daejeon 34055, Republic of Korea}
\affiliation{National University of Science and Technology (UST), Daejeon 34113, Republic of Korea}

\author[0000-0002-4355-9838]{In-Gu Shin}
\affiliation{Center for Astrophysics $|$ Harvard \& Smithsonian, 60 Garden St.,Cambridge, MA 02138, USA}

\author[0000-0003-1525-5041]{Yossi Shvartzvald}
\affiliation{Department of Particle Physics and Astrophysics, Weizmann Institute of Science, Rehovot 7610001, Israel}

\author[0000-0001-9481-7123]{Jennifer C. Yee}
\affiliation{Center for Astrophysics $|$ Harvard \& Smithsonian, 60 Garden St.,Cambridge, MA 02138, USA}


\author[0000-0002-7511-2950]{Sang-Mok Cha}
\affiliation{Korea Astronomy and Space Science Institute, Daejeon 34055, Republic of Korea}
\affiliation{School of Space Research, Kyung Hee University, Yongin, Kyeonggi 17104, Republic of Korea} 

\author{Dong-Jin Kim}
\affiliation{Korea Astronomy and Space Science Institute, Daejeon 34055, Republic of Korea}

\author{Hyoun-Woo Kim} 
\affiliation{Korea Astronomy and Space Science Institute, Daejeon 34055, Republic of Korea}

\author[0000-0003-0562-5643]{Seung-Lee Kim}
\affiliation{Korea Astronomy and Space Science Institute, Daejeon 34055, Republic of Korea}

\author[0000-0003-0043-3925]{Chung-Uk Lee}
\affiliation{Korea Astronomy and Space Science Institute, Daejeon 34055, Republic of Korea}

\author[0009-0000-5737-0908]{Dong-Joo Lee}
\affiliation{Korea Astronomy and Space Science Institute, Daejeon 34055, Republic of Korea}

\author[0000-0001-7594-8072]{Yongseok Lee}
\affiliation{Korea Astronomy and Space Science Institute, Daejeon 34055, Republic of Korea}
\affiliation{School of Space Research, Kyung Hee University, Yongin, Kyeonggi 17104, Republic of Korea}

\author[0000-0002-6982-7722]{Byeong-Gon Park}
\affiliation{Korea Astronomy and Space Science Institute, Daejeon 34055, Republic of Korea}

\author[0000-0003-1435-3053]{Richard W. Pogge}
\affiliation{Department of Astronomy, Ohio State University, 140 West 18th Ave., Columbus, OH  43210, USA}
\affiliation{Center for Cosmology and AstroParticle Physics, Ohio State University, 191 West Woodruff Ave., Columbus, OH 43210, USA}

\collaboration{(The KMTNet Collaboration)}

\author[0000-0001-7016-1692]{Przemek Mr\'{o}z}
\affiliation{Astronomical Observatory, University of Warsaw, Al. Ujazdowskie 4, 00-478 Warszawa, Poland}

\author[0000-0002-0548-8995]{Micha{\l}~K. Szyma\'{n}ski}
\affiliation{Astronomical Observatory, University of Warsaw, Al. Ujazdowskie 4, 00-478 Warszawa, Poland}

\author[0000-0002-2335-1730]{Jan Skowron}
\affiliation{Astronomical Observatory, University of Warsaw, Al. Ujazdowskie 4, 00-478 Warszawa, Poland}

\author[0000-0002-9245-6368]{Rados\l{}aw Poleski}
\affiliation{Astronomical Observatory, University of Warsaw, Al. Ujazdowskie 4, 00-478 Warszawa, Poland}

\author[0000-0002-7777-0842]{Igor Soszy\'{n}ski}
\affiliation{Astronomical Observatory, University of Warsaw, Al. Ujazdowskie 4, 00-478 Warszawa, Poland}

\author[0000-0002-2339-5899]{Pawe{\l} Pietrukowicz}
\affiliation{Astronomical Observatory, University of Warsaw, Al. Ujazdowskie 4, 00-478 Warszawa, Poland}

\author[0000-0003-4084-880X]{Szymon Koz{\l}owski}
\affiliation{Astronomical Observatory, University of Warsaw, Al. Ujazdowskie 4, 00-478 Warszawa, Poland}

\author[0000-0001-6364-408X]{Krzysztof Ulaczyk}
\affiliation{Department of Physics, University of Warwick, Gibbet Hill Road, Coventry, CV4~7AL,~UK}

\author[0000-0002-9326-9329]{Krzysztof A. Rybicki}
\affiliation{Department of Particle Physics and Astrophysics, Weizmann Institute of Science, Rehovot 76100, Israel}
\affiliation{Astronomical Observatory, University of Warsaw, Al. Ujazdowskie 4, 00-478 Warszawa, Poland}

\author[0000-0002-6212-7221]{Patryk Iwanek}
\affiliation{Astronomical Observatory, University of Warsaw, Al. Ujazdowskie 4, 00-478 Warszawa, Poland}

\author[0000-0002-3051-274X]{Marcin Wrona}
\affiliation{Astronomical Observatory, University of Warsaw, Al. Ujazdowskie 4, 00-478 Warszawa, Poland}

\author[0000-0002-1650-1518]{Mariusz Gromadzki}
\affiliation{Astronomical Observatory, University of Warsaw, Al. Ujazdowskie 4, 00-478 Warszawa, Poland}

\collaboration{(The OGLE Collaboration)}

\author{Hanyue Wang}
\affiliation{Center for Astrophysics $|$ Harvard \& Smithsonian, 60 Garden St.,Cambridge, MA 02138, USA}

\author{Jiyuan Zhang}
\affiliation{Department of Astronomy, Tsinghua University, Beijing 100084, China}

\author[0000-0003-2337-0533]{Renkun Kuang}
\affiliation{Department of Astronomy, Tsinghua University, Beijing 100084, China}
\affiliation{Department of Engineering Physics, Tsinghua University, Beijing 100084, China}

\author{Qiyue Qian}
\affiliation{Department of Astronomy, Tsinghua University, Beijing 100084, China}

\author[0000-0003-4027-4711]{Wei Zhu}
\affiliation{Department of Astronomy, Tsinghua University, Beijing 100084, China}

\collaboration{(The MAP Collaboration)}

\begin{abstract}
We report the analysis of four unambiguous planets and one possible planet from the subprime fields ($\Gamma \leq 1~{\rm hr}^{-1}$) of the 2017 Korea Microlensing Telescope Network (KMTNet) microlensing survey, to complete the KMTNet AnomalyFinder planetary sample for the 2017 subprime fields. They are KMT-2017-BLG-0849, KMT-2017-BLG-1057, OGLE-2017-BLG-0364, and KMT-2017-BLG-2331 (unambiguous), as well as KMT-2017-BLG-0958 (possible). For the four unambiguous planets, the mean planet-host mass ratios, $q$, are $(1.0, 1.2, 4.6, 13) \times 10^{-4}$, the median planetary masses are $(6.4, 24, 76, 171)~M_{\oplus}$ and the median host masses are $(0.19, 0.57, 0.49, 0.40)~M_{\odot}$ from a Bayesian analysis. We have completed the AnomalyFinder planetary sample from the first 4-year KMTNet data (2016--2019), with 112 unambiguous planets in total, which nearly tripled the microlensing planetary sample. The ``sub-Saturn desert'' ($\log q = \left[-3.6, -3.0\right]$)  found in the 2018 and 2019 KMTNet samples is confirmed by the 2016 and 2017 KMTNet samples. 

\end{abstract}

\section{Introduction}\label{intro}

The gravitational microlensing technique is most sensitive to planets around or beyond Jupiter-like orbits \citep{Shude1991, Andy1992, BennettRhie, Mao2012, Gaudi2012}. Since 2016, The Korea Microlensing Telescope Network (KMTNet, \citealt{KMT2016}) has been conducting a microlensing survey toward the Galactic bulge to search for microlensing planets. The advent of KMTNet has played a major or decisive role in about 70\% of published microlensing planetary discoveries\footnote{\url{http://exoplanetarchive.ipac.caltech.edu} as of 2024 February 13.}. Before 2021, most KMTNet planets were found using by-eye searches and published without a systematic approach. That is, most planetary detection papers did not have a strong connection with others and were published by a systematic approach based on planets' similar properties, such as mass ratios, observing seasons, or cadences. A disadvantage of this approach is difficult to form a large-scale (${\cal O}(10^2)$) homogeneous planetary sample for statistical studies. 

The KMTNet AnomalyFinder \citep{OB191053,2019_prime} adopted the KMTNet EventFinder algorithm \citep{Gould2D, KMTeventfinder} to search for anomalies from the residuals to a point-source point-lens (PSPL, \citealt{Paczynski1986}) model. The initial motivation of \cite{OB191053} for building AnomalyFinder is to solve the ``missing planetary caustics'' problem in the KMTNet planetary sample, and later \cite{OB191053} realized that AnomalyFinder can also be a new pathway toward a large-scale homogeneous KMTNet planetary sample, besides high-magnification planetary samples from the KMTNet data only \citep{OB190960} and follow-up observations \citep{KB200414}. Then, a systematic search based on AnomalyFinder, followed by systematic analyses and publications, was conducted to the 2016--2019 KMTNet data.

In total, KMTNet monitors $\sim 97~{\rm deg}^2$ of the Galactic bulge area, including $\sim 13~{\rm deg}^2$ prime fields with cadences of $\Gamma \geq 2~{\rm hr}^{-1}$ and $\sim 84~{\rm deg}^2$ subprime fields with cadences of $\Gamma \leq 1~{\rm hr}^{-1}$. See Figure 12 of \cite{KMTeventfinder} for the field locations and cadences. Prior to the construction of a complete sample for 2017  subprime fields, which is the subject of the present work, complete samples had previously been constructed for the 2019 prime fields \citep{OB191053, KB190253,2019_prime}, the 2019 subprime fields \citep{2019_subprime}, the 2018 prime fields \citep{2018_prime, KB190253, OB180383}, the 2018 subprime fields \citep{2018_subprime}, the 2017 prime fields \citep{2017_prime}, the 2016 prime fields \citep{2016_prime}, the 2016 sub-prime fields \citep{2016_subprime}, as well as all remaining KMTNet planets with the planet-host mass ratio $q < 10^{-4}$ \citep{-4planet} from 2016 to 2019. The above references are (ignoring duplicates) Papers I, II, IV, VI, V, III, VI, X, IX, XI, and VII in the AnomalyFinder paper series. In addition, based on the $\log q > -4$ AnomalyFinder planets from the 2018 and 2019 seasons and the $\log q < -4$ AnomalyFinder planets from the 2016 to 2019 seasons, \cite{OB160007} formed a homogeneously-selected statistical sample and studied the KMTNet planetary mass-ratio function. 

For this paper, we present the complete sample of the 2017 subprime-field sample, which is the twelfth paper of the AnomalyFinder paper series. From the 2017 KNTNet subprime data, the AnomalyFinder algorithm found 3315 candidate signals, and the operator (W. Zang) identified 133 anomalous events. Of them, 10 were already published using by-eye searches, including six unambiguous planets \citep{OB171140, KB171038, Han3resonant, OB171691}, two planet candidates with the binary-lens single-source/single-lens binary-source (2L1S/1L2S, \citealt{Gaudi1998}) degeneracy \citep{KB171119}, and two finite-source point-lens (FSPL, \citealt{1994ApJ...421L..75G, Shude1994, Nemiroff1994}) events \citep{OB170896, Han_3BDs}. For the remaining events, detailed light-curve analysis shows that 14 are potentially planetary with $q_{\rm online} < 0.05$, where $q_{\rm online}$ is the mass ratio from a fitting to the online data. We further investigate them with the tender-loving care (TLC) re-reduction photometry from a careful check on the reference images and other parameters of the photometric pipeline. See \cite{Yang_TLC} for an example. We find that two are clear stellar-binary events, 10 are unambiguous planetary events, one is a 2L1S/1L2S event (Gui et al. in prep), and one is a candidate planetary event with a stellar-binary alternate interpretation. \cite{-4planet} has published three of the unambiguous planetary events, while three unambiguous planetary events, OGLE-2017-BLG-1630/MOA-2017-BLG-441/KMT-2017-BLG-1237, OGLE-2017-BLG-0668/KMT-2017-BLG-1145, and KMT-2017-BLG-2197 will be published elsewhere. Here we introduce a detailed analysis of the four remaining unambiguous planetary events and the candidate planetary event.

\section{Observations }\label{obser}

\begin{table*}
    \renewcommand\arraystretch{1.5}
    \centering
    \caption{Event Names, Alert, Locations, and Cadences for the five events analyzed in this paper}
    \begin{tabular}{c c c c c c c}
    \hline
    \hline
    Event Name & First Alert Date & ${\rm RA}_{\rm J2000}$ & ${\rm Decl.}_{\rm J2000}$ & $\ell$ & $b$ & Cadence \\
    \hline
    \eventa & Post Season & 17:41:26.27 & $-$25:16:52.39 & +2.6253 & +2.7087 & $0.4~{\rm hr}^{-1}$ \\
    \hline
    \eventb & Post Season & 17:41:56.64 & $-$22:20:06.40 & +5.1968 & +4.1592 & $0.4~{\rm hr}^{-1}$ \\
    \hline
    \eventc & 20 Mar 2017 & 17:40:54.69 & $-$34:37:03.3 & $-$5.3617 & $-$2.1315 & 0.5--1 per night \\
    KMT-2017-BLG-1396 & & & & & & $0.4~{\rm hr}^{-1}$ \\
    \hline 
    \eventd & Post Season & 17:37:37.73 & $-$29:27:40.36 & $-$1.3672 & +1.2033 & $1.0~{\rm hr}^{-1}$ \\
    \hline
    \evente & Post Season & 17:48:46.65 & $-$25:44:31.20 & +3.0954 & +1.0553 & $1.0~{\rm hr}^{-1}$ \\
    \hline
    \hline
    \end{tabular}
    \label{event_info}
\end{table*}

Table \ref{event_info} shows the basic observational
information for the seven events, including the event names, the first alert dates, event coordinates in the equatorial and galactic systems, and the observing cadences from the different groups. The event names are in order of the discovery date and we designate them by their first discovery names. 

The observations of KMTNet were conducted by its three identical 1.6 m telescopes equipped with $4~{\rm deg}^2$ cameras in Chile (KMTC), South Africa (KMTS), and Australia (KMTA). The Optical Gravitational Lensing Experiment (OGLE) took data using one 1.3 m telescope equipped with a 1.4 ${\rm deg}^2$ camera in Chile \citep{OGLEIV}. The event, \eventc, was first discovered by the Early Warning System of OGLE \citep{Udalski1994, Udalski2003}, and all of the five events were found by the KMTNet post-season EventFinder algorithm \citep{KMTeventfinder}. Most KMTNet and OGLE images were taken in the $I$-band filter, and a fraction of $V$-band images were acquired for source color measurements. To the best of our knowledge, there were no follow-up observations for any of these events, and this is certainly the case for the four events with KMTNet names.

\eventa\ was located in two overlapping KMTNet fields, BLG16 and BLG19, but at the edge of the BLG19 images, so only a few KMTC19 data are useful and the effective observing cadence is still the cadence of the BLG16 field, i.e., $0.4~{\rm hr}^{-1}$. 

The data used in the light-curve analysis were reduced using the difference imaging analysis (DIA, \citealt{Tomaney1996, Alard1998}) as implemented by each group: \cite{pysis, Yang_TLC} for KMTNet and \cite{Wozniak2000} for OGLE. The photometric error bars estimated by the DIA pipelines were recalibrated using the method outlined by \cite{MB11293}, which enables the $\chi^2$ value per degree of freedom (dof) for each data set to unity.

\section{Light-curve Analysis}\label{model}

\subsection{Preamble}\label{model_preamble}
We conduct light-curve analysis in this section, following the procedures of \cite{-4planet}. To avoid redundant descriptions, we introduce the definitions of the parameter symbols here. We refer the reader to \cite{-4planet} for more details. 

All of the events are fitted by static 2L1S models, which have seven parameters, including the three PSPL parameters, $(t_0, u_0, \te)$, i.e., the time of the closest lens-source approach, the impact parameter in units of the angular Einstein radius $\thetae$, and the Einstein timescale,
\begin{equation}\label{eqn: te}
\te = \frac{\thetae}{\mu_{\rm rel}}; \qquad \thetae = \sqrt{\kappa M_{\rm L} \pi_{\rm rel}},
\end{equation}  
where $\kappa \equiv \frac{4G}{c^2\mathrm{au}} \simeq 8.144 \frac{{\rm mas}}{M_{\odot}}$, $M_{\rm L}$ is the mass of the lens system and $(\pi_{\rm rel}, \mu_{\rm rel})$ are the lens-source relative (parallax, proper motion). Three parameters describe the binary geometry, $(s, q, \alpha)$, i.e., the planet-host projected separation scaled to $\thetae$, the planet-host mass ratio, and the angle between the source trajectory and the binary axis. The last parameter, $\rho$, denotes the ratio of the angular source radius, $\an$, to $\thetae$, i.e., $\rho=\an/\thetae$. 

We use the advanced contour integration code \citep{Bozza2010, Bozza2018}, \texttt{VBBinaryLensing}, to compute the magnification of 2L1S models, $A(t)|_{(t_0, u_0, \te, \rho, q, s, \alpha)}$, at any given time $t$. For each data set $i$, we introduce two linear flux parameters, $f_{{\rm S}, i}$ and $f_{{\rm B}, i}$, to represent the flux of the source star and any blend flux, respectively. Then, the observed flux, $f_{i}(t)$, is modeled as 
\begin{equation}
    f_{i}(t) = f_{{\rm S},i}A(t)|_{(t_0, u_0, \te, \rho, q, s, \alpha)} + f_{{\rm B},i}. 
\end{equation}

We first conduct a grid search with fixed ($\log q$, $\log s, \rho$) and ($t_0, u_0, \te, \alpha$) allowed to vary. We adopt uniform priors for ($t_0, u_0, \te, \log \rho, \alpha, \log s, \log q$). We explore the models by Markov chain Monte Carlo (MCMC) $\chi^2$ minimization using the \texttt{emcee} ensemble sampler \citep{emcee} and then search for the minimum $\chi^2$ by a downhill approach from the \texttt{SciPy} package \citep{scipy}. Then, we refine one or more local minima with all parameters free. During the fitting, $f_{{\rm S}, i}$ and $f_{{\rm B}, i}$ are not MCMC parameters and are derived by linear regression. The fitting parameter values shown below are the medians and 68\% equal-tail intervals of the marginal posterior distribution.

In some cases, we also investigate whether the microlensing parallax vector \citep{Gould1992,Gould2000,Gouldpies2004},
\begin{equation}\label{equ:pie}
    \bm{\pi}_{\rm E} \equiv \frac{\pi_{\rm rel}}{\thetae} \frac{\bm{\mu}_{\rm rel}}{\mu_{\rm rel}},
\end{equation}
can be usefully constrained by the data. We fit it by two parameters, $\pi_{\rm E,N}$ and $\pi_{\rm E,E}$, the north and east components of the microlensing parallax vector in equatorial coordinates. We also consider the lens orbital motion effect \citep{MB09387, OB09020} when including the microlensing parallax and fit the $u_0 > 0$ and $u_0 < 0$ solutions for the ``ecliptic degeneracy'' \citep{Jiang2004, Poindexter2005}. 

In cases without sharp caustic-crossing features, we also check 1L2S models because 2L1S models can be mimicked by 1L2S models \citep{Gaudi1998}. For a static 1L2S model, the total effective magnification at the waveband $\lambda$, $A_{\lambda}(t)$, can be expressed as \citep{MB12486}
\begin{equation}\label{equ: 1L2S}
    A_{\lambda}(t) = \frac{A_{1}(t)f_{{\rm S}, 1,\lambda} + A_{2}(t)f_{{\rm S}, 2,\lambda}}{f_{{\rm S}, 1,\lambda} + f_{{\rm S}, 2,\lambda}} = \frac{A_{1}(t) + q_{f,\lambda}A_{2}(t)}{1 + q_{f,\lambda}}, 
\end{equation}
\begin{equation}
    q_{f,\lambda} = \frac{f_{{\rm S}, 2,\lambda}}{f_{{\rm S}, 1,\lambda}}, 
\end{equation}
where $f_{{\rm S}, j,\lambda}$ represents the source flux at the waveband $\lambda$, $A_{j}(t)$ represents magnification of each source, $q_{f,\lambda}$ is the flux ratio between the secondary and the primary sources, and $j = 1$ and $j = 2$ correspond to the primary and the secondary sources, respectively.

\cite{OB160007} considered a model as a degenerate model if it has $\Delta\chi^2 < 10$ compared to the best-fit model, and several papers in the AnomalyFinder paper series also adopted this criterion (e.g., \citealt{2016_prime}). For the present paper, we adopt a loose criterion to exclude a model with $\Delta\chi^2 > 20$. Because we will provide $\Delta\chi^2$ for each model, one can easily test different criteria when forming a planetary sample.


\subsection{\eventa}

\begin{figure}[htb] 
    \centering
    \includegraphics[width=0.95\columnwidth]{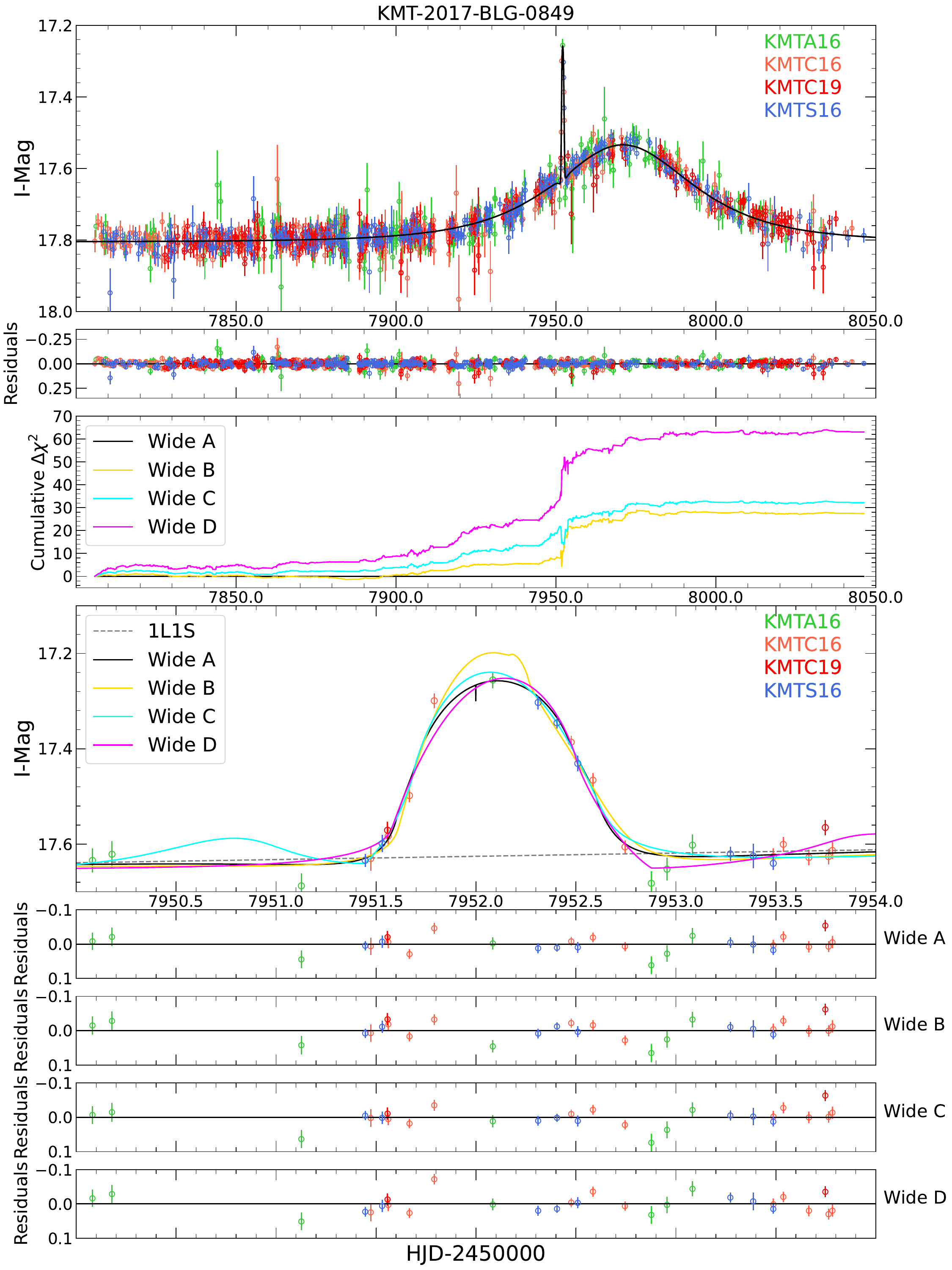}
    \caption{Observed data and the 2L1S wide (the solid black, yellow, cyan, and magenta lines) and 1L1S models (the dashed gray line) for \eventa\ and its bump-type anomaly. Different data sets are shown with different colors. The third panel shows the cumulative $\Delta\chi^2$ distribution of models relative to the ``Wide A'' model.}
    \label{lc-a-total}
\end{figure}

\begin{figure}[htb] 
    \centering
    \includegraphics[width=\columnwidth]{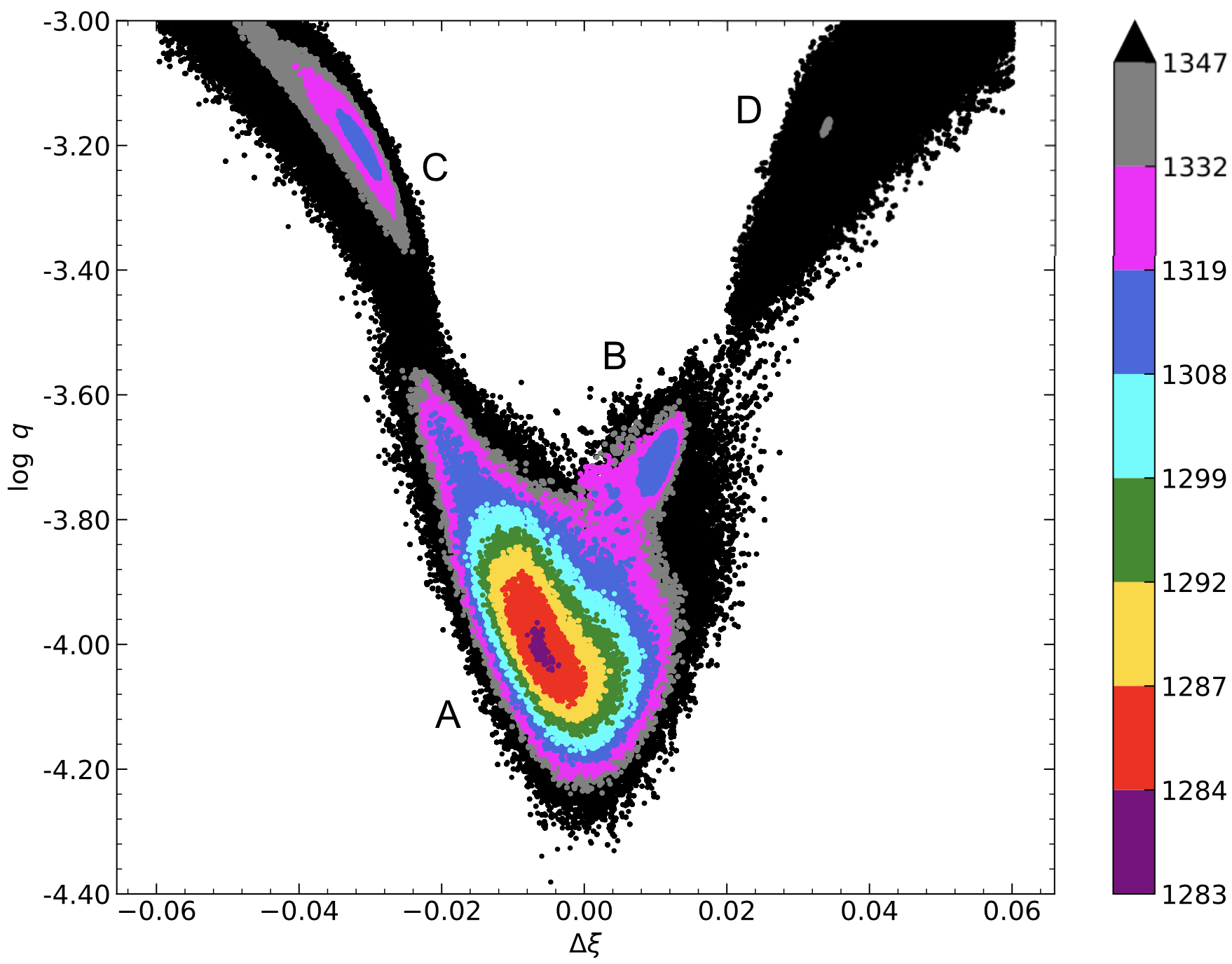}
    \caption{Scatter plot of ``hotter" MCMC of $\Delta\xi$ vs. $\log q$ of \eventa, where $\Delta\xi$ is the offset between the center of the caustic and the intersection of the source trajectory and the planet-host axis. We find $\Delta\chi^2$ barriers of $\sim 20$ between the ``Wide A'' and ``Wide B'' models, $\sim 70$ between  the ``Wide A'' and ``Wide C'' models, $\Delta\chi^2 \sim 90$ between  the ``Wide B'' and ``Wide D'' models. Color coding is (purple, red, yellow, green, cyan, blue, magenta, gray for $\Delta\chi^2<(1,\ 4,\ 9,\ 16,\ 25,\ 36,\ 49,\ 64)$. We use black dots to symbolize $\Delta\chi^2 > 64$.}
    \label{hot-a}
\end{figure}

Figure \ref{lc-a-total} displays the light curve of \eventa. The light curve exhibits a bump-type anomaly, which could, in principle, be caused by a 2L1S or a 1L2S model. We first consider the 2L1S modeling. By excluding the data over the anomaly, a PSPL fit yields ($t_0$, $u_0$, $\te$) = (7971.2, 1.10, 24.9). From Figure \ref{lc-a-total}, the anomaly occurred at $t_{\rm anom} = 7952.1$, corresponding to the lens-source offset (in units of $\thetae$) of
\begin{equation}\label{equ: uanom}
u_{\rm anom} = \sqrt{u_0^2+\left(\frac{t_{\rm anom}-t_0}{\te}\right)^2} = 1.34,
\end{equation}
and 
\begin{equation}\label{equ: alpha}
\left|\alpha\right| = \left|\sin^{-1}{\frac{u_0}{u_{\rm anom}}}\right| = 0.96~{\rm rad}.
\end{equation}
Because the planetary caustics are located at the position of $\left| s-s^{-1}\right| \sim u_{\rm anom}$, we obtain
\begin{equation}\label{equ: est_s}
s_{\pm}\sim \frac{\sqrt{u_{\rm anom}^2+4}\pm u_{\rm anom}}{2},
\end{equation}
where $s = s_{+} = 1.87$ and $s = s_{-} = 0.53$ respectively represent the major-image and minor-image planetary caustics, and we define them as the wide and close topology, respectively.

\begin{figure}[htb] 
    \centering
    \includegraphics[width=0.95\columnwidth]{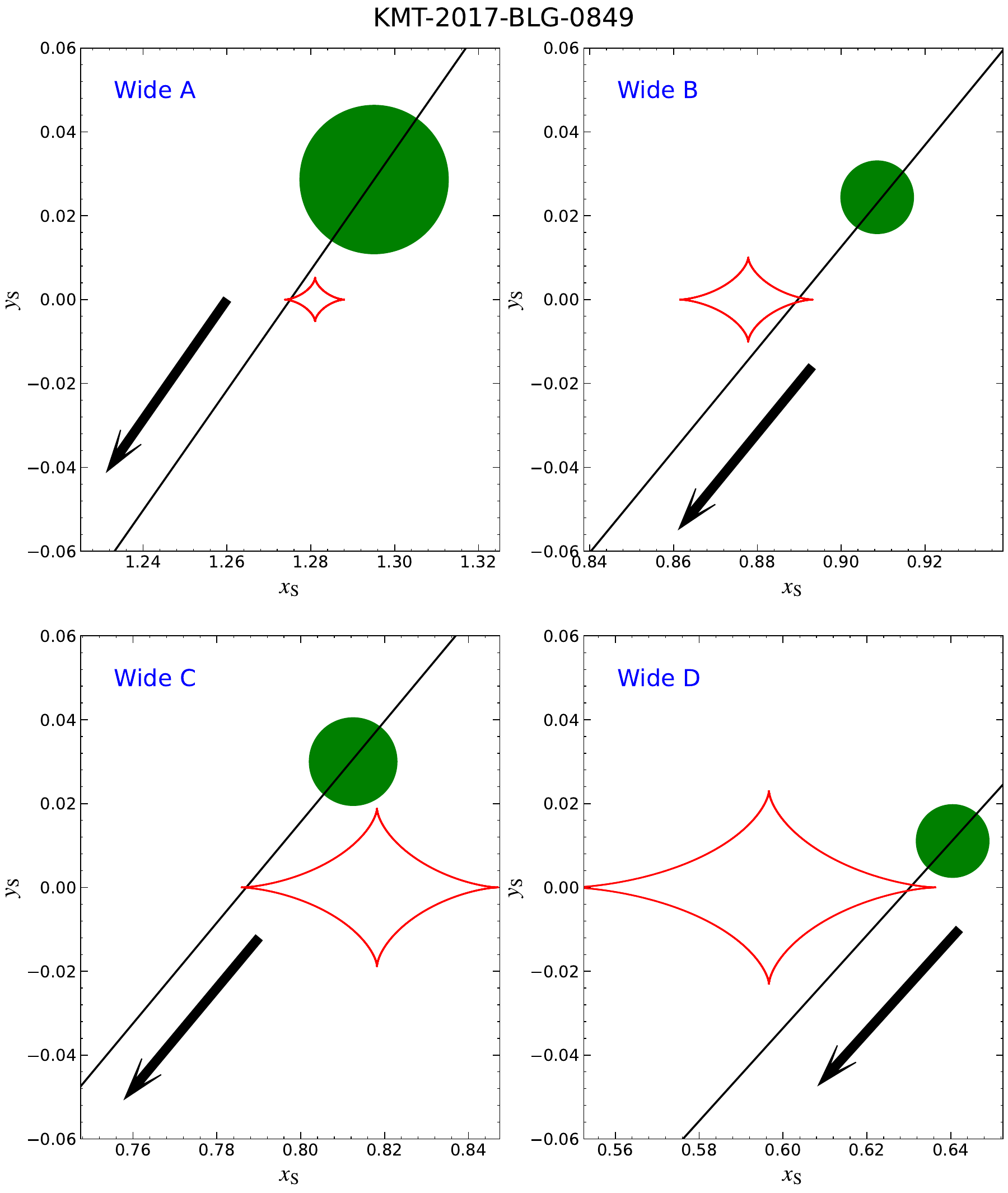}
    \includegraphics[width=0.95\columnwidth]{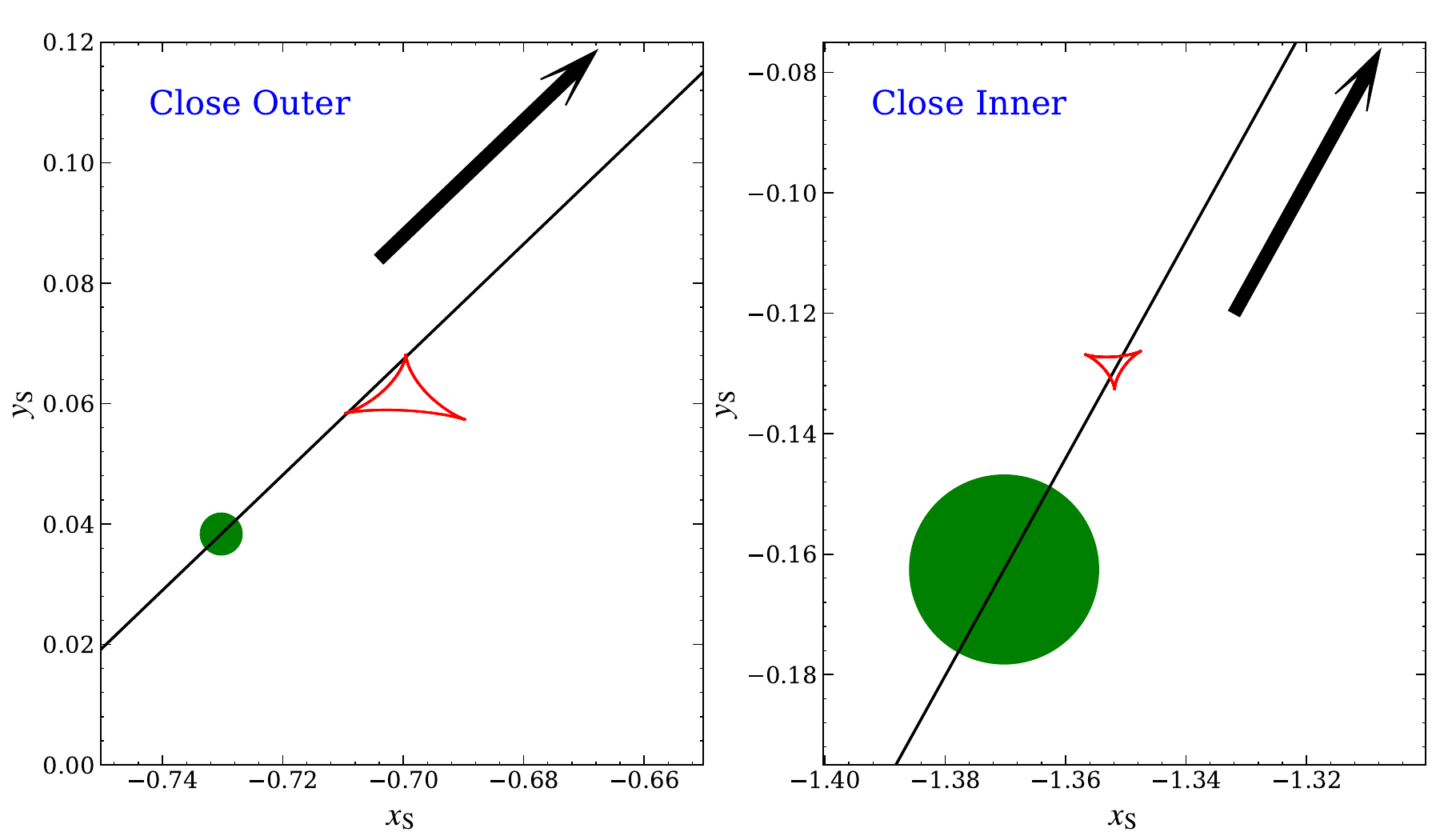}
    \caption{Caustic crossing geometries of \eventa\ models. In each panel, the red lines represent the caustics, the solid black line represents the source trajectory, the radius
    of the green dot represents the source radius, and the line with an arrow indicates the direction of the source motion.}
    \label{cau-a}
\end{figure}

\begin{table*}[htb]
    \renewcommand\arraystretch{1.10}
    \setlength{\tabcolsep}{3pt}
    \centering
    \caption{Lensing Parameters for \eventa}
    \begin{tabular}{c c c c c c c c}
    \hline
    \hline
    \multirow{2}{*}{Parameters} & \multicolumn{6}{c}{2L1S} & \multirow{2}{*}{1L2S}\\
    \cline{2-7}
    &  Wide A & Wide B & Wide C & Wide D & Close Outer & Close Inner &\\
    \hline
    $\chi^2$/dof  & $\mathbf{1283.6/1301}$ & $1310.7/1301$ & $1315.6/1301$ & $1346.3/1301$ & $1502.2/1301$ & $1603.5/1301$& $1554.2/1301$\\
    \hline
    $t_{0,1}$ (${\rm HJD}^{\prime}$)  & $\mathbf{7970.95_{-0.18}^{+0.19}}$ & $7970.95_{-0.18}^{+0.18}$ & $7970.75_{-0.19}^{+0.18}$ & $7970.75_{-0.19}^{+0.19}$ & $7971.02_{-0.18}^{+0.19}$ & $7970.87_{-0.19}^{+0.18}$ & $7972.18_{-0.17}^{+0.20}$\\
    $t_{0,2}$ (${\rm HJD}^{\prime}$)  &  &  &  &  &  &  & $7952.14_{-0.01}^{+0.01}$ \\
    $u_{0,1}$  & $\mathbf{1.00_{-0.07}^{+0.06}}$ & $0.67_{-0.03}^{+0.03}$ & $0.60_{-0.03}^{+0.03}$ & $0.47_{-0.01}^{+0.01}$ & $0.53_{-0.03}^{+0.03}$ & $1.11_{-0.03}^{+0.03}$ & $0.68_{-0.12}^{+0.17}$ \\
    $u_{0,2}$ ($10^{-3}$) &  &  &  &  &  &  & $-0.05_{{-0.68}}^{{+0.66}}$ \\
    $\te$ (days)   & $\mathbf{26.4_{-1.1}^{+1.4}}$ & $34.0_{-0.9}^{+1.1}$ & $37.2_{-1.1}^{+1.3}$ & $44.1_{-0.9}^{+0.9}$ & $40.8_{-1.4}^{+1.5}$ & $24.5_{-0.5}^{+0.6}$ & $32.8_{-4.6}^{+4.5}$\\
    $\rho_1$ ($10^{-2}$)   & $\mathbf{1.76_{-0.11}^{+0.10}}$ & $0.86_{-0.08}^{+0.08}$ & $1.06_{-0.04}^{+0.04}$ & $0.87_{-0.03}^{+0.03}$ & $0.34_{-0.02}^{+0.02}$ & $1.53_{-0.19}^{+0.07}$ &  \\
    $\rho_2$ ($10^{-2}$) &  &  &  &  &  &  & $1.09_{-0.13}^{+0.17}$ \\
    $q_{f,I} $ ($10^{-3}$) &  &  &  &  &  &  & $5.5_{-1.2}^{+1.2}$ \\
    $\alpha$ (rad)   & $\mathbf{4.095_{-0.013}^{+0.012}}$ & $4.021_{-0.012}^{+0.011}$ & $4.016_{-0.012}^{+0.012}$ & $3.978_{-0.010}^{+0.010}$ & $0.763_{-0.012}^{+0.012}$ & $1.063_{-0.008}^{+0.008}$ & \\
    $s$  & $\mathbf{1.79_{-0.06}^{+0.05}}$ & $1.52_{-0.02}^{+0.02}$ & $1.48_{-0.02}^{+0.02}$ & $1.34_{-0.01}^{+0.01}$ & $0.71_{-0.01}^{+0.01}$ & $0.53_{-0.01}^{+0.01}$ & \\
    $q (10^{-4})$  & $\mathbf{1.01_{-0.11}^{+0.15}}$ & $1.95_{-0.09}^{+0.09}$ & $6.33_{-0.42}^{+0.49}$ & $6.73_{-0.32}^{+0.32}$ & $8.81_{-0.59}^{+0.63}$ & $15.58_{-2.11}^{+1.27}$ & \\
    $\log q$ & $\mathbf{-3.996_{-0.052}^{+0.061}}$ & $-3.711_{-0.021}^{+0.019}$ & $-3.199_{-0.030}^{+0.032}$ & $-3.172_{-0.021}^{+0.020}$ & $-3.055_{-0.030}^{+0.030}$ & $-2.807_{-0.063}^{+0.034}$ & \\
    $f_{\rm S, KMTC}$ & $\mathbf{1.01_{-0.14}^{+0.14}}$ & $0.48_{-0.04}^{+0.03}$ & $0.39_{-0.03}^{+0.03}$ & $0.27_{-0.01}^{+0.01}$ & $0.32_{-0.02}^{+0.03}$ & $1.27_{-0.08}^{+0.08}$ & $0.48_{-0.13}^{+0.24}$\\
    $f_{\rm B, KMTC}$  & $\mathbf{0.19_{-0.14}^{+0.14}}$ & $0.71_{-0.03}^{+0.04}$ & $0.80_{-0.03}^{+0.03}$ & $0.92_{-0.01}^{+0.01}$ & $0.87_{-0.03}^{+0.02}$ & $-0.07_{-0.08}^{+0.08}$ & $0.72_{-0.24}^{+0.13}$\\
    \hline
    \hline
    \end{tabular}
    \tablecomments{All fluxes are on an 18th magnitude scale, e.g., $I_{\rm S} = 18 - 2.5 \log(f_{\rm S})$.}
    \label{parm-a}
\end{table*}

We first analyze the wide topology. The bump-type anomaly exhibits strong finite source effects, so the planetary caustic may be fully enveloped by a large source. According to the PSPL model and Figure \ref{lc-a-total}, the excess magnification of the anomaly is 0.63. \cite{Andy1997} showed that the excess magnification can be estimated by \begin{equation}
    \Delta A = \frac{2q}{\rho^2}. 
\end{equation}
We estimate $\rho$ using the duration of the full-width half-maximum (FWHM) of the bump, $t_{\rm fwhm} \sim 0.8$ days, and get 
\begin{equation}
    \rho \sim \frac{t_{\rm fwhm}}{2\te} \sim 0.016.
\end{equation}
Then, we obtain 
\begin{equation}
    q = \frac{\Delta A \rho^2}{2} = 8.1\times 10^{-5}.
\end{equation}
We expect that $\rho$ and $q$ are the lower limits because the caustic crossing time is $\leq2\te\rho$ for a larger source. We use the parameters obtained from the estimation above as the MCMC initial guess. 

For such a bump-type anomaly in a low-magnification event, the wide topology could have several local minima (e.g., \citealt{OB170173,OB180799}). Therefore, we investigate the wide topology by a ``hotter'' MCMC by multiplying all photometric error bars with a factor of 3.0 during MCMC. We introduce the offset between the source and the planetary caustic as the source crosses the binary axis \citep{OB170173},
\begin{equation}
    \Delta\xi = u_0\csc(\alpha) - (s - s^{-1}).
\end{equation}
Figure \ref{hot-a} shows $\Delta\xi$ vs.\ $\log q$ scatter plot for the resulting MCMC by multiplying the resulting $\chi^2$ by 9. We find four local minima and then refine all of them. We label them as ``Wide  A'', ``Wide B'', ``Wide C'', and ``Wide D'' based on $s > 1$. Table \ref{parm-a} presents the parameters for the four models, Figure \ref{lc-a-total} shows their best-fit light curves, and Figure \ref{cau-a} exhibits their caustic crossing geometries. We find that the ``Wide A'' and ``Wide C'' models cross the caustic on the left side, while the ``Wide B'' and ``Wide D'' models cross the right side. The four models follow the ``Cannae''/``von Schlieffen'' degeneracy for caustic crossing \citep{GG1997, OB170173}. That is, for the ``Wide A'' model the source fully envelops the planetary caustic (i.e., ``Cannae''), and for the other three models only one flank of the caustic is enveloped (i.e., ``von Schlieffen''). In addition, for the ``Wide C'' and ``Wide D'' models the source interacts with two ridges of the caustic separately, resulting in a small bump that abuts the observed anomaly.

For the 2L1S parameters, as expected, $\rho$ and $q$ are slightly larger than our heuristic analysis for the ``Wide A'' model because of the offset between the source trajectory from the center of the caustic, while the other three ``von Schlieffen''-type models significantly deviate from our estimate. The PSPL parameters and the source flux of the ``Wide A'' model are consistent with those obtained by excluding the anomaly, but the other three models do not exhibit such consistency. Figure \ref{lc-a-total} also shows the cumulative $\Delta\chi^2$ relative to the ``Wide A'' model of the four models, in which the other three models are disfavored not only inside the anomalous region. For most microlensing planetary events, the planetary signal itself is less significant than the microlensing signal (i.e., the host star signal), so degenerate planetary models have little influence on the PSPL parameters of the host star. However, for the present case, the maximum magnification due to the host star ($\Delta A \sim 0.3$) is only half of the maximum magnification due to the planet, so the fitting to the planetary signal affects the fitting to the host star and results in $\chi^2$ differences outside the anomalous region. 

Compared to the ``Wide A'' model, the ``Wide B'', ``Wide C'' and ``Wide D'' models are disfavored by $\Delta\chi^2 =$ 27, 32, and 63, respectively. We also try high-order effects but the three models are still disfavored by $\Delta\chi^2 > 20$, so we exclude them. 


\begin{figure}[htb] 
    \centering
    \includegraphics[width=0.95\columnwidth]{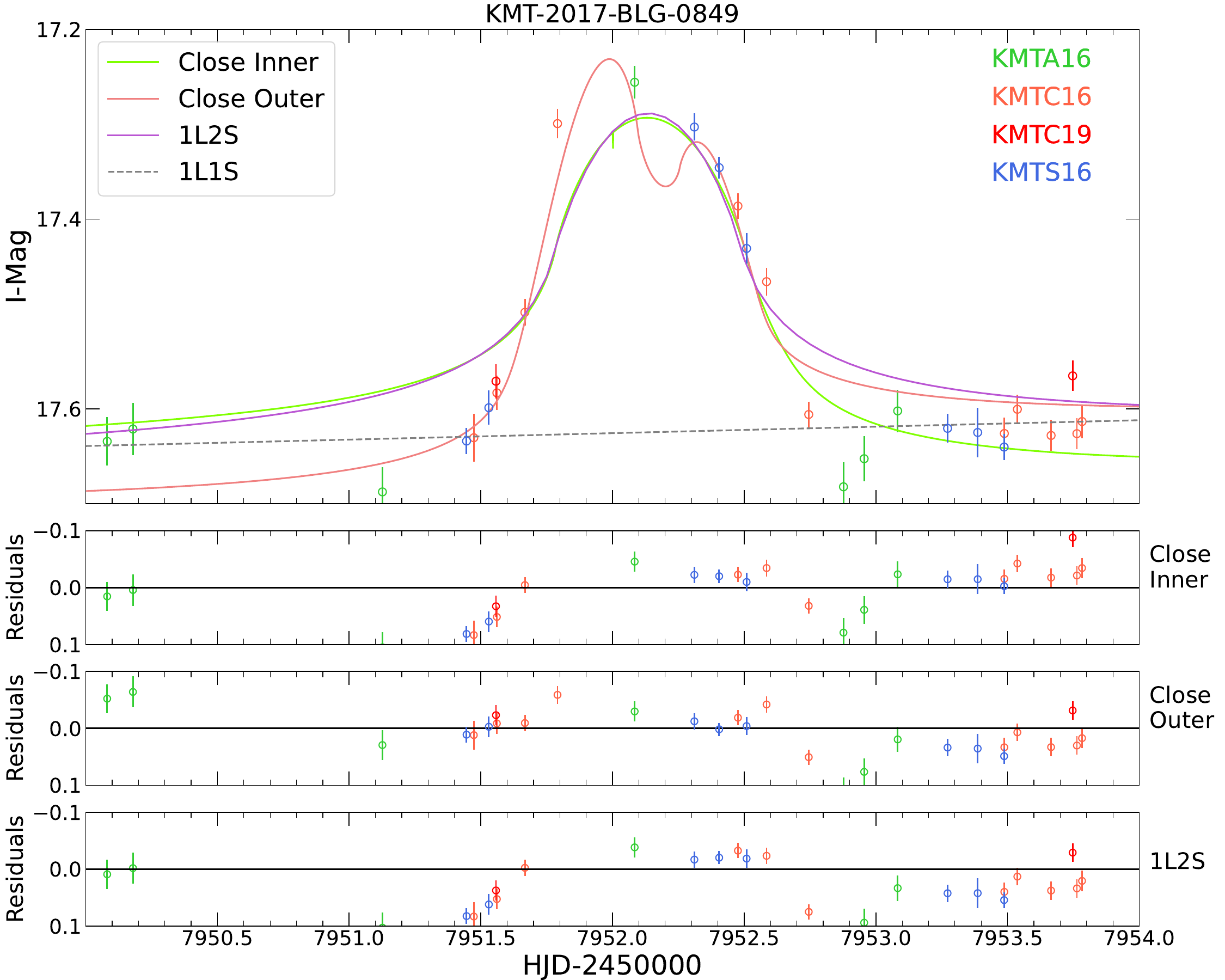}
    \caption{A close-up of the anomaly with the 2L1S close and 1L2S models of \eventa. Symbols are similar to those in Figure \ref{lc-a-total}.}
    \label{lc-a-close}
\end{figure}

For the close topology ($s < 1$), the grid search finds two models. As shown in Figure \ref{cau-a}, the source crosses the binary axis either inside or outside the planetary caustics relative to the central caustic, so we label them as the ``Close Inner'' and ``Close Outer'' models, and their parameters are given in Table \ref{parm-a}. We also try the 1L2S modeling, and the resulting parameters are shown in Table \ref{parm-a}. The two close models and the 1L2S model are disfavored by $\Delta \chi^2 > 210$ compared to the ``Wide A'' model. Figure \ref{lc-a-close} displays a close-up of the anomaly together with the three models, and the three models cannot fit the anomaly. Hence, we exclude them and only further investigate the ``Wide A'' model. The planet-host mass ratio, $q \sim 10^{-4}$, and the scaled planet-host separation, $s \sim 1.8$, indicate a low mass-ratio planet in a wide orbit. 

In addition, the inclusion of higher-order effects yields a constraint on $\pi_{\rm E, \parallel} = -0.05 \pm 0.15$, where $\pi_{\rm E, \parallel} \sim \pi_{\rm E, E}$ is the minor axes of the elliptical parallax contour and is approximately parallel with the direction of Earth’s acceleration. For the major axes of the parallax contour, $\pi_{\rm E, \bot} \sim \pi_{\rm E, N}$, there is no useful constraint because of $\sigma(\pi_{\rm E, \bot}) \sim 1$ while the typical value of $\pi_{\rm E, \bot}$ is $\sim 0.1$. We will adopt $\pi_{\rm E, \parallel}$ in the Bayesian analysis of Section \ref{lens} to estimate the lens physical parameters.

\subsection{\eventb}

\begin{figure}[htb] 
    \centering
    \includegraphics[width=0.95\columnwidth]{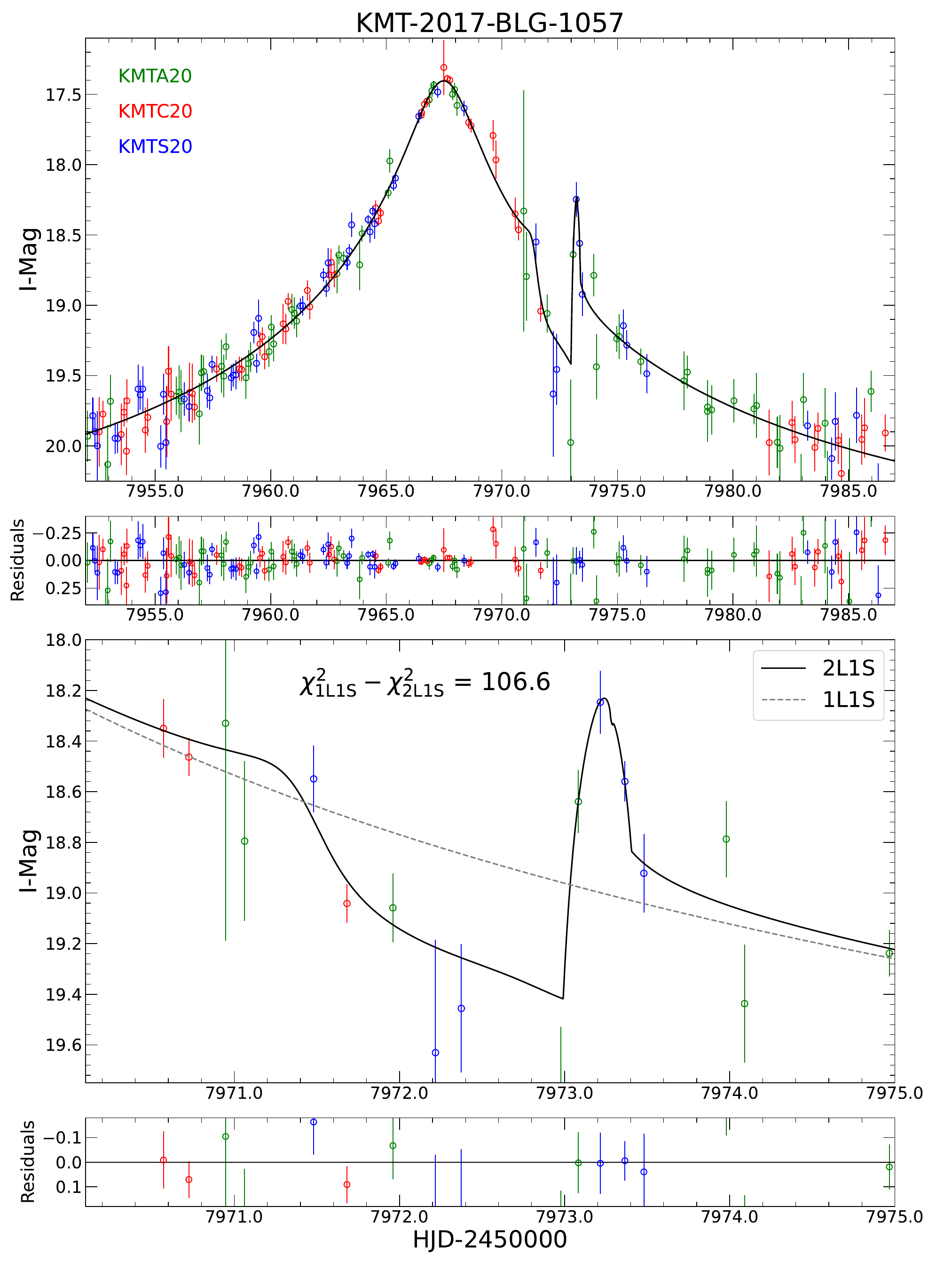}
    \caption{Observed data and the 2L1S model for \eventb. Symbols are similar to those in Figure \ref{lc-a-total}.}
    \label{lc-b}
\end{figure}

\begin{figure}[htb] 
    \centering
    \includegraphics[width=0.85\columnwidth]{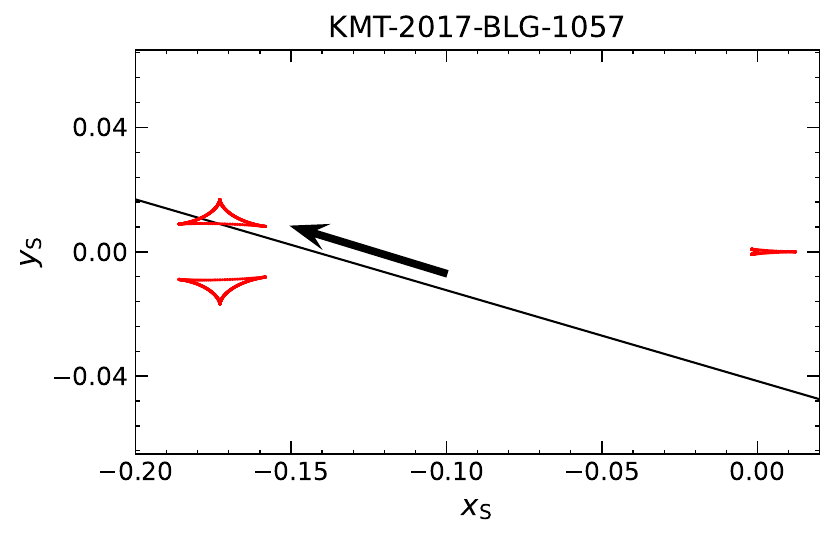}
    \includegraphics[width=0.85\columnwidth]{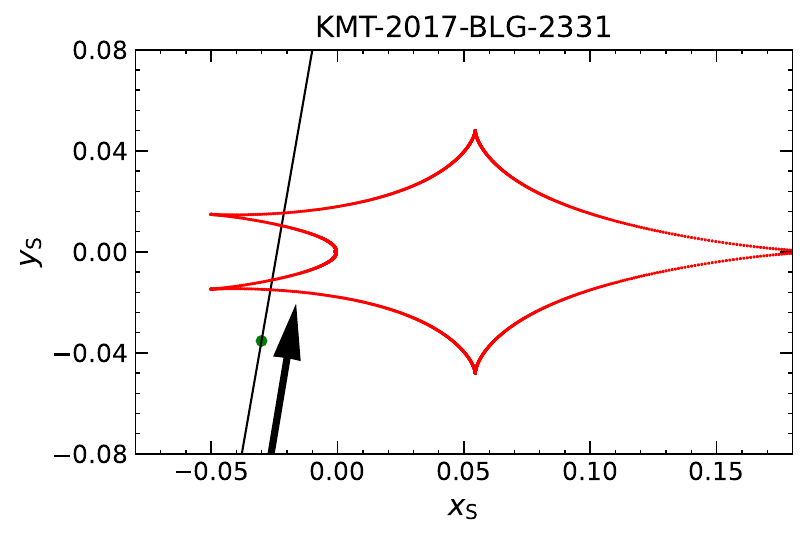}
\caption{Geometries of \eventb\ and \eventd\ models. Symbols are similar to those in Figure \ref{cau-a}.}
    \label{cau-b}
\end{figure}

Figure \ref{lc-b} shows a $\sim 1.5$-day dip five days after the peak of a PSPL model, followed by a half-day bump. Such an anomaly is likely due to a minor image perturbation and is similar to the anomaly of KMT-2017-BLG-1194 \citep{-4planet}. A grid search yields only one local minimum whose $\Delta\chi^2 < 50$ than other local minima, and further numerical analysis, including the ``hotter'' MCMC process cannot find any degenerate models. Figure \ref{cau-b} exhibits the caustic geometry and Table \ref{parm-b} presents the 2L1S parameters. The source first passes on the relatively demagnified regions between the two minor-image planetary caustics and then crosses one of the caustic, but due to the poor coverage during the caustic crossing, finite-source effects are not measured and a point-source model is consistent within $1\sigma$ level to the best-fit model. Nevertheless, we obtain a tight constraint on the upper limit of $\rho$, with $\rho < 0.0025$ at $3\sigma$, which will be used in the estimate of the lens properties.

Due to the short $\te$ and faint event, the inclusion of higher-order effects only improves the fitting by $\Delta\chi^2 < 1$ and $\sigma(\pi_{\rm E, \parallel}) > 0.4$, which is not useful for the Bayesian analysis. The mass ratio, $q = 1.15^{+0.27}_{-0.18} \times 10^{-4}$, indicates a new low-$q$ planet.

\begin{table}[htb]
    \renewcommand\arraystretch{1.10}
    \centering
    \caption{2L1S Parameters for Two Events}
    \begin{tabular}{c|c c}
    \hline
    \hline
    Parameters & \eventb & \eventd \\
    \hline
    $\chi^2$/dof & $805.29/805$ & $257.84/258$\\
    \hline
    $t_{0}$ (${\rm HJD}^{\prime}$)  & $7967.49^{+0.03}_{-0.02}$ & $8006.11 ^{+0.03}_{-0.03}$\\
    $u_{0}$  & $0.039^{+0.003}_{-0.003}$ & $0.024^{+0.003}_{-0.002}$\\
    $\te$ (days)  & $ 33.83^{+2.23}_{-2.27}$ & $ 48.09^{+3.06}_{-2.91}$\\
    $\rho$ ($10^{-3}$) & $<2.5$ & $ 2.04^{+0.32}_{-0.26}$\\
    $\alpha$ (rad) & $2.86^{+0.01}_{-0.01}$ & $1.41^{+0.03}_{-0.03}$\\
    $s$ & $0.918^{+0.005}_{-0.004}$ & $1.030^{+0.012}_{-0.010}$\\
    $q (10^{-4})$ &$1.15^{+0.27}_{-0.18}$ & $12.8^{+1.4}_{-1.1}$\\
    $\log q$ & $-3.937^{+0.091}_{-0.076}$ & $-2.893^{+0.045}_{-0.040}$\\
    $f_{\rm S, KMTC}$ & $0.068^{+0.006}_{-0.005}$ & $0.014^{+0.001}_{-0.001}$\\
    $f_{\rm B, KMTC}$ & $0.012^{+0.005}_{-0.005}$ & $0.187^{+0.001}_{-0.001}$\\
    \hline
    \hline
    \end{tabular}
    \tablecomments{The upper limit on $\rho$ is $3\sigma$.}
    \label{parm-b}
\end{table}

\subsection{\eventc}

\begin{figure}[htb] 
    \centering
    \includegraphics[width=0.95\columnwidth]{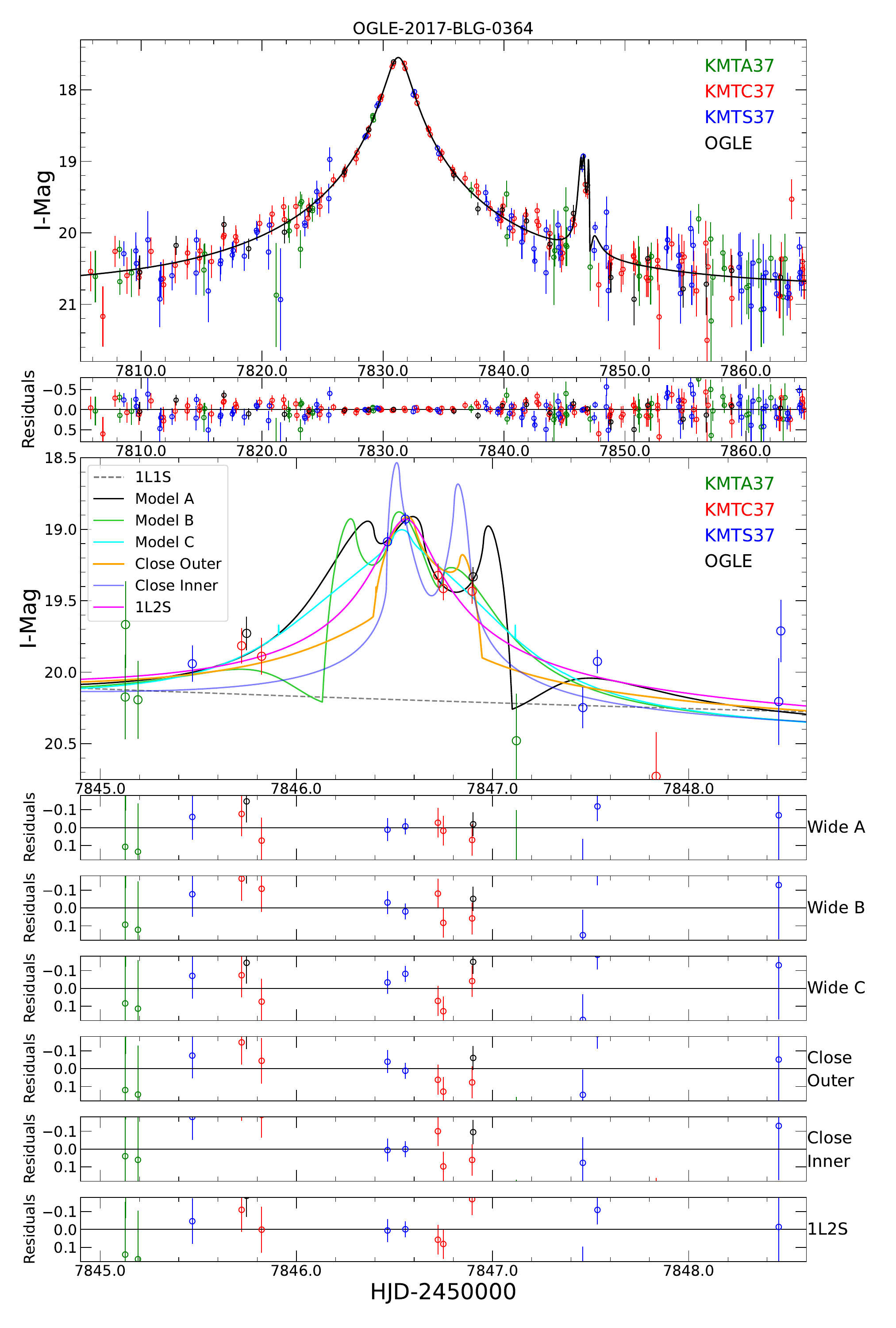}
    \caption{The observed data and the 2L1S and 1L2S models of \eventc.}
    \label{lc-c}
\end{figure}

\begin{figure}[htb] 
    \centering
    \includegraphics[width=1.05\columnwidth]{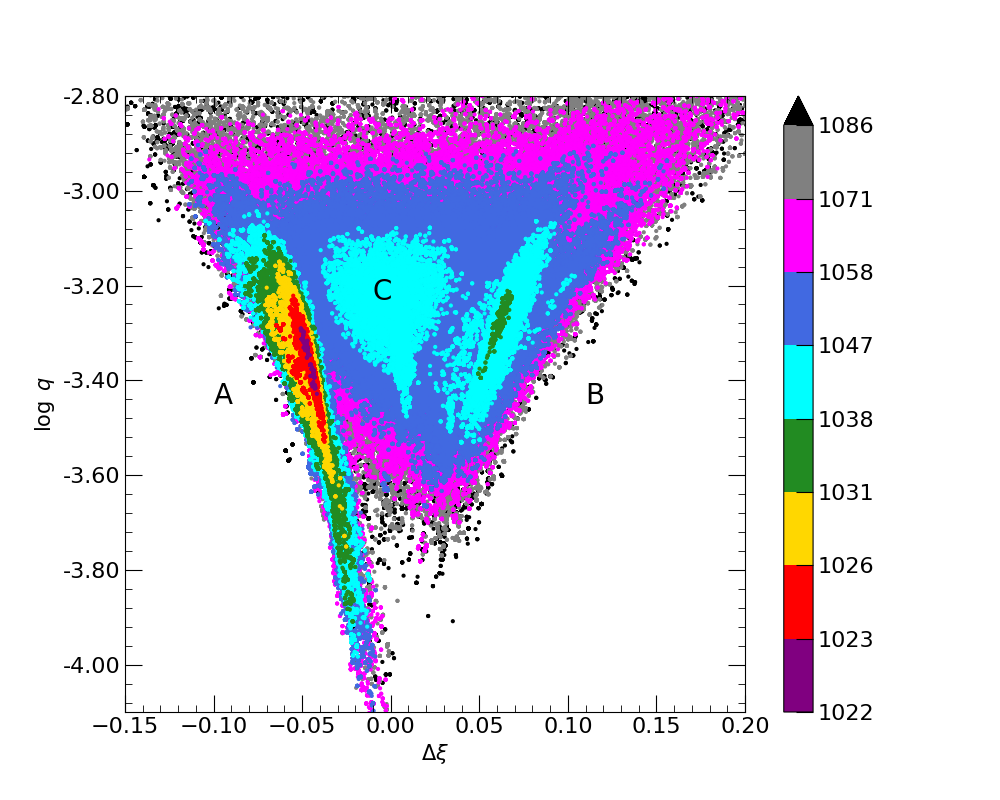}
    \caption{Scatter plot of ``hotter" MCMC of $\Delta\xi$ vs. $\log q$ for the wide models of \eventc. The distribution is derived by multiplying the photometric error bars by a factor of $\sqrt{3}$ and then multiplying the resulting $\chi^2$ by 3.0 for the plot. We find three local minima and their parameters are provided in Table \ref{parm-c}. Symbols are similar to those in Figure \ref{hot-a}.}
    \label{hot-c}
\end{figure}

\begin{figure}[htb] 
    \centering
    \includegraphics[width=1\columnwidth]{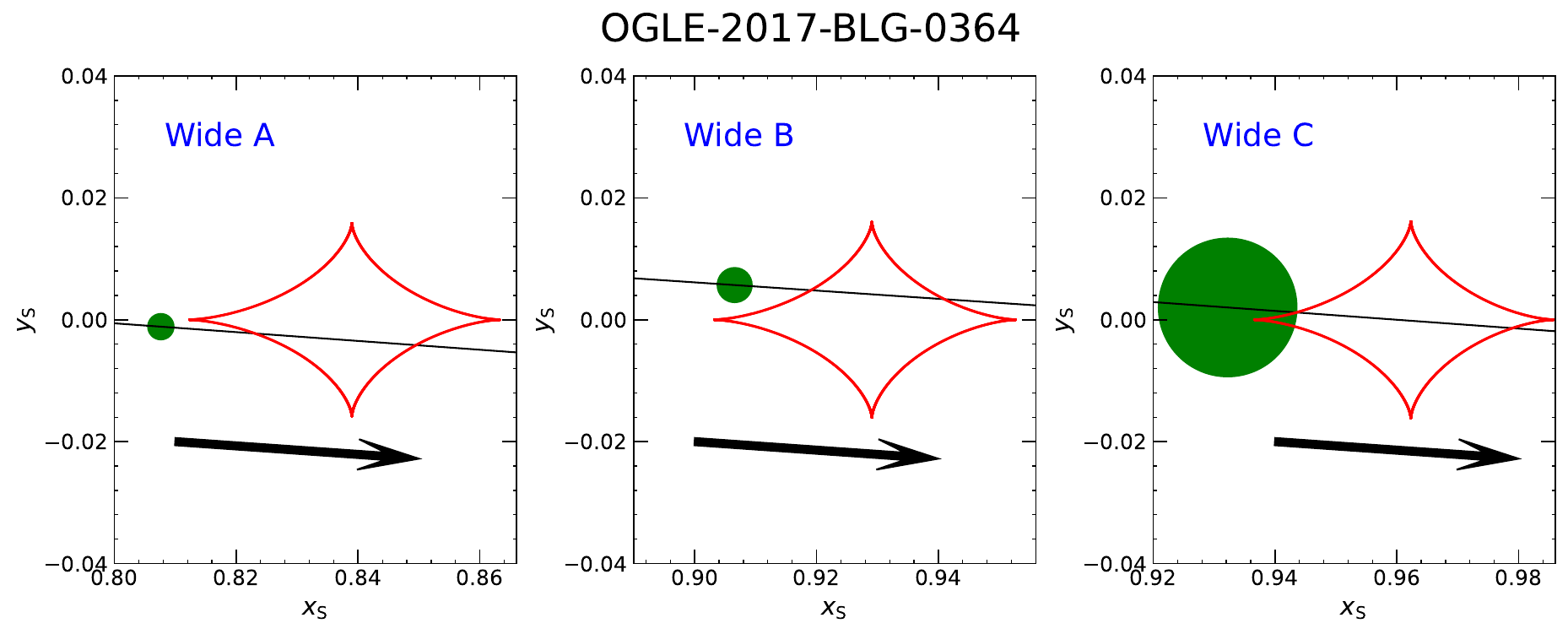}
    \includegraphics[width=0.66\columnwidth]{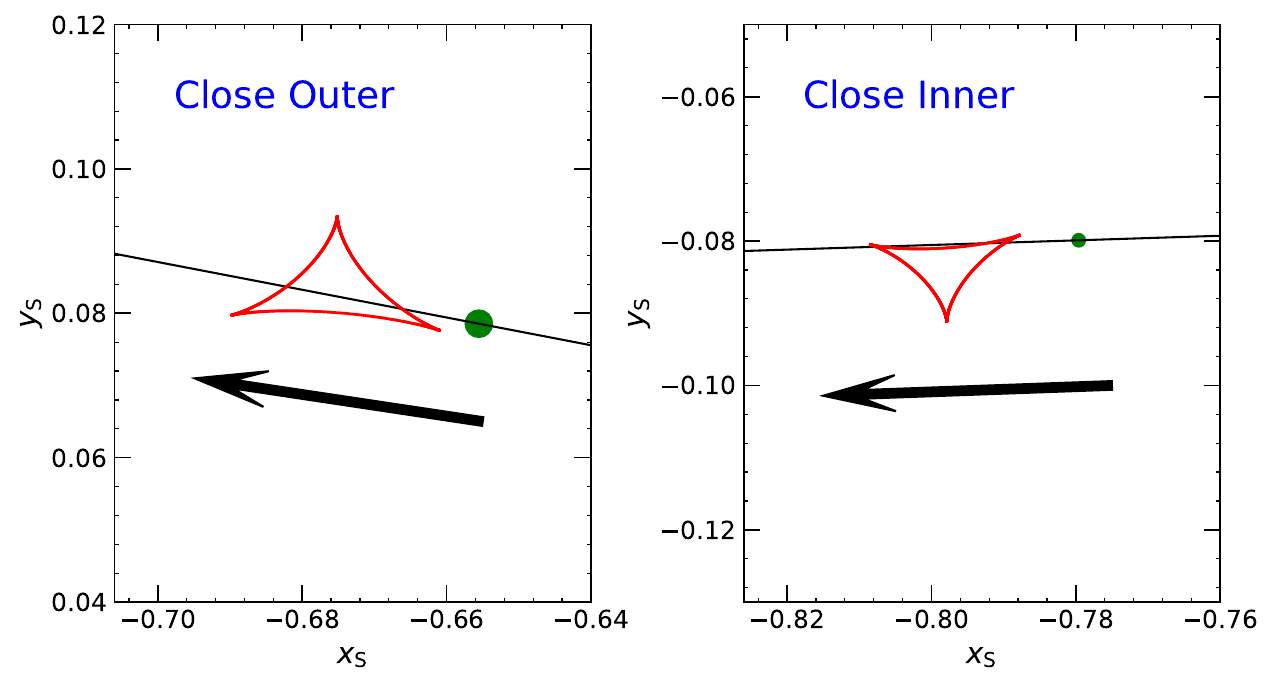}
    \caption{Caustic crossing geometries of \eventc.}
    \label{cau-c}
\end{figure}

The anomaly of \eventc, is a bump centered on $t_{\rm anom} \sim 7846.5$, as shown in Figure \ref{lc-c}. A PLPS fit by excluding the anomaly yields $(t_0,\ u_0,\ t_\mathrm{E}) = (7831.3,\ 0.05,\ 18)$. Using Equations (\ref{equ: uanom}), (\ref{equ: alpha}), and (\ref{equ: est_s}), we estimate:
\begin{equation}
    |\alpha| = 0.06~{\rm rad};\quad s_+ \sim 1.51;\quad s_- \sim 0.66.
\end{equation}
Due to the faint source, it is of low probability that the source is large enough to fully envelop the caustic, and the anomaly is not well covered, so we cannot estimate $\rho$ and $q$ from a heuristic analysis. 

A grid search identifies three local minima whose $\Delta\chi^2 < 100$ than other local minima, including two close models (i.e., ``Close Inner'' and ``Close Outer'') and one wide model. A ``hotter" MCMC analysis, as shown in Figure \ref{hot-c}, further finds three wide models, and we label them as ``Wide A'', ``Wide B'', and ``Wide C''. Their model curves are shown in Figure \ref{lc-c}, their caustic geometries are exhibited in Figure \ref{cau-c}, and the resulting parameters from the MCMC and downhill approaches are presented in Table \ref{parm-c}. The resulting $\alpha$ and $s$ are basically consistent with the estimate above. The ``Wide A'' and ``Wide B'' models are nearly symmetric, i.e., a source interacting with a ridge before and after crossing the quadrilateral caustic, respectively, so there are three peaks over the anomaly. The ``Wide C'' model passes through by the center of the caustic, with a source that is roughly the same size as the caustic, so there is only one nearly symmetric peak. For all of the models, finite-source effects are measured. 

Following our criterion of a degenerate model, we exclude the ``Close Outer'' model because of $\Delta\chi^2 = 27.2$ compared to the best-fit model, ``Wide A'' while the ``Wide B'', ``Wide C'', and ``Close Inner'' models are only disfavored by $\Delta\chi^2 = $ 14.9, 20, and 18.9, respectively. We also check the 1L2S model, and Table \ref{parm-c} lists the 1L2S parameters. We find a $\Delta\chi^2 = 32.7$ and that the 1L2S model cannot well fit the OGLE and KMTC data on the anomaly. Hence, we rule out the 1L2S possibility. For the high-order effects, there is no useful constraint, with $\sigma(\pi_{\rm E, \parallel}) > 0.3$, due to the short and faint event. 

\begin{table*}[htb]
    \renewcommand\arraystretch{1.10}
    \centering
    \caption{Lensing Parameters for \eventc}
    \begin{tabular}{c c c c c c c}
    \hline
    \hline
    \multirow{2}{*}{Parameters} & \multicolumn{5}{c}{2L1S} & \multirow{2}{*}{1L2S}\\
    \cline{2-6}
    & Wide A & Wide B & Wide C & Close Outer & Close Inner &\\
    \hline
    $\chi^2$/dof & $\mathbf{1021.8/1022}$ & $1036.7/1022$ & $1041.8/1022$ & $1049.0/1022$& $1040.7/1022$ & $1054.5/1022$\\
    \hline   
    $t_{0,1}$ (${\rm HJD}^{\prime}$)& $\mathbf{7831.27^{+0.02}_{-0.02}}$ & $7831.28^{+0.02}_{-0.02}$ & $7831.28^{+0.02}_{-0.02}$ & $7831.24^{+0.02}_{-0.02}$ & $7831.24^{+0.02}_{-0.02}$ & $7831.26^{+0.02}_{-0.02}$ \\
    $t_{0,2}$ (${\rm HJD}^{\prime}$)  &  &  &  &  &  &  $7846.56^{+0.02}_{-0.02}$ \\
    $u_{0,1}$  & $\mathbf{0.057^{+0.002}_{-0.002}}$ & $0.069^{+0.003}_{-0.004}$ & $0.072^{+0.003}_{-0.003}$ & $0.047^{+0.002}_{-0.002}$ & $0.055^{+0.002}_{-0.002}$ & $0.067^{+0.006}_{-0.006}$\\
    $u_{0,2}$  &  &  &  &  &   & $0.009^{+0.001}_{-0.001}$ \\
    $\te$ (days)  & $\mathbf{18.6^{+0.6}_{-0.4}}$ & $15.9^{+0.7}_{-0.5}$ & $15.6^{+0.5}_{-0.5}$ &$22.9^{+0.7}_{-0.7}$ & $19.3^{+0.3}_{-0.2}$ & $16.4^{+1.1}_{-1.0}$\\
    $\rho_1$ ($10^{-2}$)  & $\mathbf{0.20^{+0.03}_{-0.03}}$ & $0.31^{+0.07}_{-0.07}$ & $1.13^{+0.21}_{-0.28}$ &$0.18^{+0.03}_{-0.04}$ & $0.10^{+0.05}_{-0.02}$ &  \\
    $\rho_2$ ($10^{-2}$)  &  &  &  &  &  & $<1.9$ \\
    $q_{f,I}$ ($10^{-3}$)  &  &  &  &  &  & $25.7^{+2.7}_{-2.2}$ \\
    $\alpha$ (rad)  & $\mathbf{6.211^{+0.001}_{-0.002}}$ & $6.215^{+0.002}_{-0.001}$  & $6.211^{+0.002}_{-0.003}$& $2.951^{+0.008}_{-0.009}$ & $3.174^{+0.003}_{-0.002}$ & \\
    $s$ & $\mathbf{1.506^{+0.015}_{-0.018}}$ & $1.585^{+0.022}_{-0.030}$ & $1.603^{+0.024}_{-0.022}$ &$0.718^{+0.007}_{-0.008}$ &$0.679^{+0.003}_{-0.004}$ & \\
    $q (10^{-4})$ & $\mathbf{4.64^{+0.75}_{-0.76}}$ & $5.18^{+1.16}_{-0.92}$ & $6.05^{+1.22}_{-1.01}$ & $17.12^{+2.89}_{-2.06}$ & $13.96^{+0.97}_{-0.81}$ & \\
    $\log q$ & $\mathbf{-3.334^{+0.065}_{-0.078}}$ & $-3.286^{+0.088}_{-0.085}$ & $-3.218^{+0.080}_{-0.079}$ & $-2.767^{+0.068}_{-0.056}$ & $-2.855^{+0.029}_{-0.026}$ & \\
    $f_{\rm S, KMTC}$ & $\mathbf{0.087^{+0.002}_{-0.003}}$ & $0.106^{+0.004}_{-0.006}$ & $0.109^{+0.005}_{-0.004}$ &$0.070^{+0.003}_{-0.003}$ & $0.084^{+0.002}_{-0.002}$ & $0.103^{+0.009}_{-0.008}$\\
    $f_{\rm B, KMTC}$ & $\mathbf{-0.009^{+0.003}_{-0.002}}$ & $-0.027^{+0.006}_{-0.004}$ & $-0.030^{+0.004}_{-0.004}$ & $0.007^{+0.003}_{-0.002}$ & $-0.005^{+0.002}_{-0.002}$ & $-0.026^{+0.008}_{-0.008}$\\
    \hline
    \hline
    \end{tabular}
    \label{parm-c}
\end{table*}

\subsection{\eventd}

\begin{figure}[htb] 
    \centering
    \includegraphics[width=0.95\columnwidth]{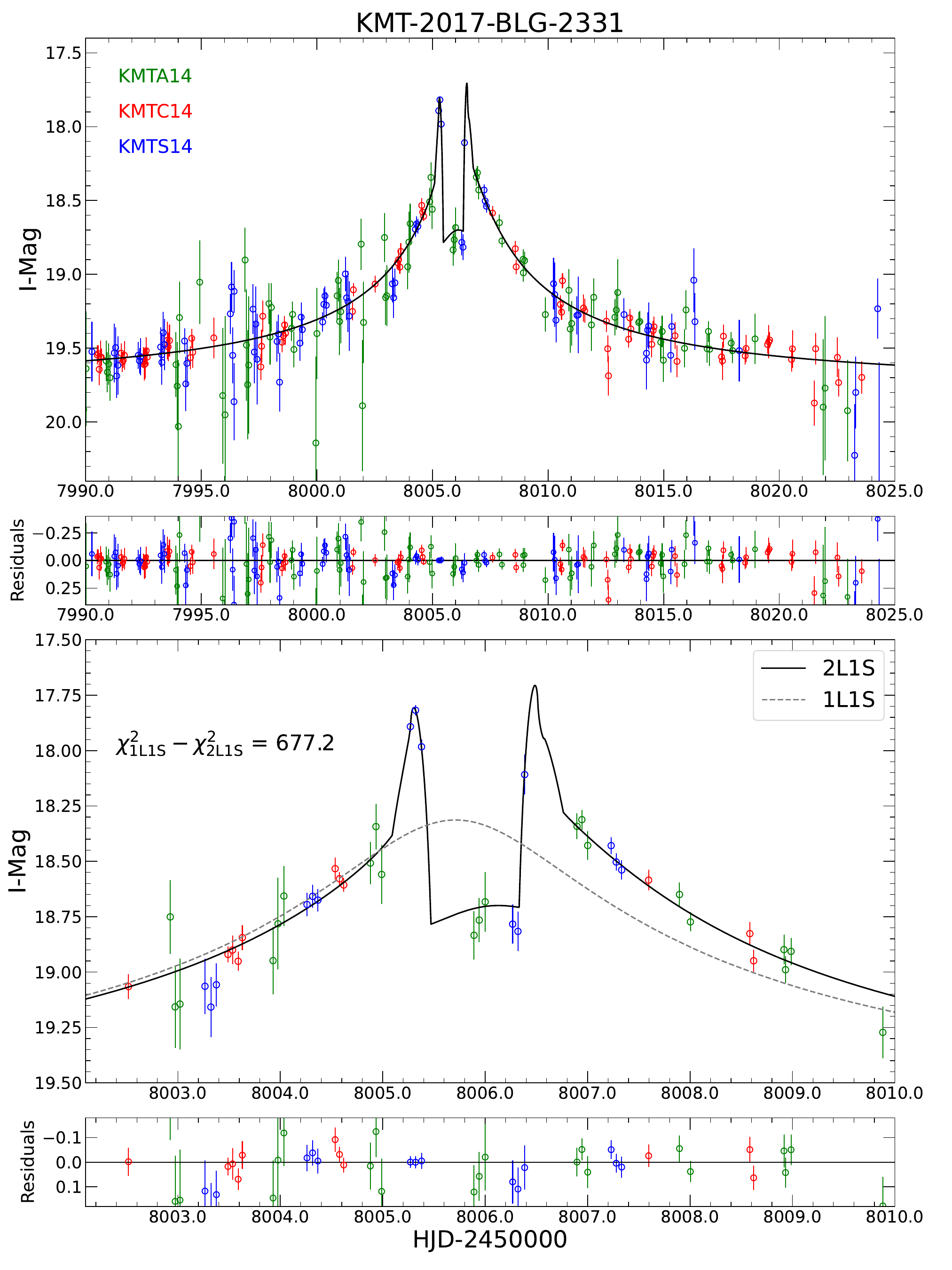}
    \caption{Observed data and the 2L1S model for \eventd.}
    \label{lc-d}
\end{figure}

As shown in Figure \ref{lc-d},  The peak of \eventd\ exhibits a double-horned profile connected by a trough, which is similar to the light curves produced by a resonant caustic (e.g., \citealt{MB13220, OB171434}) or a central caustic (e.g., \citealt{OB050071, OB050071D}). The grid search finds only one local minimum whose $\Delta\chi^2 < 100$ than other local minima and further investigation including ``hotter" MCMC does not locate any degenerate models. Figure \ref{cau-b} displays the caustic geometry for the best-fit model. As expected, the source crosses a resonant caustic. The two sharp peaks are due to the caustic crossings of two sides of the caustic, and the trough is caused by the relatively demagnified regions between them. 

Parameters from the MCMC are presented in Table \ref{parm-b}. Although there are several large gaps in the coverage of the anomaly, finite-source effects are measured, with $\rho = 2.04^{+0.32}_{-0.26} \times 10^{-3}$. This is a new Jovian mass-ratio planet. Due to the faintness of the event, we do not get a useful constraint on $\bm{\pi}_{\rm E}$.

\subsection{\evente}

\begin{figure}[htb] 
    \centering
    \includegraphics[width=0.95\columnwidth]{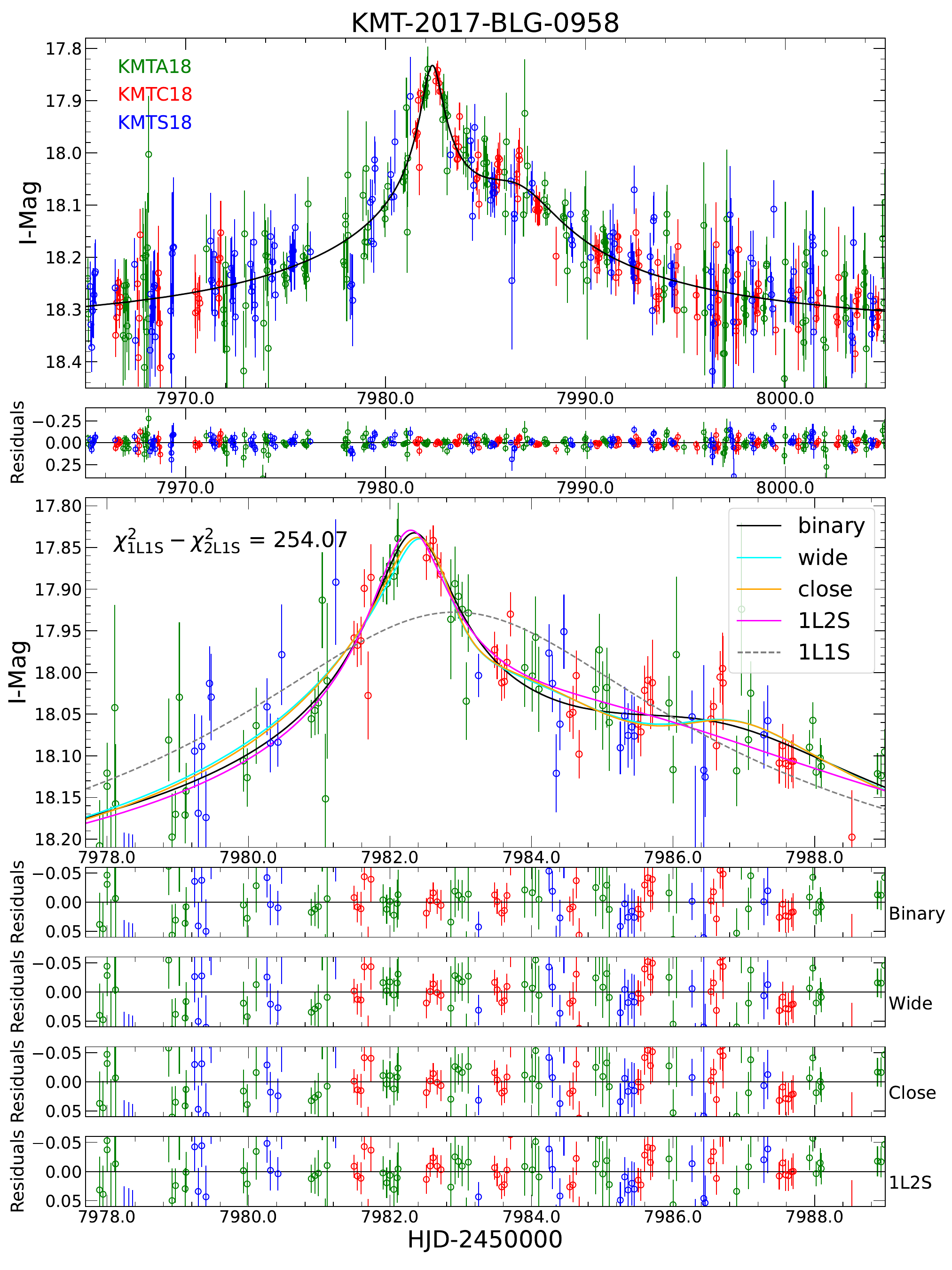}
    \caption{Observed data together with three 2L1S models and the 1L2S model for \evente. All models can fit the data well and thus this is only a candidate planetary event.}
    \label{lc-e}
\end{figure}

\begin{figure}[htb] 
    \centering
    \includegraphics[width=0.85\columnwidth]
    {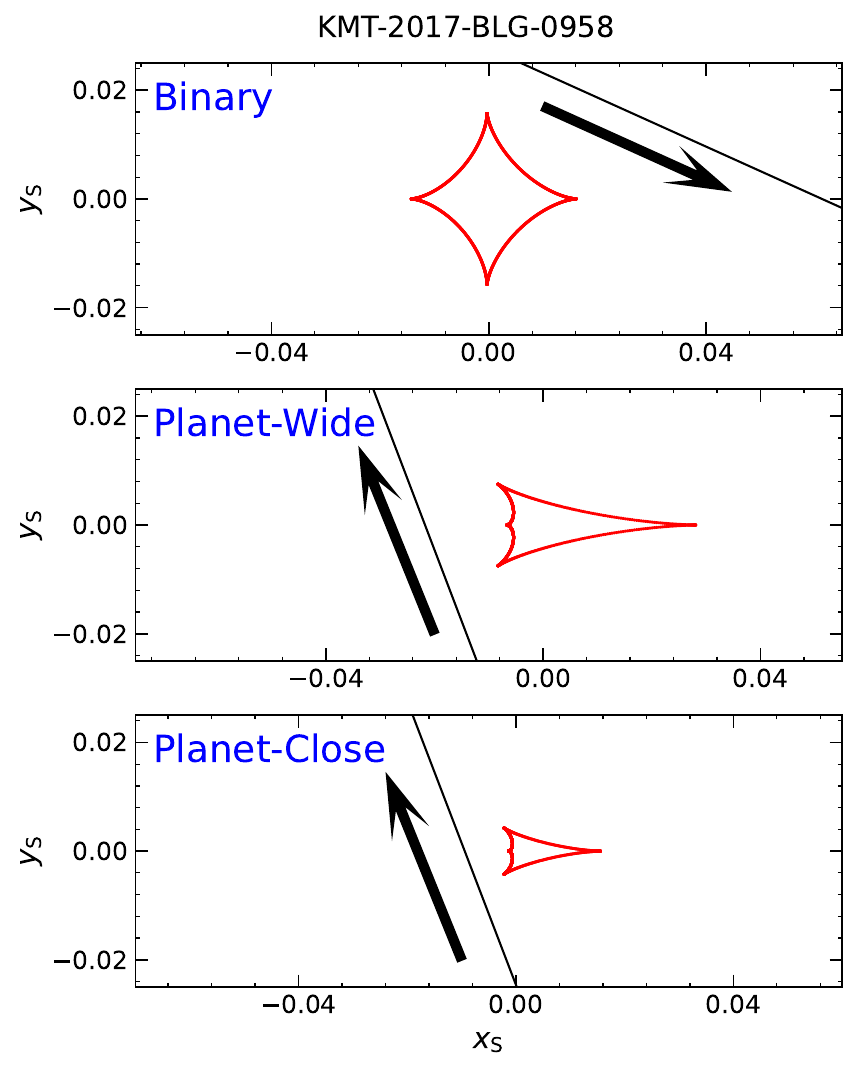}
    \caption{Caustic geometries of the candidate planetary event, \evente.}
    \label{cau-e}
\end{figure}

The anomaly of \evente\ is a two-bump feature, centered on $t = 7982$ and $t = 7987$, respectively, as shown in Figure \ref{lc-e}. Both a 2L1S and a 1L2S model can produce such an anomaly. For the 2L1S modeling, we find three degenerate models. The first model has a stellar-binary mass ratio and provides the best fit to the observed data. The other two models have a super-Jovian mass ratio and are disfavored by $\Delta\chi^2 = 6$, and we label them as ``Planet-Close'' ($s < 1$) and ``Planet-Wide'' ($s > 1$). Figure \ref{cau-e} displays the caustic geometries, and Table \ref{parm-e} presents the lensing parameters. Due to the faintness of the event, neither of the models has a good constraint on $\te$. Finite-source effects are not measured for any models. 

We also check the 1L2S model and find it disfavored by only $\Delta\chi^2 = 15$ to the best-fit 2L1S model, and even when we impose $\rho_2 = 0$ the 1L2S model is still disfavored by only $\Delta\chi^2 = 16$, so we cannot rule out the 1L2S model by a kinematic argument that the resulting $\mu_{\rm rel}$ is unlikely. Due to the severe extinction for this event, with $A_K = 0.87$ \citep{Gonzalez2012}, so the $V$-band data have no microlensing signal and we cannot exclude the 1L2S model by a color argument for different source colors \citep{Gaudi1998}.

In summary, we conclude that the lens-source system could be either 2L1S or 1L2S and if the latter a stellar-binary interpretation is preferred. Because this is a candidate planetary event, we do not conduct further analysis.

\begin{table*}[htb]
    \renewcommand\arraystretch{1.10}
    \setlength{\tabcolsep}{20pt}
    \centering
    \caption{Lensing Parameters for \evente}
    \begin{tabular}{c c c c c}
    \hline
    \hline
    \multirow{2}{*}{Parameters} & \multicolumn{3}{c}{2L1S} & \multirow{2}{*}{1L2S} \\ 
    \cline{2-4}
     & Binary & Planet-Wide & Planet-Close & \\
    \hline
    $\chi^2$/dof & $2570.03/2570$ & $2575.63/2570$& $2575.53/2570$& $2584.69/2570$\\
    \hline
    $t_{0,1}$ (${\rm HJD}^{\prime}$)  & $7983.58^{+0.09}_{-0.08}$& $7983.30^{+0.07}_{-0.07}$&$7983.56^{+0.07}_{-0.08}$&$7984.81^{+0.39}_{-0.31}$\\
    $t_{0,2}$ (${\rm HJD}^{\prime}$)  &  &  & & $7982.29^{+0.04}_{-0.04}$\\
    $u_{0,1}$  & $0.030^{+0.008}_{-0.006}$&$0.023^{+0.007}_{-0.004}$&$0.008^{+0.002}_{-0.002}$& $0.074^{+0.023}_{-0.021}$\\
    $u_{0,2}$  &  & & & $-0.009^{+0.002}_{-0.003}$\\
    $\te$ (days) & $97.10^{+19.90}_{-15.47}$ & $134.59^{+32.66}_{-29.11}$ & $291.83^{+70.11}_{-56.05}$ & $59.58^{+20.02}_{-12.46}$\\
    $\rho_1$ ($10^{-3}$) & $<11$&$<8$&$<4$ &  \\
    $\rho_2$ ($10^{-3}$) & & & & $<30$ \\
    $q_{f,I}$ &  & & & $0.17^{+0.05}_{-0.04}$\\
    $\alpha$ (rad) & $-0.44^{+0.04}_{-0.05}$& $1.94^{+0.02}_{-0.02}$& $ 1.94^{+0.03}_{-0.02}$& \\
    $s$ & $0.20^{+0.03}_{-0.02}$ & $1.50^{+0.04}_{-0.04}$ & $0.65^{+0.02}_{-0.02}$ & \\
    $q (10^{-3})$ & $0.96^{+1.29}_{-0.47}\times10^{3}$ &  $6.19^{+1.89}_{-1.24}$ & $2.89^{+0.67}_{-0.60}$& \\
    $\log q$ & $-0.020^{+0.371}_{-0.294}$&$-2.209^{+0.116}_{-0.097}$ & $-2.540^{+0.091}_{-0.100}$ & \\
    $f_{\rm S, KMTC}$ & $0.009^{+0.002}_{-0.002}$ & $0.006 ^{+0.002}_{-0.001}$ & $0.003^{+0.001}_{-0.001}$ & $0.018^{+0.006}_{-0.005}$\\
    $f_{\rm B, KMTC}$ & $0.711^{+0.001}_{-0.002}$ & $0.715^{+0.001}_{-0.001}$ & $0.716^{+0.001}_{-0.001}$ & $0.705^{+0.004}_{-0.005}$\\
    \hline
    \hline
    \end{tabular}
    \label{parm-e}
\end{table*}

\section{Lens Properties}\label{lens}

\subsection{Preamble}

\begin{figure*}[htb] 
    \centering
    \includegraphics[width=0.45\textwidth]{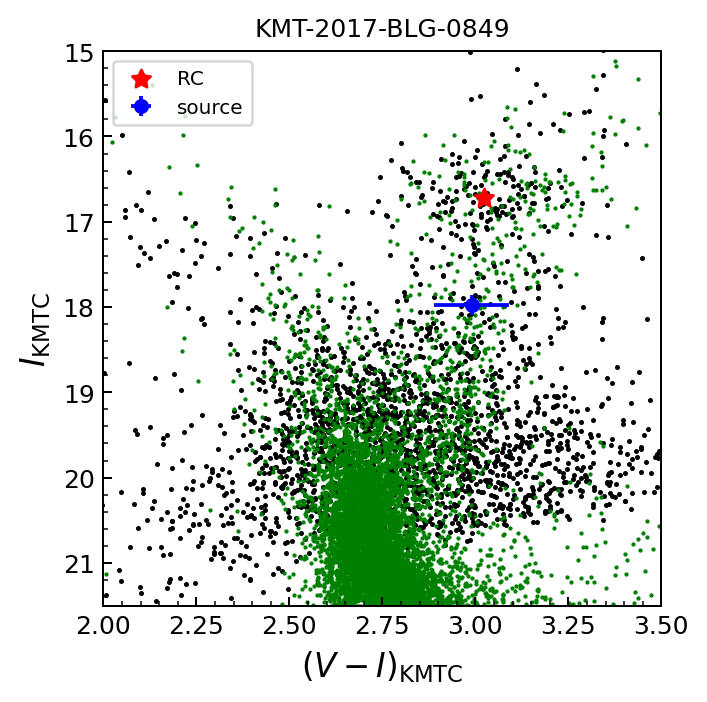}
    \includegraphics[width=0.45\textwidth]{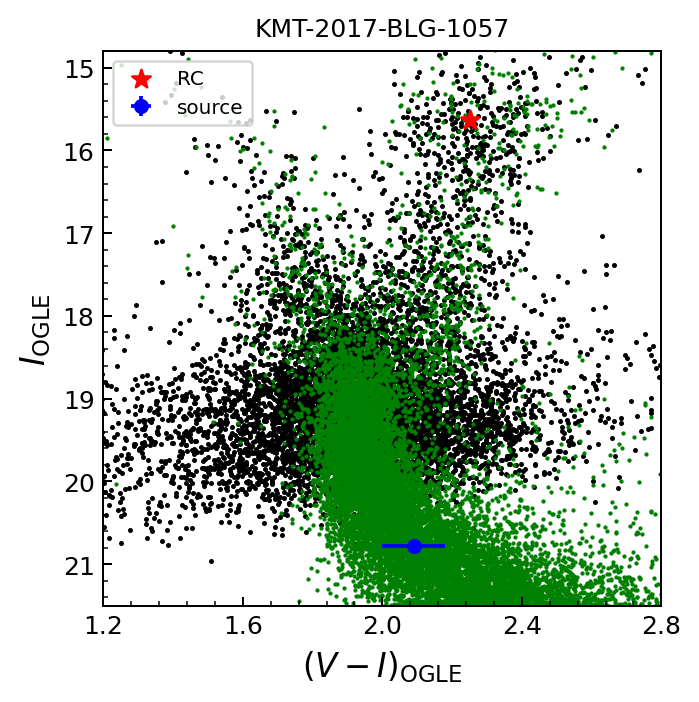}
    \includegraphics[width=0.45\textwidth]{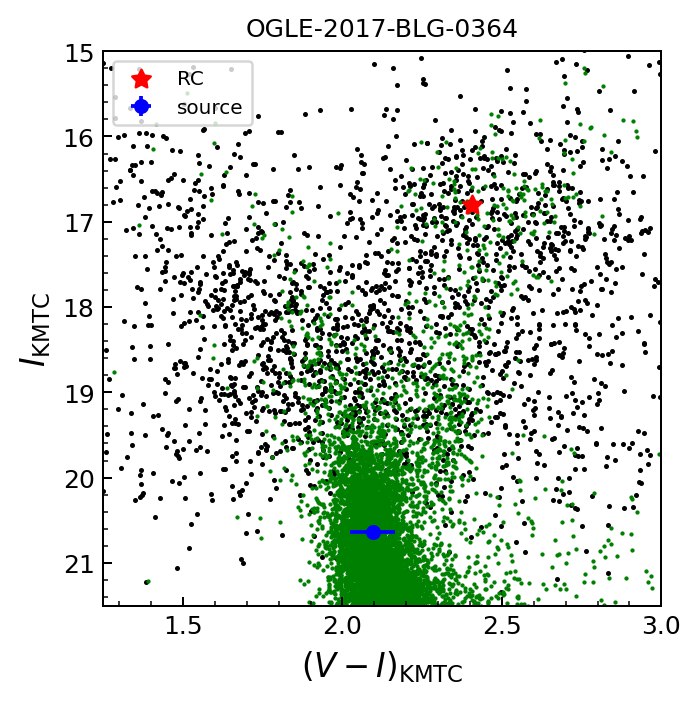}
    \includegraphics[width=0.45\textwidth]{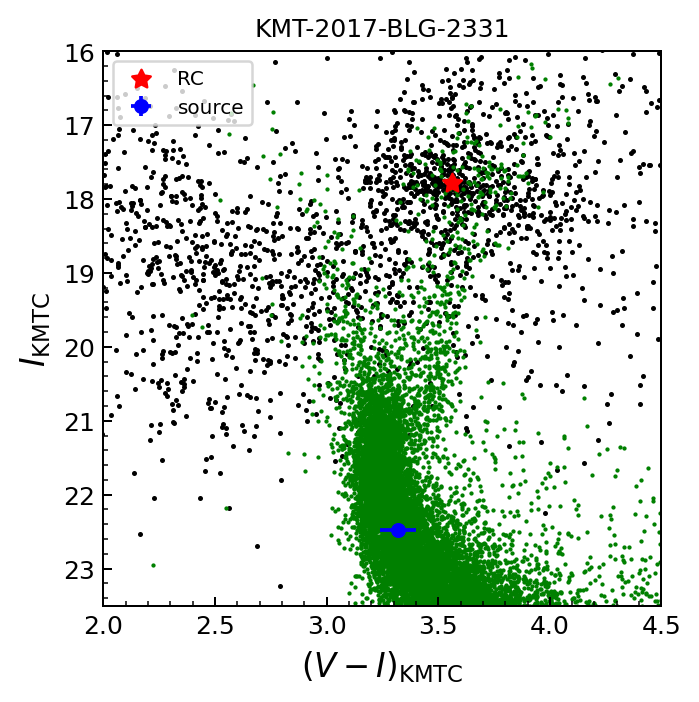}
    \caption{Color-magnitude diagram for the four unambiguous planetary events. The CMDs of \eventa, \eventb, and \eventd\ are constructed using the KMTC field stars, and the CMD of \eventc\ is built using the OGLE-III star catalog \citep{OGLEIII}. For each panel, the red asterisk and the blue dot represent the centroid of the red-giant clump and the source star, respectively. The green dots represent {\it HST} CMD of \cite{HSTCMD}, whose centroid of the red-giant clump $(V - I, I)_{\rm cl, HST} = (1.62, 15.15)$ \citep{MB07192} is matched to that of KMTC or OGLE-III.}
    \label{CMD}
\end{figure*}

We estimate the lens properties in this section. As in Section \ref{model_preamble}, we first introduce the common processes. From Equations (\ref{eqn: te}) and (\ref{equ:pie}), the lens mass, $M_\mathrm{L}$, and the lens distance, $D_\mathrm{L}$, are related to the angular Einstein radius and the microlensing parallax by \citep{Gould1992, Gould2000}:
\begin{equation}\label{eq:mass}
    M_{\rm L} = \frac{\thetae}{{\kappa}\pie};\qquad D_{\rm L} = \frac{\mathrm{au}}{\pie\thetae + \pi_{\rm S}},
\end{equation}
where $\pi_{\rm S}$ is the source parallax. 

The angular Einstein radius can be derived through $\thetae =  \an/\rho$. To estimate $\an$, we locate the source on a color–magnitude diagram (CMD, \citealt{Yoo2004}), which is constructed from the ambient stars around the event. From the CMD, we estimate the centroid of the red-giant clump as ${(V - I, I)}_{\rm cl}$, for which the de-reddened color and magnitude, $(V - I, I)_{\rm cl,0}$, are adopted from \cite{Bensby2013} and Table 1 of \cite{Nataf2013}, respectively. The source apparent magnitude is from the light-curve analysis. For the source color, which is independent of the 2L1S model, neither event has sufficient $V$-band signal-to-noise ratio to determine the source color by a regression of the KMTC $V$ versus $I$ flux. Therefore, we calibrate the {\it Hubble} Space Telescope ({\it HST}) CMD of \cite{HSTCMD} to the KMTC CMD using $I_{\rm cl}$ of red-giant clumps and then estimate the source color by the color of {\it HST} field stars within the 5-$\sigma$ brightness of the source star. Using the color–surface brightness relations of \cite{Adams2018}, we obtain the angular source radius $\theta_*$. Then, we derive $\thetae$ by $\an/\rho$ and $\mu_{\rm rel}$ by $\thetae/\te$. 

Figure \ref{CMD} displays the CMD of the four planetary events. Table \ref{source} summarizes the CMD parameters, and the resulting $\theta_*$, $\thetae$ and $\mu_{\rm rel}$. 

For the microlensing parallax, only \eventa\ has a useful constraint on $\bm{\pi}_{\rm E}$. Therefore, we estimate the physical parameters of the planetary systems from a Bayesian analysis using a Galactic model as priors. The Galactic model and the basic procedures are the same as used by \cite{Yang2021_GalacticModel}, assuming that the planetary occurrence rate is independent of the host-star properties (e.g., host mass). We refer the reader to that work for details. The only additional procedure in our Bayesian analysis is that we exclude trial events for which the lens flux exceeds the upper limits of the lens flux, $I_{\rm L, limit}$, obtained from the light-curve and imaging analysis. We adopt the mass-luminosity relation of \cite{OB171130}.

\begin{figure*}[htb] 
    \centering
    \includegraphics[width=0.7\textwidth]{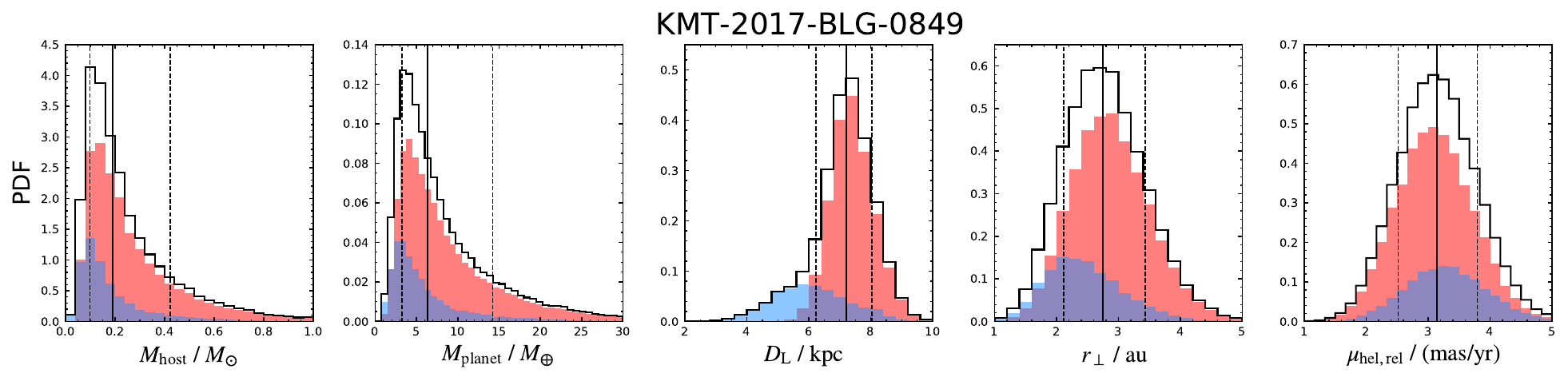}
    \includegraphics[width=0.7\textwidth]{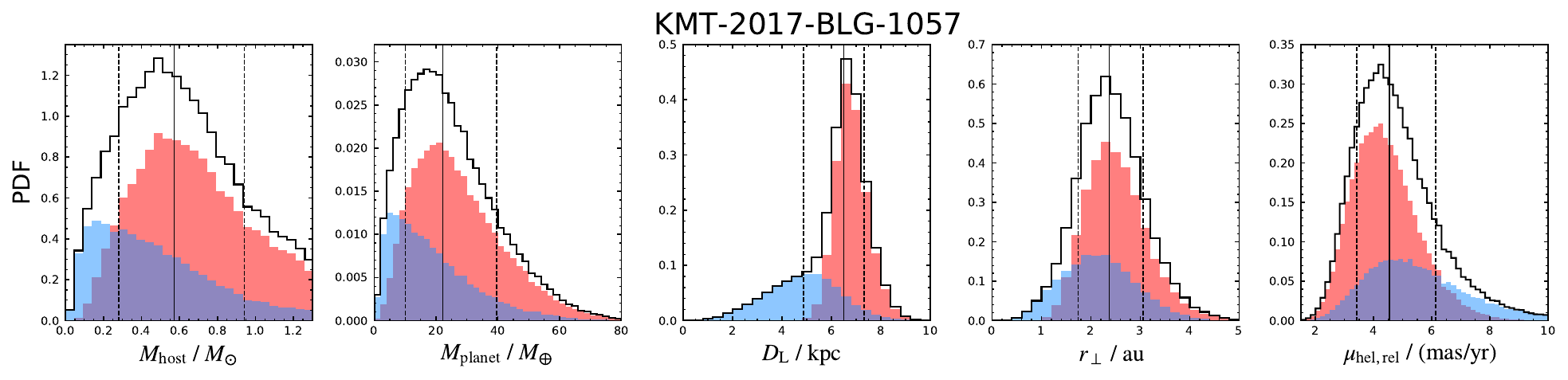}
    \includegraphics[width=0.7\textwidth]{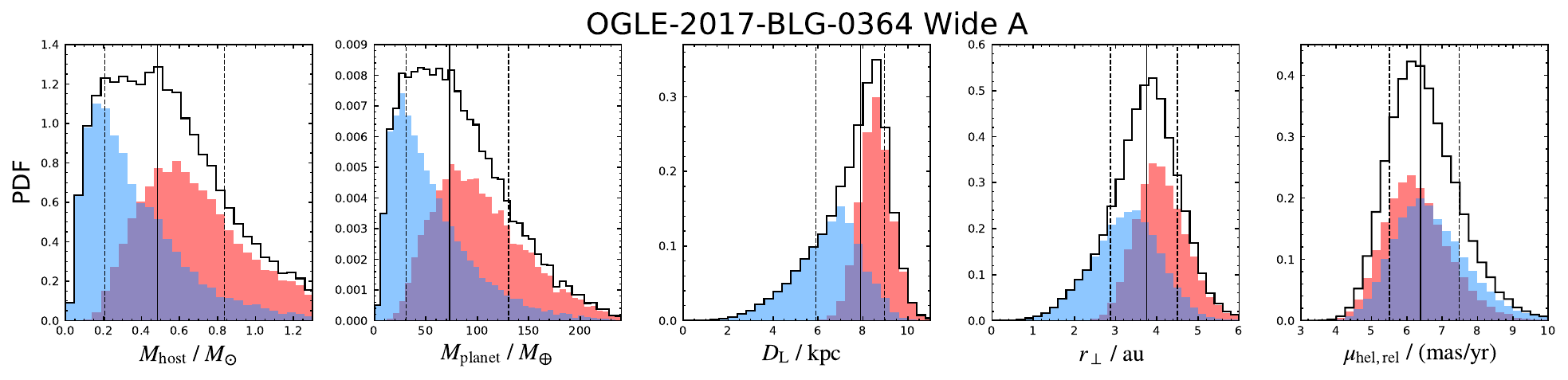}
    \includegraphics[width=0.7\textwidth]{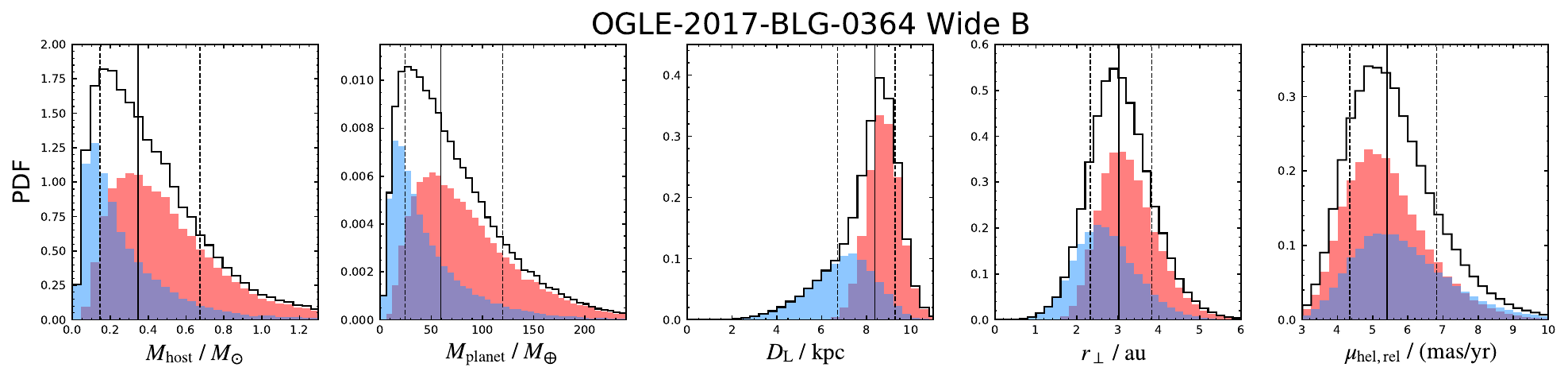}
    \includegraphics[width=0.7\textwidth]{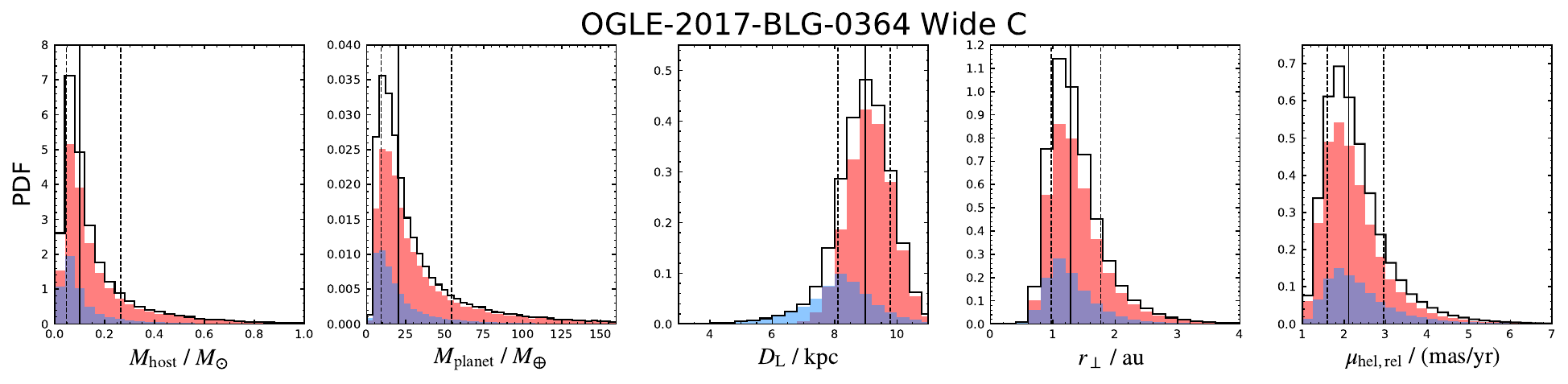}
    \includegraphics[width=0.7\textwidth]{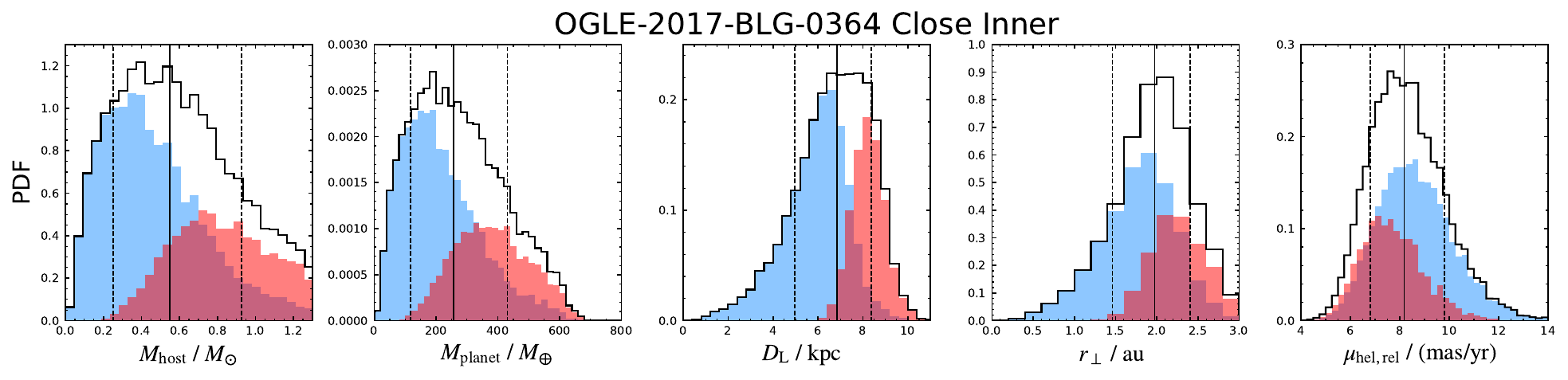}
    \includegraphics[width=0.7\textwidth]
    {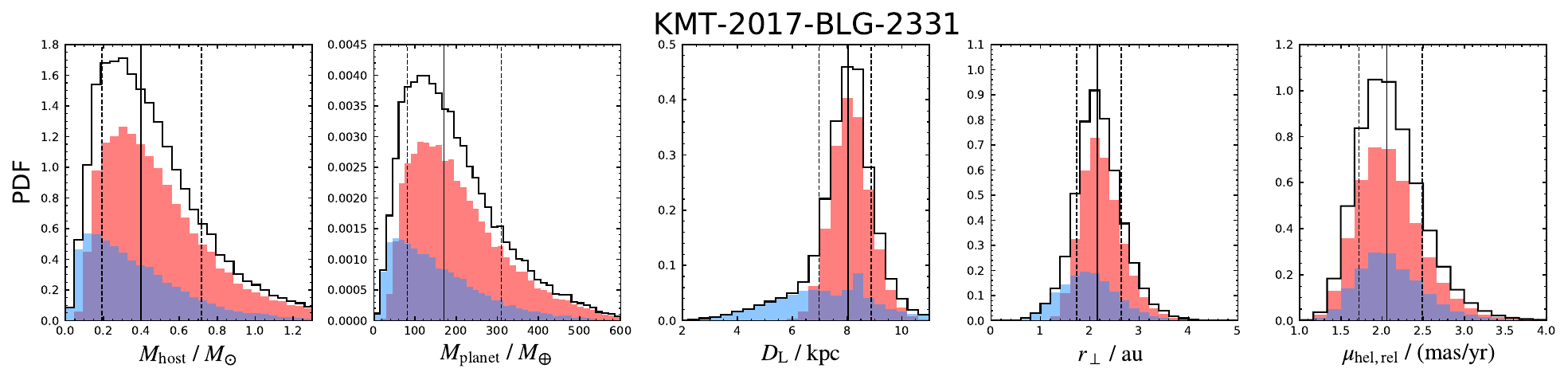}
    \caption{Bayesian posterior distributions of the mass of the host star, $M_\mathrm{host}$, the planetary mass, $M_\mathrm{planet}$, the lens distance, $D_\mathrm{L}$, the projected planet-host separation, $r_\bot$, and the lens-source relative proper motion in the heliocentric frame, $\mu_\mathrm{hel, rel}$. In each panel, the solid black line and the two dashed black lines represent the median value and the 15.9\% and 84.1\% percentages of the distribution. The solid black histogram shows the total distribution, and the bulge and disk lens distributions are shown in red and blue, respectively.}
    \label{baysfigure}
\end{figure*}

Table \ref{baysparm} and Figure \ref{baysfigure} show the posterior distributions from the Bayesian analysis, including the host mass, $M_{\rm host}$, the planetary mass, $M_{\rm planet}$, the lens distance, $D_{\rm L}$, the projected planet-host separation, $r_\perp$, and the heliocentric lens-source relative proper motion, $\mu_{\rm hel, rel}$. For \eventc, we also provide the relative probability from the Galactic model for each degenerate model. 

\begin{table*}
    \renewcommand\arraystretch{1.5}
    \centering
    \caption{CMD Parameters, $\theta_*$, $\thetae$ and $\mu_{\rm rel}$ for the Four Planetary Events}
    \begin{tabular}{c|c|c|c c c c|c}
    \hline
    \hline
    \multirow{2}{*}{Parameter} & \multirow{2}{*}{KB170849} & \multirow{2}{*}{KB171057} & \multicolumn{4}{c|}{OB170364} & \multirow{2}{*}{KB172331} \\
      &  &  & Wide A & Wide B & Wide C & Close Inner &  \\
    \hline
    $(V - I)_{\rm cl}$ & N.A. & N.A. & N.A. & $\longleftarrow$ & $\longleftarrow$ & $\longleftarrow$ & N.A.\\
    $I_{\rm cl}$ & $16.72\pm0.04$ & $15.63\pm0.03$ & $16.81\pm0.09$ & $\longleftarrow$ & $\longleftarrow$ & $\longleftarrow$ & $17.78\pm0.03$\\
    $I_{\rm cl,0}$ & $14.30\pm0.04$ & $14.30\pm0.04$ & $14.62\pm0.04$ & $\longleftarrow$ & $\longleftarrow$ & $\longleftarrow$ & $14.52\pm0.04$\\
    $(V - I)_{\rm S}$  & N.A. & N.A. & N.A. & $\longleftarrow$ & $\longleftarrow$ & $\longleftarrow$ & N.A.\\
    $I_{\rm S}$ & $17.98\pm0.12$ & $20.90\pm0.08$ & $20.64\pm0.03$ & $20.47\pm0.05$ & $20.43\pm0.05$ & $20.69\pm0.02$ & $22.48\pm0.08$\\
    $(V - I)_{\rm S,0}$ & $1.03\pm0.10$ & $0.92 \pm 0.09$ & $0.75\pm0.07$ & $0.75\pm0.07$ & $0.74\pm0.07$ & $0.75\pm0.07$ & $0.82\pm0.08$\\
    $I_{\rm S,0}$ & $15.55\pm0.13$ & $19.57\pm0.09$ & $18.45\pm0.10$ & $18.28\pm0.11$ & $18.24\pm0.11$ & $18.50\pm0.10$ & $19.22\pm0.09$\\
    $\theta_*$ ($\mu$as) & $3.59 \pm 0.81$ & $0.473\pm0.045$ & $0.681\pm0.055$ & $0.739\pm0.062$ & $0.746\pm0.063$ & $0.668\pm0.054$ & $0.509\pm0.043$\\
    $\thetae$ (mas) & $ 0.201 \pm 0.047$ & $>0.19$ & $0.341\pm0.058$ & $0.238\pm0.065$ & $0.067\pm0.017$ & $0.607\pm0.173$ & $0.246\pm0.041$\\ 
    $\mu_{\rm rel}$ (${\rm mas\,yr^{-1}}$) & $2.80 \pm 0.65$ & $>2.0$ & $6.70 \pm 1.15$ & $5.43 \pm 1.50$ & $1.57 \pm 0.40$ & $11.49 \pm 3.28$ & $1.86\pm0.34$\\
    \hline
    \hline
    \end{tabular}
    \tablecomments{$(V - I)_{\rm cl,0} = 1.06 \pm 0.03$ \citep{Bensby2013}. Event names are abbreviations, e.g., \eventa\ to KB170849. The upper limits on $\thetae$ and $\mu_{\rm rel}$ are $3\sigma$.}
    \label{source}
\end{table*}

\begin{table*}[htb]
    \renewcommand\arraystretch{1.25}
    \centering
    \caption{Lensing Physical Parameters for the Four Planetary Events from a Bayesian Analysis.}
    \begin{tabular}{c c |c c c c c |c }
    \hline
    \hline
    Event & Model &\multicolumn{5}{c|}{Physical Properties} & Relative Weight  \\
     & & $M_\mathrm{host}(M_\odot)$ & $M_\mathrm{planet}(M_\oplus)$ & $D_\mathrm{L}(\mathrm{kpc})$ & $r_\bot(\mathrm{au})$ & $\mu_\mathrm{hel,rel}(\mathrm{mas\ yr^{-1}})$ & Gal.Mod. \\
    \hline
    KB170849 &  & $0.19_{-0.09}^{+0.23}$ & $6.39_{-3.09}^{+7.80}$ & $7.22_{-0.99}^{+0.82}$ & $2.73_{-0.62}^{+0.68}$ & $3.14_{-0.62}^{+0.65}$ &  \\
    \hline
    KB171057 &  & $0.57_{-0.29}^{+0.37}$ & $23.5_{-12.0}^{+15.0}$ & $6.50_{-1.62}^{+0.81}$ & $2.36_{-0.62}^{+0.69}$ & $4.55_{-1.14}^{+1.58}$ & \\
    \hline
    \multirow{4}{*}{OB170364} & Wide A & $0.49_{-0.28}^{+0.35}$ & $75.8_{-42.9}^{+54.5}$ & $7.86_{-1.99}^{+1.08}$ & $3.79_{-0.89}^{+0.76}$ & $6.53_{-0.94}^{+1.14}$ & $0.60$  \\
    & Wide B & $0.35_{-0.20}^{+0.33}$ & $60.9_{-35.4}^{+57.9}$ & $8.38_{-1.66}^{+0.92}$ & $3.00_{-0.69}^{+0.80}$ & $5.41_{-1.05}^{+1.41}$ & 1.00 \\
    & Wide C & $0.10_{-0.05}^{+0.16}$ & $20.4_{-10.7}^{+33.8}$ & $8.98_{-0.88}^{+0.80}$ & $1.27_{-0.30}^{+0.49}$ & $2.11_{-0.51}^{+0.84}$ & 0.18 \\
    & Close Inner & $0.55_{-0.30}^{+0.38}$ & $256_{-139}^{+176}$ & $6.84_{-1.86}^{+1.54}$ & $1.96_{-0.51}^{+0.43}$ & $8.18_{-1.37}^{+1.63}$ & 0.11 \\
    \hline
    KB172331 &  & $0.40_{-0.21}^{+0.32}$ & $171_{-88}^{+137}$ & $8.03_{-1.05}^{+0.84}$ & $2.14_{-0.42}^{+0.48}$ & $2.06_{-0.34}^{+0.43}$ & \\
    \hline
    \hline
    \end{tabular}
    \label{baysparm}
\end{table*}

\subsection{\eventa}

We use stars in the $2' \times 2'$ square centered on the event to construct the CMD. The source is probably a red giant and the color, $(V - I)_{\rm S, 0} = 1.03 \pm 0.10$, is derived by matching the {\it HST} CMD to the KMTC CMD. We adopt the 3-$\sigma$ upper limit of the blended light as the upper limits of the lens flux, i.e., $I_{\rm L, limit, KMTC} = 18.7$. The low relative proper motion, $\mu_{\rm rel} = 2.80 \pm 0.65~\mathrm{mas\ yr^{-1}}$, indicates a bulge lensing system. 

The host star from the Bayesian analysis prefers a low-mass M dwarf. The planet is likely a super-Earth/mini-Neptune. The preferred projected planet-host separation, $r_{\bot} \sim 2.$ au, favors that this planet is located well beyond the snow line of the planetary system \citep{snowline}.

\subsection{\eventb}

The CMD is constructed from the OGLE-III field stars \citep{OGLEIII} within $2.5'$ centered on the event, and we calibrate the KMTC flux to the OGLE flux by matching bright field stars. The source is probably a G dwarf. The $3\sigma$ upper limit of the blended light is $I_{\rm B, KMTC} = 20.62$. Considering the mottled background \citep{MB03037} of the crowded stellar field, we adopt $I_{\rm L, limit, KMTC} = 20.0$. With the constraint on $\rho$ from the light-curve analysis, we obtain $\thetae > 0.19$ mas and $\mu_{\rm rel} > 2.0~\mathrm{mas\ yr^{-1}}$ at $3\sigma$. The likelihood distribution of $\thetae$ used for the Bayesian analysis is derived by the minimum $\chi^2$ for the lower envelope of the ($\chi^2$ vs. $\rho$) diagram and the $\theta_*$ distribution. 

According to the Bayesian analysis, the host prefers a K or M dwarf, and the planetary mass prefers a super-Neptune mass. The lensing system can be located in either the bulge or the disk.

\subsection{\eventc}

We build the CMD using the KMTC stars within a $4' \times 4'$ square centered on the event. The source color is also estimated by the {\it HST} CMD and slightly varies among different models because of the different source brightness. The blended flux is consistent with zero and we adopt $I_{\rm L, limit, KMTC} = 20.0$ for considering the mottled background. 

Because of $\ell = -5.3617$, $D_{\rm S} = 10.0_{-0.8}^{+0.9}$ kpc according to the Bayesian analysis, and thus the resulting lens distances are farther than most microlensing events. The ``Wide A'' and ``Wide B'' models are favored. The ``Wide C'' and ``Close Inner'' models are disfavored because of the small $\thetae$ and the high $\mu_{\rm rel}$, respectively. The ``Wide A'' and ``Wide B'' models have almost the same preferred lensing properties, i.e., a sub-Saturn orbiting an $M$ dwarf at a projected separation of $\sim 3$ au. The nature of the lens could be resolved by future high-resolution imaging because the ``Wide C'' and ``Close Inner'' models have different $\mu_{\rm rel}$ than the ``Wide A'' and ``Wide B'' models.This can certainly be done at first light of ELTs (roughly 2030) when the separation will be about 80 mas (five times the imaging FWHM for EELT K-band). And it may be possible with Keck as early as 2027, provided that the ``Wide A'' and ``Wide B'' models are correct and the host is sufficiently bright.

\subsection{\eventd}

The CMD size for \eventd\ is a $4' \times 4'$ square centered on the event. Because of the high extinction $A_I \sim 3.8$ and the low stellar surface density, the mottled background is negligible (fluctuations $> 21$ mag). We use the 3-$\sigma$ upper limit of the blended light, $I_{\rm B, KMTC} = 19.8$, as $I_{\rm L, limit, KMTC}$.

The Bayesian analysis prefers a sub-Jupiter-mass planet orbiting an M dwarf. The lensing system is probably located in the Galactic bulge, consistent with the low proper motion, $\mu_{\rm rel} \sim 2 ~\mathrm{mas\ yr^{-1}}$.

\section{Discussion: A Complete Sample from the First 4 yr KMTNet Survey}\label{dis}

\begin{figure*}[htb] 
    \centering
    \includegraphics[width=0.98\columnwidth]{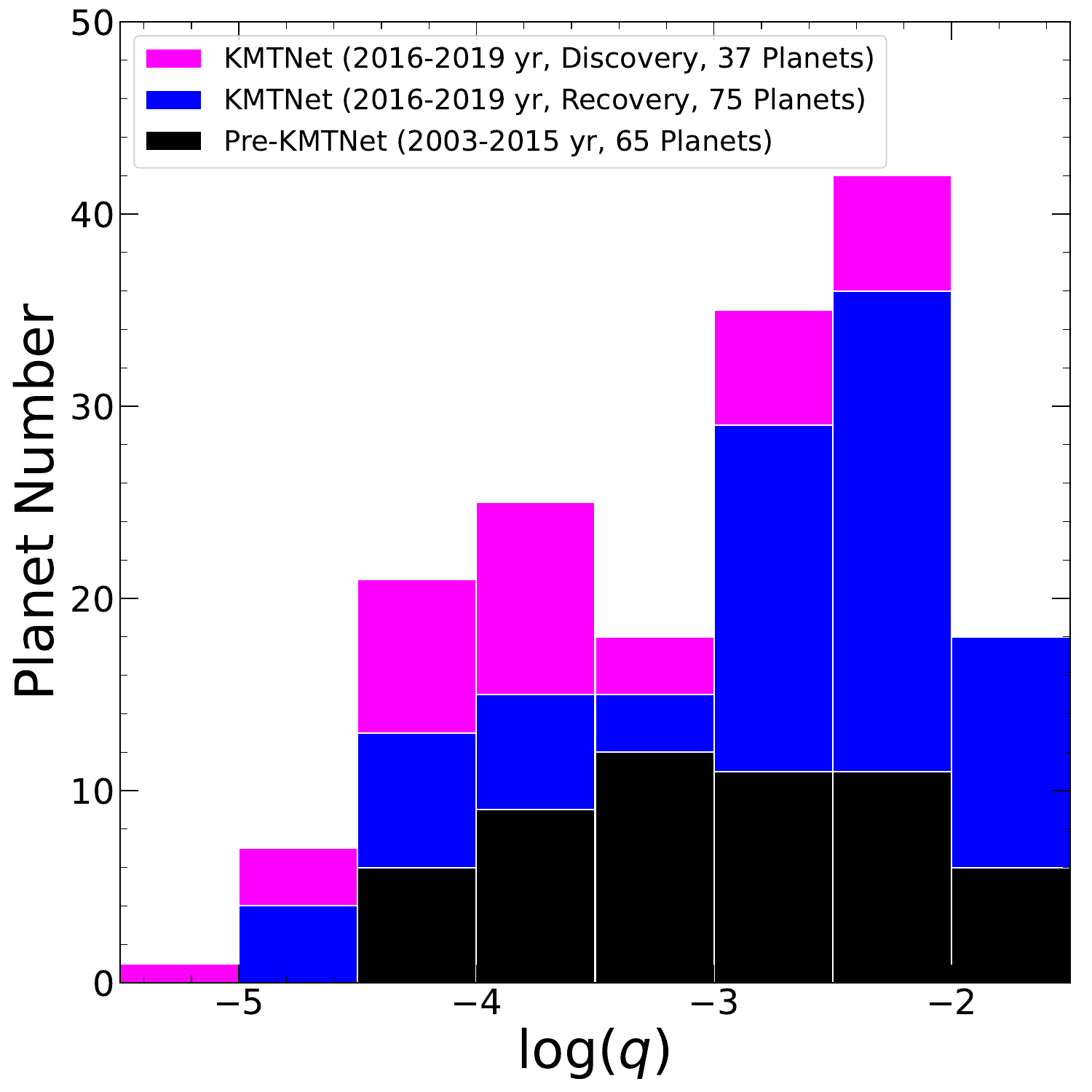}
    \includegraphics[width=0.98\columnwidth]{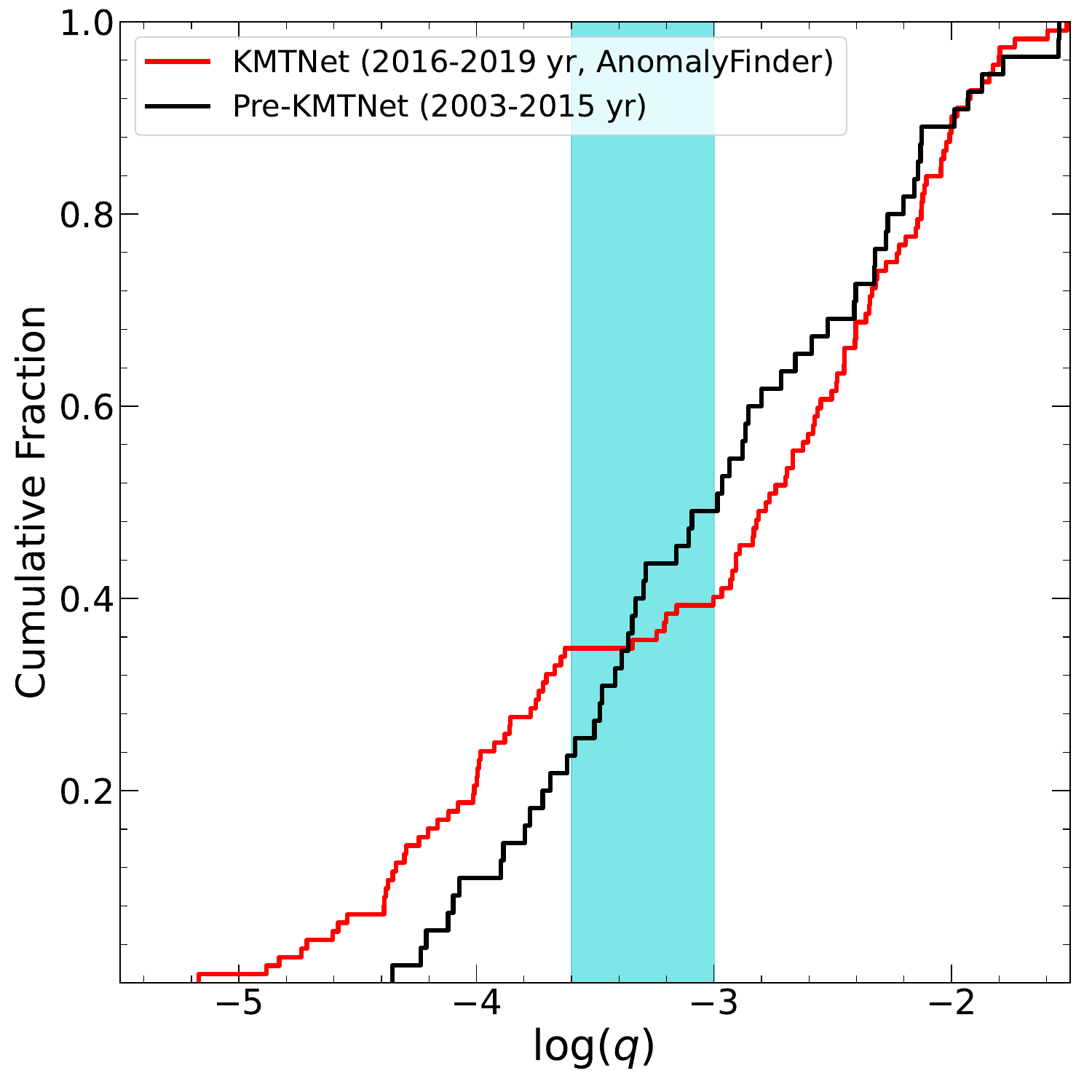}
    \caption{{\it Left}: Histogram distributions of $\log q$ for the 2016--2019 KMTNet AnomalyFinder-discovery (magenta) and AnomlyFinder-recovery (blue) planets, and the planets detected before KMTNet's regular survey (black). {\it Right}: Cumulative distributions of $\log q$ for the pre-KMTNet planets and the 2016--2019 KMTNet AnomalyFinder planets, which are a sum of AnomalyFinder-discovery and AnomalyFinder-recovery planets shown in the left panel. The dark turquoise region indicates the ``sub-Saturn desert'' ($\log q = \left[-3.6, -3.0\right]$) in the KMTNet sample.}
    \label{sample}
\end{figure*}

In this paper, we have presented the analysis of four new planets. Together with nine that are already published and three that will be published elsewhere, there are 16 clear planets from the 2017 KNTNet subprime data. Among these, the anomaly of OGLE-2017-BLG-0668/KMT-2017-BLG-1145 was generated primarily by two sources. The current AnomalyFinder algorithm cannot yield a complete sample for such events \citep{OB191470}, so we exclude this event from the AnomalyFinder planetary sample of the 2017 subprime fields. Table \ref{tab:sample_info} summarizes the 15 planets, including $\log q$, $s$, discovery method, and $\Delta\chi^2$ compared to the best-fit models for degeneracy. Among them, five were discovered by AnomlyFinder and ten were first discovered using by-eye searches and then recovered by AnomalyFinder. A striking feature of this sample is that all AnomlyFinder-discovery planets have $\log q < -3.0$, and these comprise 5/7 of the $\log q < -3.0$ planets, demonstrating AnomalyFinder's important role in the detection of low-$q$ planets, while by-eye searches mainly detected massive planets. 

With the complete planetary sample of the 2017 subprime field, we have completed the planetary sample from the first 4-year KMTNet data (2016--2019). Below we briefly review this sample, to understand the impact of KMTNet and AnomalyFinder. We follow the criteria of \cite{OB160007} for the definition of a planetary event. That is, we exclude planets in binary-star systems, planets with degenerate stellar-binary models ($\log q > -1.5$ with $\Delta\chi^2 < 10$), and planets with the 2L1S/1L2S degeneracy (a 1L2S model with $\Delta\chi^2 < 16$). The difference is that we keep planets with large uncertainties in $\log q$ because we do not attempt to study the mass-ratio function here.

In total, we form a sample of 112 planets, with 28 planets per year on average. The seasonal distribution, (23, 30, 35, 24) for 2016--2019, is consistent with Poisson variations. Of them, 37 were AnomlyFinder-discovery planets, so AnomalyFinder increases the KMTNet planetary detections by 50\%. In 2015, KMTNet conducted commissioning observations toward four fields with a cadence of $\Gamma = 6~{\rm hr}^{-1}$. Since 2016, KMTNet has devoted about half of its time to subprime fields, and 52 (i.e, 46\%) planets are from subprime fields. Below we mainly focus on the discussion of the distribution of $q$, because the distributions of other parameters (e.g., caustic crossing, anomaly type) have been investigated by \cite{2019_subprime,-4planet} in detail using the subgroups of this sample and we do not find a big difference.

We adopt the best-fit model of each planet, and the left panel of Figure \ref{sample} shows the mass-ratio distributions for the AnomlyFinder-discovery and AnomlyFinder-recovery planets, respectively. Similar to the 2017 subprime sample, most (25) AnomlyFinder-discovery planets have $\log q < -3$, increasing the number of KMTNet $\log q < -3$ planets by 125\%. The fundamental reason is that for most cases AnomlyFinder can find significantly more subtle signals than by-eye searches, as previously found by \cite{KB190253,2019_prime} for smaller samples. For massive planets ($\log q > -3$), by-eye searches are basically sufficient and AnomlyFinder increases the number by only 22\%.  

Using the same criteria above that are listed, we form the microlensing planetary sample from 2003 (i.e., the year of the first microlensing planet) to 2015 (i.e., the year before KMTNet's regular survey). The only difference is that we keep three planets in binary binary-star systems \citep{OB07349,OB08092,OB130341}. We exclude the planet, OGLE-2015-BLG-1771Lb, for which the planetary signal was detected by the commissioning data of KMTNet. For the other two cases with the KMTNet data, OGLE-2015-BLG-0954Lb \citep{OB150954,OB150954moa} and OGLE-2015-BLG-1670Lb \cite{OB151670}, the planetary signal can be well covered by the OGLE and MOA data. In total, this sample contains 65 planets, and the mass-ratio distribution is also shown in the left panel of Figure \ref{sample}. Overall, the first 4-year KMTNet microlensing survey nearly tripled the microlensing planetary sample. The most significant advance occurs at the low end of the mass-ratio distribution. The KMTNet sample reaches mass ratios about an order of magnitude lower than the pre-KMTNet sample and expands the $\log q < -4$ sample five times. Furthermore, the pre-KMTNet sample consists of planets from several groups and the resulting statistical samples have relatively small sizes, with six planets in the sample from the Microlensing Follow Up Network ($\mu$FUN, \citealt{mufun}), three planets in the sample from the Probing Lensing Anomalies NETwork (PLANET) follow-up network \citep{Cassan2012}, 22 planets in the sample from the Microlensing Observations in Astrophysics (MOA) group \citep{Suzuki2016}, and eight planets in the sample from a combination of four-year OGLE + MOA + Wise data \citep{Wise}. Combined with planets after 2020, KMTNet will form a statistical sample an order of magnitude larger than any previous samples. 

The right panel of figure \ref{sample} displays the cumulative mass-ratio distribution for the KMTNet and pre-KMTNet samples. Besides the low-$q$ planets, another striking difference between the two samples is that the KMTNet sample shows a plateau between $\log q = \left[-3.6, -3.0\right]$, while the pre-KMTNet does not \footnote{This plateau still exists if we follow the criteria of \cite{OB160007} to remove planets with large uncertainties in $\log q$.}. This mass-ratio desert is consistent with the prediction from the standard core accretion runaway growth scenario \citep{Ida2004,Mordasini2009}. We refer to this desert as the ``sub-Saturn desert'', considering that the typical hosts are expected to be M and K dwarfs for microlensing planets. 

The ``sub-Saturn desert'' in the KMTNet sample was first noticed by \cite{KB161836} mostly using a sample of 2016 and 2017 KMTNet planets from by-eye searches. However, because the sample of \cite{KB161836} was incomplete and not homogeneously selected and the pre-KMTNet sample does not show such a desert, \cite{KB161836} concluded that the discrepancy between the two samples was most likely caused by publication bias. However, the complete and homogeneously-selected sample of 2018 and 2019 KMTNet planets, together corrected by the KMTNet and AnomalyFinder detection efficiency, strongly supports the existence of ``sub-Saturn desert'' in the KMTNet sample \citep{OB160007}. And now, the complete sample of 2016 and 2017 KMTNet planets confirms the ``sub-Saturn desert''. A more detailed study of the ``sub-Saturn desert'' will be presented together with the 2021 AnomalyFinder planetary sample (Shin et al. in prep).

\acknowledgments
We appreciate the anonymous referee for helping to improve the paper. Y.G., W.Zang, R.Z., H.Y., S.M., J.Z., R.K., Q.Q., and W.Zhu acknowledge support by the National Natural Science Foundation of China (Grant No. 12133005). W.Zang acknowledges the support from the Harvard-Smithsonian Center for Astrophysics through the CfA Fellowship. This research has made use of the KMTNet system operated by the Korea Astronomy and Space Science Institute (KASI) at three host sites of CTIO in Chile, SAAO in South Africa, and SSO in Australia. Data transfer from the host site to KASI was supported by the Korea Research Environment Open NETwork (KREONET). This research was supported by KASI under the R\&D program (project No. 2024-1-832-01) supervised by the Ministry of Science and ICT. Work by R.P. and J.S. was supported by Polish National Agency for Academic Exchange grant ``Polish Returns 2019.'' Work by J.C.Y. and I.-G.S. acknowledge support from N.S.F Grant No. AST-2108414. Work by C.H. was supported by the grants of National Research Foundation of Korea (2019R1A2C2085965 and 2020R1A4A2002885). Y.S. acknowledges support from BSF Grant No. 2020740. The authors acknowledge the Tsinghua Astrophysics High-Performance Computing platform at Tsinghua University for providing computational and data storage resources that have contributed to the research results reported within this paper.

\software{pySIS \citep{pysis,Yang_TLC}, OGLE DIA pipeline \citep{Wozniak2000}, numpy \citep{numpy}, emcee \citep{emcee2,emcee}, Matplotlib \citep{Matplotlib}, SciPy \citep{scipy}}

\bibliography{Zang.bib}

\begin{thebibliography}{}
\expandafter\ifx\csname natexlab\endcsname\relax\def\natexlab#1{#1}\fi
\providecommand{\url}[1]{\href{#1}{#1}}
\providecommand{\dodoi}[1]{doi:~\href{http://doi.org/#1}{\nolinkurl{#1}}}
\providecommand{\doeprint}[1]{\href{http://ascl.net/#1}{\nolinkurl{http://ascl.net/#1}}}
\providecommand{\doarXiv}[1]{\href{https://arxiv.org/abs/#1}{\nolinkurl{https://arxiv.org/abs/#1}}}

\bibitem[{{Adams} {et~al.}(2018){Adams}, {Boyajian}, \& {von
  Braun}}]{Adams2018}
{Adams}, A.~D., {Boyajian}, T.~S., \& {von Braun}, K. 2018, \mnras, 473, 3608,
  \dodoi{10.1093/mnras/stx2367}

\bibitem[{{Alard} \& {Lupton}(1998)}]{Alard1998}
{Alard}, C., \& {Lupton}, R.~H. 1998, \apj, 503, 325, \dodoi{10.1086/305984}

\bibitem[{{Albrow} {et~al.}(2009){Albrow}, {Horne}, {Bramich}, {Fouqu{\'e}},
  {Miller}, {Beaulieu}, {Coutures}, {Menzies}, {Williams}, {Batista},
  {Bennett}, {Brillant}, {Cassan}, {Dieters}, {Dominis Prester}, {Donatowicz},
  {Greenhill}, {Kains}, {Kane}, {Kubas}, {Marquette}, {Pollard}, {Sahu},
  {Tsapras}, {Wambsganss}, \& {Zub}}]{pysis}
{Albrow}, M.~D., {Horne}, K., {Bramich}, D.~M., {et~al.} 2009, \mnras, 397,
  2099, \dodoi{10.1111/j.1365-2966.2009.15098.x}

\bibitem[{{Batista} {et~al.}(2011){Batista}, {Gould}, {Dieters}, {Dong},
  {Bond}, {Beaulieu}, {Maoz}, {Monard}, {Christie}, {McCormick}, {Albrow},
  {Horne}, {Tsapras}, {Burgdorf}, {Calchi Novati}, {Skottfelt}, {Caldwell},
  {Koz{\l}owski}, {Kubas}, {Gaudi}, {Han}, {Bennett}, {An}, {MOA
  Collaboration}, {Abe}, {Botzler}, {Douchin}, {Freeman}, {Fukui}, {Furusawa},
  {Hearnshaw}, {Hosaka}, {Itow}, {Kamiya}, {Kilmartin}, {Korpela}, {Lin},
  {Ling}, {Makita}, {Masuda}, {Matsubara}, {Miyake}, {Muraki}, {Nagaya},
  {Nishimoto}, {Ohnishi}, {Okumura}, {Perrott}, {Rattenbury}, {Saito},
  {Sullivan}, {Sumi}, {Sweatman}, {Tristram}, {von Seggern}, {Yock}, {PLANET
  Collaboration}, {Brillant}, {Calitz}, {Cassan}, {Cole}, {Cook}, {Coutures},
  {Dominis Prester}, {Donatowicz}, {Greenhill}, {Hoffman}, {Jablonski}, {Kane},
  {Kains}, {Marquette}, {Martin}, {Martioli}, {Meintjes}, {Menzies},
  {Pedretti}, {Pollard}, {Sahu}, {Vinter}, {Wambsganss}, {Watson}, {Williams},
  {Zub}, {FUN Collaboration}, {Allen}, {Bolt}, {Bos}, {DePoy}, {Drummond},
  {Eastman}, {Gal-Yam}, {Gorbikov}, {Higgins}, {Janczak}, {Kaspi}, {Lee},
  {Mallia}, {Maury}, {Monard}, {Moorhouse}, {Morgan}, {Natusch}, {Ofek},
  {Park}, {Pogge}, {Polishook}, {Santallo}, {Shporer}, {Spector}, {Thornley},
  {Yee}, {MiNDSTEp Consortium}, {Bozza}, {Browne}, {Dominik}, {Dreizler},
  {Finet}, {Glitrup}, {Grundahl}, {Harps{\o}e}, {Hessman}, {Hinse},
  {Hundertmark}, {J{\o}rgensen}, {Liebig}, {Maier}, {Mancini}, {Mathiasen},
  {Rahvar}, {Ricci}, {Scarpetta}, {Southworth}, {Surdej}, {Zimmer}, {RoboNet
  Collaboration}, {Allan}, {Bramich}, {Snodgrass}, {Steele}, \&
  {Street}}]{MB09387}
{Batista}, V., {Gould}, A., {Dieters}, S., {et~al.} 2011, \aap, 529, A102,
  \dodoi{10.1051/0004-6361/201016111}

\bibitem[{{Bennett} \& {Rhie}(1996)}]{BennettRhie}
{Bennett}, D.~P., \& {Rhie}, S.~H. 1996, \apj, 472, 660, \dodoi{10.1086/178096}

\bibitem[{{Bennett} {et~al.}(2008){Bennett}, {Bond}, {Udalski}, {Sumi}, {Abe},
  {Fukui}, {Furusawa}, {Hearnshaw}, {Holderness}, {Itow}, {Kamiya}, {Korpela},
  {Kilmartin}, {Lin}, {Ling}, {Masuda}, {Matsubara}, {Miyake}, {Muraki},
  {Nagaya}, {Okumura}, {Ohnishi}, {Perrott}, {Rattenbury}, {Sako}, {Saito},
  {Sato}, {Skuljan}, {Sullivan}, {Sweatman}, {Tristram}, {Yock}, {Kubiak},
  {Szyma{\'n}ski}, {Pietrzy{\'n}ski}, {Soszy{\'n}ski}, {Szewczyk},
  {Wyrzykowski}, {Ulaczyk}, {Batista}, {Beaulieu}, {Brillant}, {Cassan},
  {Fouqu{\'e}}, {Kervella}, {Kubas}, \& {Marquette}}]{MB07192}
{Bennett}, D.~P., {Bond}, I.~A., {Udalski}, A., {et~al.} 2008, \apj, 684, 663,
  \dodoi{10.1086/589940}

\bibitem[{{Bennett} {et~al.}(2016){Bennett}, {Rhie}, {Udalski}, {Gould},
  {Tsapras}, {Kubas}, {Bond}, {Greenhill}, {Cassan}, {Rattenbury}, {Boyajian},
  {Luhn}, {Penny}, {Anderson}, {Abe}, {Bhattacharya}, {Botzler}, {Donachie},
  {Freeman}, {Fukui}, {Hirao}, {Itow}, {Koshimoto}, {Li}, {Ling}, {Masuda},
  {Matsubara}, {Muraki}, {Nagakane}, {Ohnishi}, {Oyokawa}, {Perrott}, {Saito},
  {Sharan}, {Sullivan}, {Sumi}, {Suzuki}, {Tristram}, {Yonehara}, {Yock}, {MOA
  Collaboration}, {Szyma{\'n}ski}, {Soszy{\'n}ski}, {Ulaczyk}, {Wyrzykowski},
  {OGLE Collaboration}, {Allen}, {DePoy}, {Gal-Yam}, {Gaudi}, {Han}, {Monard},
  {Ofek}, {Pogge}, {{\ensuremath{\mu}}FUN Collaboration}, {Street}, {Bramich},
  {Dominik}, {Horne}, {Snodgrass}, {Steele}, {Robonet Collaboration}, {Albrow},
  {Bachelet}, {Batista}, {Beaulieu}, {Brillant}, {Caldwell}, {Cole},
  {Coutures}, {Dieters}, {Dominis Prester}, {Donatowicz}, {Fouqu{\'e}},
  {Hundertmark}, {J{\o}rgensen}, {Kains}, {Kane}, {Marquette}, {Menzies},
  {Pollard}, {Ranc}, {Sahu}, {Wambsganss}, {Williams}, {Zub}, \& {PLANET
  Collaboration}}]{OB07349}
{Bennett}, D.~P., {Rhie}, S.~H., {Udalski}, A., {et~al.} 2016, \aj, 152, 125,
  \dodoi{10.3847/0004-6256/152/5/125}

\bibitem[{{Bennett} {et~al.}(2017){Bennett}, {Bond}, {Abe}, {Asakura}, {Barry},
  {Bhattacharya}, {Donachie}, {Evans}, {Fukui}, {Hirao}, {Itow}, {Koshimoto},
  {Li}, {Ling}, {Masuda}, {Matsubara}, {Muraki}, {Nagakane}, {Ohnishi}, {Ranc},
  {Rattenbury}, {Saito}, {Sharan}, {Sullivan}, {Sumi}, {Suzuki}, {Tristram},
  {Yamada}, {Yamada}, {Yonehara}, \& {MOA Collaboration}}]{OB150954moa}
{Bennett}, D.~P., {Bond}, I.~A., {Abe}, F., {et~al.} 2017, \aj, 154, 68,
  \dodoi{10.3847/1538-3881/aa7aee}

\bibitem[{{Bensby} {et~al.}(2013){Bensby}, {Yee}, {Feltzing}, {Johnson},
  {Gould}, {Cohen}, {Asplund}, {Mel{\'e}ndez}, {Lucatello}, {Han}, {Thompson},
  {Gal-Yam}, {Udalski}, {Bennett}, {Bond}, {Kohei}, {Sumi}, {Suzuki}, {Suzuki},
  {Takino}, {Tristram}, {Yamai}, \& {Yonehara}}]{Bensby2013}
{Bensby}, T., {Yee}, J.~C., {Feltzing}, S., {et~al.} 2013, \aap, 549, A147,
  \dodoi{10.1051/0004-6361/201220678}

\bibitem[{{Bozza}(2010)}]{Bozza2010}
{Bozza}, V. 2010, \mnras, 408, 2188, \dodoi{10.1111/j.1365-2966.2010.17265.x}

\bibitem[{{Bozza} {et~al.}(2018){Bozza}, {Bachelet}, {Bartoli{\'c}}, {Heintz},
  {Hoag}, \& {Hundertmark}}]{Bozza2018}
{Bozza}, V., {Bachelet}, E., {Bartoli{\'c}}, F., {et~al.} 2018, \mnras, 479,
  5157, \dodoi{10.1093/mnras/sty1791}

\bibitem[{{Calchi Novati} {et~al.}(2018){Calchi Novati}, {Skowron}, {Jung},
  {Beichman}, {Bryden}, {Carey}, {Gaudi}, {Henderson}, {Shvartzvald}, {Yee},
  {Zhu}, {Spitzer Team}, {Udalski}, {Szyma{\'n}ski}, {Mr{\'o}z}, {Poleski},
  {Soszy{\'n}ski}, {Koz{\l}owski}, {Pietrukowicz}, {Ulaczyk}, {Pawlak},
  {Rybicki}, {Iwanek}, {OGLE Collaboration}, {Albrow}, {Chung}, {Gould}, {Han},
  {Hwang}, {Ryu}, {Shin}, {Zang}, {Cha}, {Kim}, {Kim}, {Kim}, {Lee}, {Lee},
  {Lee}, {Park}, {Pogge}, \& {KMTNet Collaboration}}]{OB171140}
{Calchi Novati}, S., {Skowron}, J., {Jung}, Y.~K., {et~al.} 2018, \aj, 155,
  261, \dodoi{10.3847/1538-3881/aac21c}

\bibitem[{{Cassan} {et~al.}(2012){Cassan}, {Kubas}, {Beaulieu}, {Dominik},
  {Horne}, {Greenhill}, {Wambsganss}, {Menzies}, {Williams}, {J{\o}rgensen},
  {Udalski}, {Bennett}, {Albrow}, {Batista}, {Brillant}, {Caldwell}, {Cole},
  {Coutures}, {Cook}, {Dieters}, {Dominis Prester}, {Donatowicz}, {Fouqu{\'e}},
  {Hill}, {Kains}, {Kane}, {Marquette}, {Martin}, {Pollard}, {Sahu}, {Vinter},
  {Warren}, {Watson}, {Zub}, {Sumi}, {Szyma{\'n}ski}, {Kubiak}, {Poleski},
  {Soszynski}, {Ulaczyk}, {Pietrzy{\'n}ski}, \& {Wyrzykowski}}]{Cassan2012}
{Cassan}, A., {Kubas}, D., {Beaulieu}, J.~P., {et~al.} 2012, \nat, 481, 167,
  \dodoi{10.1038/nature10684}

\bibitem[{{Dong} {et~al.}(2009){Dong}, {Gould}, {Udalski}, {Anderson},
  {Christie}, {Gaudi}, {OGLE Collaboration}, {Jaroszy{\'n}ski}, {Kubiak},
  {Szyma{\'n}ski}, {Pietrzy{\'n}ski}, {Soszy{\'n}ski}, {Szewczyk}, {Ulaczyk},
  {Wyrzykowski}, {{$\mu$}FUN Collaboration}, {DePoy}, {Fox}, {Gal-Yam}, {Han},
  {L{\'e}pine}, {McCormick}, {Ofek}, {Park}, {Pogge}, {MOA Collaboration},
  {Abe}, {Bennett}, {Bond}, {Britton}, {Gilmore}, {Hearnshaw}, {Itow},
  {Kamiya}, {Kilmartin}, {Korpela}, {Masuda}, {Matsubara}, {Motomura},
  {Muraki}, {Nakamura}, {Ohnishi}, {Okada}, {Rattenbury}, {Saito}, {Sako},
  {Sasaki}, {Sullivan}, {Sumi}, {Tristram}, {Yanagisawa}, {Yock}, {Yoshoika},
  {PLANET/RoboNet Collaborations}, {Albrow}, {Beaulieu}, {Brillant}, {Calitz},
  {Cassan}, {Cook}, {Coutures}, {Dieters}, {Dominis Prester}, {Donatowicz},
  {Fouqu{\'e}}, {Greenhill}, {Hill}, {Hoffman}, {Horne}, {J{\o}rgensen},
  {Kane}, {Kubas}, {Marquette}, {Martin}, {Meintjes}, {Menzies}, {Pollard},
  {Sahu}, {Vinter}, {Wambsganss}, {Williams}, {Bode}, {Bramich}, {Burgdorf},
  {Snodgrass}, {Steele}, {Doublier}, \& {Foellmi}}]{OB050071D}
{Dong}, S., {Gould}, A., {Udalski}, A., {et~al.} 2009, \apj, 695, 970,
  \dodoi{10.1088/0004-637X/695/2/970}

\bibitem[{{Foreman-Mackey} {et~al.}(2013){Foreman-Mackey}, {Hogg}, {Lang}, \&
  {Goodman}}]{emcee}
{Foreman-Mackey}, D., {Hogg}, D.~W., {Lang}, D., \& {Goodman}, J. 2013, \pasp,
  125, 306, \dodoi{10.1086/670067}

\bibitem[{{Gaudi}(1998)}]{Gaudi1998}
{Gaudi}, B.~S. 1998, \apj, 506, 533, \dodoi{10.1086/306256}

\bibitem[{{Gaudi}(2012)}]{Gaudi2012}
---. 2012, \araa, 50, 411, \dodoi{10.1146/annurev-astro-081811-125518}

\bibitem[{{Gaudi} \& {Gould}(1997)}]{GG1997}
{Gaudi}, B.~S., \& {Gould}, A. 1997, \apj, 486, 85, \dodoi{10.1086/304491}

\bibitem[{{Gonzalez} {et~al.}(2012){Gonzalez}, {Rejkuba}, {Zoccali}, {Valenti},
  {Minniti}, {Schultheis}, {Tobar}, \& {Chen}}]{Gonzalez2012}
{Gonzalez}, O.~A., {Rejkuba}, M., {Zoccali}, M., {et~al.} 2012, \aap, 543, A13,
  \dodoi{10.1051/0004-6361/201219222}

\bibitem[{{Goodman} \& {Weare}(2010)}]{emcee2}
{Goodman}, J., \& {Weare}, J. 2010, Communications in Applied Mathematics and
  Computational Science, 5, 65, \dodoi{10.2140/camcos.2010.5.65}

\bibitem[{{Gould}(1992)}]{Gould1992}
{Gould}, A. 1992, \apj, 392, 442, \dodoi{10.1086/171443}

\bibitem[{{Gould}(1994)}]{1994ApJ...421L..75G}
---. 1994, \apjl, 421, L75, \dodoi{10.1086/187191}

\bibitem[{{Gould}(1996)}]{Gould2D}
---. 1996, \apj, 470, 201, \dodoi{10.1086/177861}

\bibitem[{{Gould}(2000)}]{Gould2000}
---. 2000, \apj, 542, 785, \dodoi{10.1086/317037}

\bibitem[{{Gould}(2004)}]{Gouldpies2004}
---. 2004, \apj, 606, 319, \dodoi{10.1086/382782}

\bibitem[{{Gould} \& {Gaucherel}(1997)}]{Andy1997}
{Gould}, A., \& {Gaucherel}, C. 1997, \apj, 477, 580, \dodoi{10.1086/303751}

\bibitem[{{Gould} \& {Loeb}(1992)}]{Andy1992}
{Gould}, A., \& {Loeb}, A. 1992, \apj, 396, 104, \dodoi{10.1086/171700}

\bibitem[{{Gould} {et~al.}(2010){Gould}, {Dong}, {Gaudi}, {Udalski}, {Bond},
  {Greenhill}, {Street}, {Dominik}, {Sumi}, {Szyma{\'n}ski}, {Han}, {Allen},
  {Bolt}, {Bos}, {Christie}, {DePoy}, {Drummond}, {Eastman}, {Gal-Yam},
  {Higgins}, {Janczak}, {Kaspi}, {Koz{\l}owski}, {Lee}, {Mallia}, {Maury},
  {Maoz}, {McCormick}, {Monard}, {Moorhouse}, {Morgan}, {Natusch}, {Ofek},
  {Park}, {Pogge}, {Polishook}, {Santallo}, {Shporer}, {Spector}, {Thornley},
  {Yee}, {{$\mu$}FUN Collaboration}, {Kubiak}, {Pietrzy{\'n}ski},
  {Soszy{\'n}ski}, {Szewczyk}, {Wyrzykowski}, {Ulaczyk}, {Poleski}, {OGLE
  Collaboration}, {Abe}, {Bennett}, {Botzler}, {Douchin}, {Freeman}, {Fukui},
  {Furusawa}, {Hearnshaw}, {Hosaka}, {Itow}, {Kamiya}, {Kilmartin}, {Korpela},
  {Lin}, {Ling}, {Makita}, {Masuda}, {Matsubara}, {Miyake}, {Muraki}, {Nagaya},
  {Nishimoto}, {Ohnishi}, {Okumura}, {Perrott}, {Philpott}, {Rattenbury},
  {Saito}, {Sako}, {Sullivan}, {Sweatman}, {Tristram}, {von Seggern}, {Yock},
  {MOA Collaboration}, {Albrow}, {Batista}, {Beaulieu}, {Brillant}, {Caldwell},
  {Calitz}, {Cassan}, {Cole}, {Cook}, {Coutures}, {Dieters}, {Dominis Prester},
  {Donatowicz}, {Fouqu{\'e}}, {Hill}, {Hoffman}, {Jablonski}, {Kane}, {Kains},
  {Kubas}, {Marquette}, {Martin}, {Martioli}, {Meintjes}, {Menzies},
  {Pedretti}, {Pollard}, {Sahu}, {Vinter}, {Wambsganss}, {Watson}, {Williams},
  {Zub}, {PLANET Collaboration}, {Allan}, {Bode}, {Bramich}, {Burgdorf},
  {Clay}, {Fraser}, {Hawkins}, {Horne}, {Kerins}, {Lister}, {Mottram},
  {Saunders}, {Snodgrass}, {Steele}, {Tsapras}, {RoboNet Collaboration},
  {J{\o}rgensen}, {Anguita}, {Bozza}, {Calchi Novati}, {Harps{\o}e}, {Hinse},
  {Hundertmark}, {Kj{\ae}rgaard}, {Liebig}, {Mancini}, {Masi}, {Mathiasen},
  {Rahvar}, {Ricci}, {Scarpetta}, {Southworth}, {Surdej}, {Th{\"o}ne}, \&
  {MiNDSTEp Consortium}}]{mufun}
{Gould}, A., {Dong}, S., {Gaudi}, B.~S., {et~al.} 2010, \apj, 720, 1073,
  \dodoi{10.1088/0004-637X/720/2/1073}

\bibitem[{{Gould} {et~al.}(2014){Gould}, {Udalski}, {Shin}, {Porritt},
  {Skowron}, {Han}, {Yee}, {Koz{\l}owski}, {Choi}, {Poleski}, {Wyrzykowski},
  {Ulaczyk}, {Pietrukowicz}, {Mr{\'o}z}, {Szyma{\'n}ski}, {Kubiak},
  {Soszy{\'n}ski}, {Pietrzy{\'n}ski}, {Gaudi}, {Christie}, {Drummond},
  {McCormick}, {Natusch}, {Ngan}, {Tan}, {Albrow}, {DePoy}, {Hwang}, {Jung},
  {Lee}, {Park}, {Pogge}, {Abe}, {Bennett}, {Bond}, {Botzler}, {Freeman},
  {Fukui}, {Fukunaga}, {Itow}, {Koshimoto}, {Larsen}, {Ling}, {Masuda},
  {Matsubara}, {Muraki}, {Namba}, {Ohnishi}, {Philpott}, {Rattenbury}, {Saito},
  {Sullivan}, {Sumi}, {Suzuki}, {Tristram}, {Tsurumi}, {Wada}, {Yamai}, {Yock},
  {Yonehara}, {Shvartzvald}, {Maoz}, {Kaspi}, \& {Friedmann}}]{OB130341}
{Gould}, A., {Udalski}, A., {Shin}, I.~G., {et~al.} 2014, Science, 345, 46,
  \dodoi{10.1126/science.1251527}

\bibitem[{{Gould} {et~al.}(2022){Gould}, {Han}, {Zang}, {Yang}, {Hwang},
  {Udalski}, {Bond}, {Albrow}, {Chung}, {Jung}, {Ryu}, {Shin}, {Shvartzvald},
  {Yee}, {Cha}, {Kim}, {Kim}, {Kim}, {Lee}, {Lee}, {Lee}, {Park}, {Pogge},
  {KMTNet Collaboration}, {Mr{\'o}z}, {Szyma{\'n}ski}, {Skowron}, {Poleski},
  {Soszy{\'n}ski}, {Pietrukowicz}, {Koz{\l}owski}, {Ulaczyk}, {Rybicki},
  {Iwanek}, {Wrona}, {OGLE Collaboration}, {Abe}, {Barry}, {Bennett},
  {Bhattacharya}, {Fujii}, {Fukui}, {Hirao}, {Silva}, {Kirikawa}, {Kondo},
  {Koshimoto}, {Matsubara}, {Matsumoto}, {Miyazaki}, {Muraki}, {Okamura},
  {Olmschenk}, {Ranc}, {Rattenbury}, {Satoh}, {Sumi}, {Suzuki}, {Toda},
  {Tristram}, {Vandorou}, {Yama}, {Moa Collaboration}, {Beichman}, {Bryden},
  {Novati}, {Gaudi}, {Henderson}, {Penny}, {Jacklin}, {Stassun}, \& {Ukirt
  Microlensing Team}}]{2018_prime}
{Gould}, A., {Han}, C., {Zang}, W., {et~al.} 2022, \aap, 664, A13,
  \dodoi{10.1051/0004-6361/202243744}

\bibitem[{{Han} {et~al.}(2020){Han}, {Lee}, {Udalski}, {Gould}, {Bond},
  {Bozza}, {Albrow}, {Chung}, {Hwang}, {Jung}, {Ryu}, {Shin}, {Shvartzvald},
  {Yee}, {Zang}, {Cha}, {Kim}, {Kim}, {Kim}, {Lee}, {Lee}, {Park}, {Pogge},
  {Jee}, {Kim}, {KMTNet Collaboration}, {Mr{\'o}z}, {Szyma{\'n}ski}, {Skowron},
  {Poleski}, {Soszy{\'n}ski}, {Pietrukowicz}, {Koz{\l}owski}, {Ulaczyk},
  {Rybicki}, {Iwanek}, {Wrona}, {OGLE Collaboration}, {Abe}, {Barry},
  {Bennett}, {Bhattacharya}, {Donachie}, {Fujii}, {Fukui}, {Itow}, {Hirao},
  {Kamei}, {Kondo}, {Koshimoto}, {Li}, {Matsubara}, {Muraki}, {Miyazaki},
  {Nagakane}, {Ranc}, {Rattenbury}, {Satoh}, {Shoji}, {Suematsu}, {Sullivan},
  {Sumi}, {Suzuki}, {Tristram}, {Yamakawa}, {Yamawaki}, {Yonehara}, \& {MOA
  Collaboration}}]{Han_3BDs}
{Han}, C., {Lee}, C.-U., {Udalski}, A., {et~al.} 2020, \aj, 159, 134,
  \dodoi{10.3847/1538-3881/ab6f66}

\bibitem[{{Han} {et~al.}(2021){Han}, {Udalski}, {Kim}, {Ryu}, {Bozza},
  {Albrow}, {Chung}, {Gould}, {Hwang}, {Jung}, {Lee}, {Shin}, {Shvartzvald},
  {Yee}, {Zang}, {Cha}, {Kim}, {Kim}, {Kim}, {Lee}, {Lee}, {Park}, {Pogge},
  {Mr{\'o}z}, {Szyma{\'n}ski}, {Skowron}, {Poleski}, {Soszy{\'n}ski},
  {Pietrukowicz}, {Koz{\l}owski}, {Ulaczyk}, {Rybicki}, {Iwanek}, {Wrona}, \&
  {Gromadzki}}]{Han3resonant}
{Han}, C., {Udalski}, A., {Kim}, D., {et~al.} 2021, \aap, 655, A21,
  \dodoi{10.1051/0004-6361/202141517}

\bibitem[{{Han} {et~al.}(2022){Han}, {Kim}, {Gould}, {Udalski}, {Bond},
  {Bozza}, {Jung}, {Albrow}, {Chung}, {Hwang}, {Ryu}, {Shin}, {Shvartzvald},
  {Yee}, {Zang}, {Cha}, {Kim}, {Kim}, {Lee}, {Lee}, {Lee}, {Park}, {Pogge},
  {KMTNet Collaboration}, {Mr{\'o}z}, {Szyma{\'n}ski}, {Skowron}, {Poleski},
  {Soszy{\'n}ski}, {Pietrukowicz}, {Koz{\l}owski}, {Ulaczyk}, {Rybicki},
  {Iwanek}, {OGLE Collaboration}, {Abe}, {Barry}, {Bennett}, {Bhattacharya},
  {Fujii}, {Fukui}, {Hirao}, {Itow}, {Kirikawa}, {Koshimoto}, {Kondo},
  {Matsubara}, {Matsumoto}, {Miyazaki}, {Muraki}, {Olmschenk}, {Okamura},
  {Ranc}, {Rattenbury}, {Satoh}, {Silva}, {Sumi}, {Suzuki}, {Toda}, {Tristram},
  {Vandorou}, {Yama}, \& {MOA Collaboration}}]{OB171691}
{Han}, C., {Kim}, D., {Gould}, A., {et~al.} 2022, \aap, 664, A33,
  \dodoi{10.1051/0004-6361/202243484}

\bibitem[{Harris {et~al.}(2020)Harris, Millman, van~der Walt, Gommers,
  Virtanen, Cournapeau, Wieser, Taylor, Berg, Smith, Kern, Picus, Hoyer, van
  Kerkwijk, Brett, Haldane, del R{\'{i}}o, Wiebe, Peterson,
  G{\'{e}}rard-Marchant, Sheppard, Reddy, Weckesser, Abbasi, Gohlke, \&
  Oliphant}]{numpy}
Harris, C.~R., Millman, K.~J., van~der Walt, S.~J., {et~al.} 2020, Nature, 585,
  357, \dodoi{10.1038/s41586-020-2649-2}

\bibitem[{{Holtzman} {et~al.}(1998){Holtzman}, {Watson}, {Baum}, {Grillmair},
  {Groth}, {Light}, {Lynds}, \& {O'Neil}}]{HSTCMD}
{Holtzman}, J.~A., {Watson}, A.~M., {Baum}, W.~A., {et~al.} 1998, \aj, 115,
  1946, \dodoi{10.1086/300336}

\bibitem[{Hunter(2007)}]{Matplotlib}
Hunter, J.~D. 2007, Computing in Science \& Engineering, 9, 90,
  \dodoi{10.1109/MCSE.2007.55}

\bibitem[{{Hwang} {et~al.}(2013){Hwang}, {Choi}, {Bond}, {Sumi}, {Han},
  {Gaudi}, {Gould}, {Bozza}, {Beaulieu}, {Tsapras}, {Abe}, {Bennett},
  {Botzler}, {Chote}, {Freeman}, {Fukui}, {Fukunaga}, {Harris}, {Itow},
  {Koshimoto}, {Ling}, {Masuda}, {Matsubara}, {Muraki}, {Namba}, {Ohnishi},
  {Rattenbury}, {Saito}, {Sullivan}, {Sweatman}, {Suzuki}, {Tristram}, {Wada},
  {Yamai}, {Yock}, {Yonehara}, {MOA Collaboration}, {de Almeida}, {DePoy},
  {Dong}, {Jablonski}, {Jung}, {Kavka}, {Lee}, {Park}, {Pogge}, {}, {Yee},
  {{$\mu$}FUN Collaboration}, {Albrow}, {Bachelet}, {Batista}, {Brillant},
  {Caldwell}, {Cassan}, {Cole}, {Corrales}, {Coutures}, {Dieters}, {Dominis
  Prester}, {Donatowicz}, {Fouqu{\'e}}, {Greenhill}, {J{\o}rgensen}, {Kane},
  {Kubas}, {Marquette}, {Martin}, {Meintjes}, {Menzies}, {Pollard}, {Williams},
  {Wouters}, {PLANET Collaboration}, {Bramich}, {Dominik}, {Horne}, {Browne},
  {Hundertmark}, {Ipatov}, {Kains}, {Snodgrass}, {Steele}, {Street}, \&
  {RoboNet Collaboration}}]{MB12486}
{Hwang}, K.-H., {Choi}, J.-Y., {Bond}, I.~A., {et~al.} 2013, \apj, 778, 55,
  \dodoi{10.1088/0004-637X/778/1/55}

\bibitem[{{Hwang} {et~al.}(2018){Hwang}, {Udalski}, {Shvartzvald}, {Ryu},
  {Albrow}, {Chung}, {Gould}, {Han}, {Jung}, {}, {Yee}, {Zhu}, {Cha}, {Kim},
  {Kim}, {Kim}, {Lee}, {Lee}, {Lee}, {Park}, {Pogge}, {KMTNet Collaboration},
  {Skowron}, {Mr{\'o}z}, {Poleski}, {Koz{\l}owski}, {Soszy{\'n}ski},
  {Pietrukowicz}, {Szyma{\'n}ski}, {Ulaczyk}, {Pawlak}, {OGLE Collaboration},
  {Bryden}, {Beichman}, {Calchi Novati}, {Gaudi}, {Henderson}, {Jacklin},
  {Penny}, \& {UKIRT Microlensing Team}}]{OB170173}
{Hwang}, K.-H., {Udalski}, A., {Shvartzvald}, Y., {et~al.} 2018, \aj, 155, 20,
  \dodoi{10.3847/1538-3881/aa992f}

\bibitem[{{Hwang} {et~al.}(2022){Hwang}, {Zang}, {Gould}, {Udalski}, {Bond},
  {Yang}, {Mao}, {Mao}, {Albrow}, {Chung}, {Han}, {Kil Jung}, {Ryu}, {Shin},
  {Shvartzvald}, {Yee}, {Cha}, {Kim}, {Kim}, {Kim}, {Lee}, {Lee}, {Lee},
  {Park}, {Pogge}, {Pogge}, {Mr{\'o}z}, {Poleski}, {Skowron}, {Szyma{\'n}ski},
  {Soszy{\'n}ski}, {Pietrukowicz}, {Koz{\l}owski}, {Ulaczyk}, {Rybicki},
  {Iwanek}, {Wrona}, {Gromadzki}, {Gromadzki}, {Abe}, {Barry}, {Bennett},
  {Bhattacharya}, {Fujii}, {Fukui}, {Hirao}, {Itow}, {Kirikawa}, {Kondo},
  {Koshimoto}, {Munford}, {Matsubara}, {Miyazaki}, {Muraki}, {Olmschenk},
  {Ranc}, {Rattenbury}, {Satoh}, {Shoji}, {Ishitani Silva}, {Sumi}, {Suzuki},
  {Tristram}, {Yonehara}, {Yonehara}, {Zhang}, {Zhu}, {Penny}, {Fouqu{\'e}}, \&
  {Fouqu{\'e}}}]{KB190253}
{Hwang}, K.-H., {Zang}, W., {Gould}, A., {et~al.} 2022, \aj, 163, 43,
  \dodoi{10.3847/1538-3881/ac38ad}

\bibitem[{{Ida} \& {Lin}(2004)}]{Ida2004}
{Ida}, S., \& {Lin}, D.~N.~C. 2004, \apj, 604, 388, \dodoi{10.1086/381724}

\bibitem[{{Jiang} {et~al.}(2004){Jiang}, {DePoy}, {Gal-Yam}, {Gaudi}, {Gould},
  {Han}, {Lipkin}, {Maoz}, {Ofek}, {Park}, {Pogge}, {MuFun Collaboration},
  {Udalski}, {Kubiak}, {Szyma{\'n}ski}, {Szewczyk}, {{\.Z}ebru{\'n}},
  {Wyrzykowski}, {Soszy{\'n}ski}, {Pietrzy{\'n}ski}, {OGLE Collaboration},
  {Albrow}, {Beaulieu}, {Caldwell}, {Cassan}, {Coutures}, {Dominik},
  {Donatowicz}, {Fouqu{\'e}}, {Greenhill}, {Hill}, {Horne}, {J{\o}rgensen},
  {J{\o}rgensen}, {Kane}, {Kubas}, {Martin}, {Menzies}, {Pollard}, {Sahu},
  {Wambsganss}, {Watson}, {Williams}, \& {PLANET Collaboration}}]{Jiang2004}
{Jiang}, G., {DePoy}, D.~L., {Gal-Yam}, A., {et~al.} 2004, \apj, 617, 1307,
  \dodoi{10.1086/425678}

\bibitem[{{Jung} {et~al.}(2022){Jung}, {Zang}, {Han}, {Gould}, {Udalski},
  {Albrow}, {Chung}, {Hwang}, {Ryu}, {Shin}, {Shvartzvald}, {Yang}, {Yee},
  {Cha}, {Kim}, {Kim}, {Lee}, {Lee}, {Lee}, {Park}, {Pogge}, {KMTNet
  Collaboration}, {Mr{\'o}z}, {Szyma{\'n}ski}, {Skowron}, {Poleski},
  {Soszy{\'n}ski}, {Pietrukowicz}, {Koz{\l}owski}, {Ulaczyk}, {Rybicki},
  {Iwanek}, {Wrona}, \& {OGLE Collaboration}}]{2018_subprime}
{Jung}, Y.~K., {Zang}, W., {Han}, C., {et~al.} 2022, \aj, 164, 262,
  \dodoi{10.3847/1538-3881/ac9c5c}

\bibitem[{{Jung} {et~al.}(2023){Jung}, {Zang}, {Wang}, {Han}, {Gould},
  {Udalski}, {Albrow}, {Chung}, {Hwang}, {Ryu}, {Shin}, {Shvartzvald}, {Yang},
  {Yee}, {Cha}, {Kim}, {Kim}, {Lee}, {Lee}, {Lee}, {Park}, {Pogge}, {KMTNet
  Collaboration}, {Szyma{\'n}ski}, {Skowron}, {Poleski}, {Soszy{\'n}ski},
  {Pietrukowicz}, {Koz{\l}owski}, {Ulaczyk}, {Rybicki}, {Iwanek}, {Wrona},
  {OGLE Collaboration}, {Green}, {Hennerley}, {Marmont}, {Mao}, {Maoz},
  {McCormick}, {Natusch}, {Penny}, {Porritt}, {Zhu}, {Tsinghua Team}, \& {FUN
  Follow-Up Team}}]{2019_subprime}
{Jung}, Y.~K., {Zang}, W., {Wang}, H., {et~al.} 2023, \aj, 165, 226,
  \dodoi{10.3847/1538-3881/accb8f}

\bibitem[{{Kennedy} \& {Kenyon}(2008)}]{snowline}
{Kennedy}, G.~M., \& {Kenyon}, S.~J. 2008, \apj, 673, 502,
  \dodoi{10.1086/524130}

\bibitem[{{Kim} {et~al.}(2018){Kim}, {Kim}, {Hwang}, {Albrow}, {Chung},
  {Gould}, {Han}, {Jung}, {Ryu}, {}, {Yee}, {Zhu}, {Cha}, {Kim}, {Lee}, {Lee},
  {Lee}, {Park}, {Pogge}, \& {The KMTNet Collaboration}}]{KMTeventfinder}
{Kim}, D.-J., {Kim}, H.-W., {Hwang}, K.-H., {et~al.} 2018, \aj, 155, 76,
  \dodoi{10.3847/1538-3881/aaa47b}

\bibitem[{{Kim} {et~al.}(2016){Kim}, {Lee}, {Park}, {Kim}, {Cha}, {Lee}, {Han},
  {Chun}, \& {Yuk}}]{KMT2016}
{Kim}, S.-L., {Lee}, C.-U., {Park}, B.-G., {et~al.} 2016, Journal of Korean
  Astronomical Society, 49, 37, \dodoi{10.5303/JKAS.2016.49.1.037}

\bibitem[{{Kuang} {et~al.}(2022){Kuang}, {Zang}, {Jung}, {Udalski}, {Yang},
  {Mao}, {Albrow}, {Chung}, {Gould}, {Han}, {Hwang}, {Ryu}, {Shin},
  {Shvartzvald}, {Yee}, {Cha}, {Kim}, {Kim}, {Kim}, {Lee}, {Lee}, {Lee},
  {Park}, {Pogge}, {Mr{\'o}z}, {Skowron}, {Poleski}, {Szyma{\'n}ski},
  {Soszy{\'n}ski}, {Pietrukowicz}, {Koz{\l}owski}, {Ulaczyk}, {Rybicki},
  {Iwanek}, {Wrona}, {Gromadzki}, {Wang}, {Huang}, \& {Zhu}}]{OB191470}
{Kuang}, R., {Zang}, W., {Jung}, Y.~K., {et~al.} 2022, \mnras, 516, 1704,
  \dodoi{10.1093/mnras/stac2315}

\bibitem[{{Mao}(2012)}]{Mao2012}
{Mao}, S. 2012, Research in Astronomy and Astrophysics, 12, 947,
  \dodoi{10.1088/1674-4527/12/8/005}

\bibitem[{{Mao} \& {Paczynski}(1991)}]{Shude1991}
{Mao}, S., \& {Paczynski}, B. 1991, \apjl, 374, L37, \dodoi{10.1086/186066}

\bibitem[{{Mordasini} {et~al.}(2009){Mordasini}, {Alibert}, \&
  {Benz}}]{Mordasini2009}
{Mordasini}, C., {Alibert}, Y., \& {Benz}, W. 2009, \aap, 501, 1139,
  \dodoi{10.1051/0004-6361/200810301}

\bibitem[{{Nataf} {et~al.}(2013){Nataf}, {Gould}, {Fouqu{\'e}}, {Gonzalez},
  {Johnson}, {Skowron}, {}, {Szyma{\'n}ski}, {Kubiak}, {Pietrzy{\'n}ski},
  {Soszy{\'n}ski}, {Ulaczyk}, {Wyrzykowski}, \& {Poleski}}]{Nataf2013}
{Nataf}, D.~M., {Gould}, A., {Fouqu{\'e}}, P., {et~al.} 2013, \apj, 769, 88,
  \dodoi{10.1088/0004-637X/769/2/88}

\bibitem[{{Nemiroff} \& {Wickramasinghe}(1994)}]{Nemiroff1994}
{Nemiroff}, R.~J., \& {Wickramasinghe}, W.~A.~D.~T. 1994, \apjl, 424, L21,
  \dodoi{10.1086/187265}

\bibitem[{{Paczy{\'n}ski}(1986)}]{Paczynski1986}
{Paczy{\'n}ski}, B. 1986, \apj, 304, 1, \dodoi{10.1086/164140}

\bibitem[{{Park} {et~al.}(2004){Park}, {DePoy}, {Gaudi}, {Gould}, {Han},
  {Pogge}, {muFun Collaboration}, {Abe}, {Bennett}, {Bond}, {Eguchi}, {Furuta},
  {Hearnshaw}, {Kamiya}, {Kilmartin}, {Kurata}, {Masuda}, {Matsubara},
  {Muraki}, {Noda}, {Okajima}, {Rattenbury}, {Sako}, {Sekiguchi}, {Sullivan},
  {Sumi}, {Tristram}, {Yanagisawa}, {Yock}, \& {MOA Collaboration}}]{MB03037}
{Park}, B.~G., {DePoy}, D.~L., {Gaudi}, B.~S., {et~al.} 2004, \apj, 609, 166,
  \dodoi{10.1086/420926}

\bibitem[{{Poindexter} {et~al.}(2005){Poindexter}, {Afonso}, {Bennett},
  {Glicenstein}, {Gould}, {Szyma{\'n}ski}, \& {Udalski}}]{Poindexter2005}
{Poindexter}, S., {Afonso}, C., {Bennett}, D.~P., {et~al.} 2005, \apj, 633,
  914, \dodoi{10.1086/468182}

\bibitem[{{Poleski} {et~al.}(2014){Poleski}, {Skowron}, {Udalski}, {Han},
  {Koz{\l}owski}, {Wyrzykowski}, {Dong}, {Szyma{\'n}ski}, {Kubiak},
  {Pietrzy{\'n}ski}, {Soszy{\'n}ski}, {Ulaczyk}, {Pietrukowicz}, \&
  {Gould}}]{OB08092}
{Poleski}, R., {Skowron}, J., {Udalski}, A., {et~al.} 2014, \apj, 795, 42,
  \dodoi{10.1088/0004-637X/795/1/42}

\bibitem[{{Ranc} {et~al.}(2019){Ranc}, {Bennett}, {Hirao}, {Udalski}, {Han},
  {Bond}, {Yee}, {and}, {Albrow}, {Chung}, {Gould}, {Hwang}, {Jung}, {Ryu},
  {Shin}, {Shvartzvald}, {Zang}, {Zhu}, {Cha}, {Kim}, {Kim}, {Kim}, {Lee},
  {Lee}, {Lee}, {Park}, {Pogge}, {KMTNet Collaboration}, {Abe}, {Barry},
  {Bhattacharya}, {Donachie}, {Fukui}, {Itow}, {Kawasaki}, {Kondo},
  {Koshimoto}, {Li}, {Matsubara}, {Miyazaki}, {Muraki}, {Nagakane},
  {Rattenbury}, {Suematsu}, {Sullivan}, {Sumi}, {Suzuki}, {Tristram},
  {Yonehara}, {MOA Collaboration}, {Poleski}, {Mr{\'o}z}, {Skowron},
  {Szyma{\'n}ski}, {Soszy{\'n}ski}, {Koz{\l}owski}, {Pietrukowicz}, {Ulaczyk},
  \& {OGLE Collaboration}}]{OB151670}
{Ranc}, C., {Bennett}, D.~P., {Hirao}, Y., {et~al.} 2019, \aj, 157, 232,
  \dodoi{10.3847/1538-3881/ab141b}

\bibitem[{{Ryu} {et~al.}(2024){Ryu}, {Udalski}, {Yee}, {Zang}, {Shvartzvald},
  {Han}, {Gould}, {Albrow}, {Chung}, {Hwang}, {Jung}, {Shin}, {Yang}, {Cha},
  {Kim}, {Kim}, {Lee}, {Lee}, {Lee}, {Park}, {Pogge}, {Wang}, {KMTNet
  Collaboration}, {Mr{\'o}z}, {Szyma{\'n}ski}, {Skowron}, {Poleski},
  {Soszy{\'n}ski}, {Pietrukowicz}, {Koz{\l}owski}, {Ulaczyk}, {Rybicki},
  {Iwanek}, {Wrona}, {OGLE Collaboration}, {Beichman}, {Bryden}, {Carey},
  {Henderson}, {Calchi Novati}, {Zhu}, {SPITZER Team}, {Jacklin}, {Penny}, \&
  {UKIRT Team}}]{2017_prime}
{Ryu}, Y.-H., {Udalski}, A., {Yee}, J.~C., {et~al.} 2024, \aj, 167, 88,
  \dodoi{10.3847/1538-3881/ad1888}

\bibitem[{{Shin} {et~al.}(2016){Shin}, {Ryu}, {Udalski}, {Albrow}, {Cha},
  {Choi}, {Chung}, {Han}, {Hwang}, {Jung}, {Kim}, {Kim}, {Lee}, {Lee}, {Park},
  {Park}, {Pogge}, {Yee}, {Pietrukowicz}, {Mroz}, {Kozlowski}, {Poleski},
  {Skowron}, {Soszynski}, {Szymanski}, {Ulaczyk}, {Wyrzykowski}, {Pawlak}, \&
  {Gould}}]{OB150954}
{Shin}, I.~G., {Ryu}, Y.~H., {Udalski}, A., {et~al.} 2016, Journal of Korean
  Astronomical Society, 49, 73, \dodoi{10.5303/JKAS.2016.49.3.73}

\bibitem[{{Shin} {et~al.}(2019{\natexlab{a}}){Shin}, {Ryu}, {Yee}, {Gould},
  {Albrow}, {Chung}, {Han}, {Hwang}, {Jung}, {Shvartzvald}, {Zang}, {Lee},
  {Cha}, {Kim}, {Kim}, {Kim}, {Lee}, {Lee}, {Park}, \& {Pogge}}]{KB171038}
{Shin}, I.~G., {Ryu}, Y.~H., {Yee}, J.~C., {et~al.} 2019{\natexlab{a}}, \aj,
  157, 146, \dodoi{10.3847/1538-3881/ab07c2}

\bibitem[{{Shin} {et~al.}(2019{\natexlab{b}}){Shin}, {Yee}, {Gould}, {Penny},
  {Bond}, {Albrow}, {Chung}, {Han}, {Hwang}, {Jung}, {Ryu}, {Shvartzvald},
  {Cha}, {Kim}, {Kim}, {Kim}, {Lee}, {Lee}, {Lee}, {Park}, {Pogge}, {(KMTNet
  Collaboration}, {Abe}, {Barry}, {Bennett}, {Bhattacharya}, {Donachie},
  {Fujii}, {Fukui}, {Hirao}, {Itow}, {Kamei}, {Kondo}, {Koshimoto}, {Li},
  {Matsubara}, {Miyazaki}, {Muraki}, {Nagakane}, {Ranc}, {Rattenbury},
  {Suematsu}, {Sullivan}, {Sumi}, {Suzuki}, {Tristram}, {Yamakawa}, {Yonehara},
  {(MOA Collaboration}, {Fouqu{\'e}}, {Zang}, \& {(CFHT-K2C9 Microlensing
  Collaboration}}]{KB171119}
{Shin}, I.~G., {Yee}, J.~C., {Gould}, A., {et~al.} 2019{\natexlab{b}}, \aj,
  158, 199, \dodoi{10.3847/1538-3881/ab46a5}

\bibitem[{{Shin} {et~al.}(2023){Shin}, {Yee}, {Zang}, {Yang}, {Hwang}, {Han},
  {Gould}, {Udalski}, {Bond}, {Albrow}, {Chung}, {Jung}, {Ryu}, {Shvartzvald},
  {Cha}, {Kim}, {Kim}, {Lee}, {Lee}, {Lee}, {Park}, {Pogge}, {Mr{\'o}z},
  {Szyma{\'n}ski}, {Skowron}, {Poleski}, {Soszy{\'n}ski}, {Pietrukowicz},
  {Koz{\l}owski}, {Rybicki}, {Iwanek}, {Ulaczyk}, {Wrona}, {Gromadzki}, {Abe},
  {Barry}, {Bennett}, {Bhattacharya}, {Fujii}, {Fukui}, {Hamada}, {Hirao},
  {Silva}, {Itow}, {Kirikawa}, {Kondo}, {Koshimoto}, {Matsubara}, {Miyazaki},
  {Muraki}, {Olmschenk}, {Ranc}, {Rattenbury}, {Satoh}, {Sumi}, {Suzuki},
  {Tomoyoshi}, {Tristram}, {Vandorou}, {Yama}, \& {Yamashita}}]{2016_prime}
{Shin}, I.-G., {Yee}, J.~C., {Zang}, W., {et~al.} 2023, \aj, 166, 104,
  \dodoi{10.3847/1538-3881/ace96d}

\bibitem[{{Shin} {et~al.}(2024){Shin}, {Yee}, {Zang}, {Han}, {Yang}, {Gould},
  {Lee}, {Udalski}, {Sumi}, {Albrow}, {Chung}, {Hwang}, {Jung}, {Ryu},
  {Shvartzvald}, {Cha}, {Kim}, {Kim}, {Kim}, {Lee}, {Lee}, {Park}, {Pogge},
  {Mr{\'o}z}, {Szyma{\'n}ski}, {Skowron}, {Poleski}, {Soszy{\'n}ski},
  {Pietrukowicz}, {Koz{\l}owski}, {Rybicki}, {Iwanek}, {Ulaczyk}, {Wrona},
  {Gromadzki}, {Abe}, {Bando}, {Barry}, {Bennett}, {Bhattacharya}, {Bond},
  {Fujii}, {Fukui}, {Hamada}, {Hamada}, {Hamasaki}, {Hirao}, {Ishitani Silva},
  {Itow}, {Kirikawa}, {Koshimoto}, {Matsubara}, {Miyazaki}, {Muraki}, {Nagai},
  {Nunota}, {Olmschenk}, {Ranc}, {Rattenbury}, {Satoh}, {Suzuki}, {Tomoyoshi},
  {Tristram}, {Vandorou}, {Yama}, \& {Yamashita}}]{2016_subprime}
---. 2024, arXiv e-prints, arXiv:2401.04256, \dodoi{10.48550/arXiv.2401.04256}

\bibitem[{{Shvartzvald} {et~al.}(2016){Shvartzvald}, {Maoz}, {Udalski}, {Sumi},
  {Friedmann}, {Kaspi}, {Poleski}, {Szyma{\'n}ski}, {Skowron}, {Koz{\l}owski},
  {Wyrzykowski}, {Mr{\'o}z}, {Pietrukowicz}, {Pietrzy{\'n}ski},
  {Soszy{\'n}ski}, {Ulaczyk}, {Abe}, {Barry}, {Bennett}, {Bhattacharya},
  {Bond}, {Freeman}, {Inayama}, {Itow}, {Koshimoto}, {Ling}, {Masuda}, {Fukui},
  {Matsubara}, {Muraki}, {Ohnishi}, {Rattenbury}, {Saito}, {Sullivan},
  {Suzuki}, {Tristram}, {Wakiyama}, \& {Yonehara}}]{Wise}
{Shvartzvald}, Y., {Maoz}, D., {Udalski}, A., {et~al.} 2016, \mnras, 457, 4089,
  \dodoi{10.1093/mnras/stw191}

\bibitem[{{Shvartzvald} {et~al.}(2019){Shvartzvald}, {Yee}, {Skowron}, {Lee},
  {Udalski}, {Calchi Novati}, {Bozza}, {Beichman}, {Bryden}, {Carey}, {Gaudi},
  {Henderson}, {Zhu}, {Spitzer team}, {Bachelet}, {Bolt}, {Christie}, {Maoz},
  {Natusch}, {Pogge}, {Street}, {Tan}, {Tsapras}, {LCO}, {{\ensuremath{\mu}}FUN
  Follow-up Teams}, {Pietrukowicz}, {Soszy{\'n}ski}, {Szyma{\'n}ski},
  {Mr{\'o}z}, {Poleski}, {Koz{\l}owski}, {Ulaczyk}, {Pawlak}, {Rybicki},
  {Iwanek}, {OGLE Collaboration}, {Albrow}, {Cha}, {Chung}, {Gould}, {Han},
  {Hwang}, {Jung}, {Kim}, {Kim}, {Kim}, {Lee}, {Lee}, {Park}, {Ryu}, {Shin},
  {Zang}, {KMTNet Collaboration}, {Dominik}, {Helling}, {Hundertmark},
  {J{\o}rgensen}, {Longa-Pe{\~n}a}, {Lowry}, {Sajadian}, {Burgdorf},
  {Campbell-White}, {Ciceri}, {Evans}, {Fujii}, {Hinse}, {Rahvar}, {Rabus},
  {Skottfelt}, {Snodgrass}, {Southworth}, \& {MiNDSTEp
  Collaboration}}]{OB170896}
{Shvartzvald}, Y., {Yee}, J.~C., {Skowron}, J., {et~al.} 2019, \aj, 157, 106,
  \dodoi{10.3847/1538-3881/aafe12}

\bibitem[{{Skowron} {et~al.}(2011){Skowron}, {Udalski}, {Gould}, {Dong},
  {Monard}, {Han}, {Nelson}, {McCormick}, {Moorhouse}, {Thornley}, {Maury},
  {Bramich}, {Greenhill}, {Koz{\l}owski}, {Bond}, {Poleski}, {Wyrzykowski},
  {Ulaczyk}, {Kubiak}, {Szyma{\'n}ski}, {Pietrzy{\'n}ski}, {Soszy{\'n}ski},
  {OGLE Collaboration}, {Gaudi}, {Yee}, {Hung}, {Pogge}, {DePoy}, {Lee},
  {Park}, {Allen}, {Mallia}, {Drummond}, {Bolt}, {{$\mu$}FUN Collaboration},
  {Allan}, {Browne}, {Clay}, {Dominik}, {Fraser}, {Horne}, {Kains}, {Mottram},
  {Snodgrass}, {Steele}, {Street}, {Tsapras}, {RoboNet Collaboration}, {Abe},
  {Bennett}, {Botzler}, {Douchin}, {Freeman}, {Fukui}, {Furusawa}, {Hayashi},
  {Hearnshaw}, {Hosaka}, {Itow}, {Kamiya}, {Kilmartin}, {Korpela}, {Lin},
  {Ling}, {Makita}, {Masuda}, {Matsubara}, {Muraki}, {Nagayama}, {Miyake},
  {Nishimoto}, {Ohnishi}, {Perrott}, {Rattenbury}, {Saito}, {Skuljan},
  {Sullivan}, {Sumi}, {Suzuki}, {Sweatman}, {Tristram}, {Wada}, {Yock}, {MOA
  Collaboration}, {Beaulieu}, {Fouqu{\'e}}, {Albrow}, {Batista}, {Brillant},
  {Caldwell}, {Cassan}, {Cole}, {Cook}, {Coutures}, {Dieters}, {Dominis
  Prester}, {Donatowicz}, {Kane}, {Kubas}, {Marquette}, {Martin}, {Menzies},
  {Sahu}, {Wambsganss}, {Williams}, {Zub}, \& {PLANET Collaboration}}]{OB09020}
{Skowron}, J., {Udalski}, A., {Gould}, A., {et~al.} 2011, \apj, 738, 87,
  \dodoi{10.1088/0004-637X/738/1/87}

\bibitem[{{Suzuki} {et~al.}(2016){Suzuki}, {Bennett}, {Sumi}, {Bond}, {Rogers},
  {Abe}, {Asakura}, {Bhattacharya}, {Donachie}, {Freeman}, {Fukui}, {Hirao},
  {Itow}, {Koshimoto}, {Li}, {Ling}, {Masuda}, {Matsubara}, {Muraki},
  {Nagakane}, {Onishi}, {Oyokawa}, {Rattenbury}, {Saito}, {Sharan}, {Shibai},
  {Sullivan}, {Tristram}, {Yonehara}, \& {MOA Collaboration}}]{Suzuki2016}
{Suzuki}, D., {Bennett}, D.~P., {Sumi}, T., {et~al.} 2016, \apj, 833, 145,
  \dodoi{10.3847/1538-4357/833/2/145}

\bibitem[{{Szyma{\'n}ski} {et~al.}(2011){Szyma{\'n}ski}, {Udalski},
  {Soszy{\'n}ski}, {Kubiak}, {Pietrzy{\'n}ski}, {Poleski}, {Wyrzykowski}, \&
  {Ulaczyk}}]{OGLEIII}
{Szyma{\'n}ski}, M.~K., {Udalski}, A., {Soszy{\'n}ski}, I., {et~al.} 2011,
  \actaa, 61, 83.
\newblock \doarXiv{1107.4008}

\bibitem[{{Tomaney} \& {Crotts}(1996)}]{Tomaney1996}
{Tomaney}, A.~B., \& {Crotts}, A. P.~S. 1996, \aj, 112, 2872,
  \dodoi{10.1086/118228}

\bibitem[{{Udalski}(2003)}]{Udalski2003}
{Udalski}, A. 2003, \actaa, 53, 291

\bibitem[{{Udalski} {et~al.}(1994){Udalski}, {Szymanski}, {Kaluzny}, {Kubiak},
  {Mateo}, {Krzeminski}, \& {Paczynski}}]{Udalski1994}
{Udalski}, A., {Szymanski}, M., {Kaluzny}, J., {et~al.} 1994, \actaa, 44, 227

\bibitem[{{Udalski} {et~al.}(2015){Udalski}, {Szyma{\'n}ski}, \&
  {Szyma{\'n}ski}}]{OGLEIV}
{Udalski}, A., {Szyma{\'n}ski}, M.~K., \& {Szyma{\'n}ski}, G. 2015, \actaa, 65,
  1.
\newblock \doarXiv{1504.05966}

\bibitem[{{Udalski} {et~al.}(2005){Udalski}, {Jaroszy{\'n}ski},
  {Paczy{\'n}ski}, {Kubiak}, {Szyma{\'n}ski}, {Soszy{\'n}ski},
  {Pietrzy{\'n}ski}, {Ulaczyk}, {Szewczyk}, {Wyrzykowski}, {OGLE
  Collaboration}, {Christie}, {DePoy}, {Dong}, {Gal-Yam}, {Gaudi}, {Gould},
  {Han}, {L{\'e}pine}, {McCormick}, {Park}, {Pogge}, {{$\mu$}FUN
  Collaboration}, {Bennett}, {Bond}, {Muraki}, {Tristram}, {Yock}, {MOA
  Collaboration}, {Beaulieu}, {Bramich}, {Dieters}, {Greenhill}, {Hill},
  {Horne}, {Kubas}, \& {PLANET/ROBONET Collaboration}}]{OB050071}
{Udalski}, A., {Jaroszy{\'n}ski}, M., {Paczy{\'n}ski}, B., {et~al.} 2005,
  \apjl, 628, L109, \dodoi{10.1086/432795}

\bibitem[{{Udalski} {et~al.}(2018){Udalski}, {Ryu}, {Sajadian}, {Gould},
  {Mr{\'o}z}, {Poleski}, {Szyma{\'n}ski}, {Skowron}, {Soszy{\'n}ski},
  {Koz{\l}owski}, {Pietrukowicz}, {Ulaczyk}, {Pawlak}, {Rybicki}, {Iwanek},
  {Albrow}, {Chung}, {Han}, {Hwang}, {Jung}, {}, {Shvartzvald}, {Yee}, {Zang},
  {Zhu}, {Cha}, {Kim}, {Kim}, {Kim}, {Lee}, {Lee}, {Lee}, {Park}, {Pogge},
  {Bozza}, {Dominik}, {Helling}, {Hundertmark}, {J{\o}rgensen},
  {Longa-Pe{\~n}a}, {Lowry}, {Burgdorf}, {Campbell-White}, {Ciceri}, {Evans},
  {Figuera Jaimes}, {Fujii}, {Haikala}, {Henning}, {Hinse}, {Mancini},
  {Peixinho}, {Rahvar}, {Rabus}, {Skottfelt}, {Snodgrass}, {Southworth}, \&
  {von Essen}}]{OB171434}
{Udalski}, A., {Ryu}, Y.-H., {Sajadian}, S., {et~al.} 2018, \actaa, 68, 1.
\newblock \doarXiv{1802.02582}

\bibitem[{{Virtanen} {et~al.}(2020){Virtanen}, {Gommers}, {Oliphant},
  {Haberland}, {Reddy}, {Cournapeau}, {Burovski}, {Peterson}, {Weckesser},
  {Bright}, {van der Walt}, {Brett}, {Wilson}, {Millman}, {Mayorov}, {Nelson},
  {Jones}, {Kern}, {Larson}, {Carey}, {Polat}, {Feng}, {Moore}, {VanderPlas},
  {Laxalde}, {Perktold}, {Cimrman}, {Henriksen}, {Quintero}, {Harris},
  {Archibald}, {Ribeiro}, {Pedregosa}, {van Mulbregt}, \& {SciPy 1. 0
  Contributors}}]{scipy}
{Virtanen}, P., {Gommers}, R., {Oliphant}, T.~E., {et~al.} 2020, Nature
  Methods, 17, 261, \dodoi{10.1038/s41592-019-0686-2}

\bibitem[{{Wang} {et~al.}(2022){Wang}, {Zang}, {Zhu}, {Hwang}, {Udalski},
  {Gould}, {Han}, {Albrow}, {Chung}, {Jung}, {Kim}, {Ryu}, {Shin},
  {Shvartzvald}, {Yee}, {Cha}, {Kim}, {Kim}, {Kim}, {Lee}, {Lee}, {Lee},
  {Park}, {Pogge}, {Poleski}, {Mr{\'o}z}, {Skowron}, {Szyma{\'n}ski},
  {Soszy{\'n}ski}, {Pietrukowicz}, {Koz{\l}owski}, {Ulaczyk}, {Rybicki},
  {Iwanek}, {Wrona}, {Gromadzki}, {Yang}, {Mao}, \& {Zhang}}]{OB180383}
{Wang}, H., {Zang}, W., {Zhu}, W., {et~al.} 2022, \mnras, 510, 1778,
  \dodoi{10.1093/mnras/stab3581}

\bibitem[{{Wang} {et~al.}(2018){Wang}, {Calchi Novati}, {Udalski}, {Gould},
  {Mao}, {Zang}, {Beichman}, {Bryden}, {Carey}, {Gaudi}, {Henderson},
  {Shvartzvald}, {Yee}, {Spitzer Team}, {Mr{\'o}z}, {Poleski}, {Skowron},
  {Szyma{\'n}ski}, {Soszy{\'n}ski}, {Koz{\l}owski}, {Pietrukowicz}, {Ulaczyk},
  {Pawlak}, {OGLE Collaboration}, {Albrow}, {Chung}, {Han}, {Hwang}, {Jung},
  {Ryu}, {Shin}, {Zhu}, {Cha}, {Kim}, {Kim}, {Kim}, {Lee}, {Lee}, {Lee},
  {Park}, {Pogge}, \& {KMTNet Collaboration}}]{OB171130}
{Wang}, T., {Calchi Novati}, S., {Udalski}, A., {et~al.} 2018, \apj, 860, 25,
  \dodoi{10.3847/1538-4357/aabcd2}

\bibitem[{{Witt} \& {Mao}(1994)}]{Shude1994}
{Witt}, H.~J., \& {Mao}, S. 1994, \apj, 430, 505, \dodoi{10.1086/174426}

\bibitem[{{Wozniak}(2000)}]{Wozniak2000}
{Wozniak}, P.~R. 2000, \actaa, 50, 421

\bibitem[{{Yang} {et~al.}(2021){Yang}, {Mao}, {Zang}, \&
  {Zhang}}]{Yang2021_GalacticModel}
{Yang}, H., {Mao}, S., {Zang}, W., \& {Zhang}, X. 2021, \mnras, 502, 5631,
  \dodoi{10.1093/mnras/stab441}

\bibitem[{{Yang} {et~al.}(2020){Yang}, {Zhang}, {Hwang}, {Zang}, {Gould},
  {Wang}, {Mao}, {Albrow}, {Chung}, {Han}, {Jung}, {Ryu}, {Shin},
  {Shvartzvald}, {Yee}, {Zhu}, {Penny}, {Fouqu{\'e}}, {Cha}, {Kim}, {Kim},
  {Kim}, {Lee}, {Lee}, {Lee}, {Park}, \& {Pogge}}]{KB161836}
{Yang}, H., {Zhang}, X., {Hwang}, K.-H., {et~al.} 2020, \aj, 159, 98,
  \dodoi{10.3847/1538-3881/ab660e}

\bibitem[{{Yang} {et~al.}(2024){Yang}, {Yee}, {Hwang}, {Qian}, {Bond}, {Gould},
  {Hu}, {Zhang}, {Mao}, {Zhu}, {Albrow}, {Chung}, {Kim}, {Park}, {Han}, {Jung},
  {Ryu}, {Shin}, {Shvartzvald}, {Cha}, {Kim}, {Kim}, {Lee}, {Lee}, {Lee},
  {Pogge}, {Zang}, {Abe}, {Barry}, {Bennett}, {Bhattacharya}, {Donachie},
  {Fujii}, {Fukui}, {Hirao}, {Itow}, {Kirikawa}, {Kondo}, {Koshimoto}, {Silva},
  {Li}, {Matsubara}, {Muraki}, {Suzuki}, {Tristram}, {Yonehara}, {Ranc},
  {Miyazaki}, {Olmschenk}, {Rattenbury}, {Satoh}, {Shoji}, {Sumi}, {Tanaka}, \&
  {Yamawaki}}]{Yang_TLC}
{Yang}, H., {Yee}, J.~C., {Hwang}, K.-H., {et~al.} 2024, \mnras, 528, 11,
  \dodoi{10.1093/mnras/stad3672}

\bibitem[{{Yee} {et~al.}(2012){Yee}, {Shvartzvald}, {Gal-Yam}, {Bond},
  {Udalski}, {Koz{\l}owski}, {Han}, {Gould}, {Skowron}, {Suzuki}, {Abe},
  {Bennett}, {Botzler}, {Chote}, {Freeman}, {Fukui}, {Furusawa}, {Itow},
  {Kobara}, {Ling}, {Masuda}, {Matsubara}, {Miyake}, {Muraki}, {Ohmori},
  {Ohnishi}, {Rattenbury}, {Saito}, {Sullivan}, {Sumi}, {Suzuki}, {Sweatman},
  {Takino}, {Tristram}, {Wada}, {MOA Collaboration}, {Szyma{\'n}ski}, {Kubiak},
  {Pietrzy{\'n}ski}, {Soszy{\'n}ski}, {Poleski}, {Ulaczyk}, {Wyrzykowski},
  {Pietrukowicz}, {OGLE Collaboration}, {Allen}, {Almeida}, {Batista}, {Bos},
  {Christie}, {DePoy}, {Dong}, {Drummond}, {Finkelman}, {Gaudi}, {Gorbikov},
  {Henderson}, {Higgins}, {Jablonski}, {Kaspi}, {Manulis}, {Maoz}, {McCormick},
  {McGregor}, {Monard}, {Moorhouse}, {Mu{\~n}oz}, {Natusch}, {Ngan}, {Ofek},
  {Pogge}, {Santallo}, {Tan}, {Thornley}, {Shin}, {Choi}, {Park}, {Lee}, {Koo},
  \& {{\ensuremath{\mu}}FUN Collaboration}}]{MB11293}
{Yee}, J.~C., {Shvartzvald}, Y., {Gal-Yam}, A., {et~al.} 2012, \apj, 755, 102,
  \dodoi{10.1088/0004-637X/755/2/102}

\bibitem[{{Yee} {et~al.}(2014){Yee}, {Han}, {Gould}, {Skowron}, {Bond},
  {Udalski}, {Hundertmark}, {Monard}, {Porritt}, {Nelson}, {Bozza}, {Albrow},
  {Choi}, {Christie}, {DePoy}, {Gaudi}, {Hwang}, {Jung}, {Lee}, {McCormick},
  {Natusch}, {Ngan}, {Park}, {Pogge}, {Shin}, {Tan}, {{\ensuremath{\mu}}FUN
  Collaboration}, {Abe}, {Bennett}, {Botzler}, {Freeman}, {Fukui}, {Fukunaga},
  {Itow}, {Koshimoto}, {Larsen}, {Ling}, {Masuda}, {Matsubara}, {Muraki},
  {Namba}, {Ohnishi}, {Philpott}, {Rattenbury}, {Saito}, {Sullivan}, {Sumi},
  {Sweatman}, {Suzuki}, {Tristram}, {Tsurumi}, {Wada}, {Yamai}, {Yock},
  {Yonehara}, {MOA Collaboration}, {Szyma{\'n}ski}, {Ulaczyk}, {Koz{\l}owski},
  {Poleski}, {Wyrzykowski}, {Kubiak}, {Pietrukowicz}, {Pietrzy{\'n}ski},
  {Soszy{\'n}ski}, {OGLE Collaboration}, {Bramich}, {Browne}, {Figuera Jaimes},
  {Horne}, {Ipatov}, {Kains}, {Snodgrass}, {Steele}, {Street}, {Tsapras}, \&
  {RoboNet Collaboration}}]{MB13220}
{Yee}, J.~C., {Han}, C., {Gould}, A., {et~al.} 2014, \apj, 790, 14,
  \dodoi{10.1088/0004-637X/790/1/14}

\bibitem[{{Yee} {et~al.}(2021){Yee}, {Zang}, {Udalski}, {Ryu}, {Green},
  {Hennerley}, {Marmont}, {Sumi}, {Mao}, {Gromadzki}, {Mr{\'o}z}, {Skowron},
  {Poleski}, {Szyma{\'n}ski}, {Soszy{\'n}ski}, {Pietrukowicz}, {Koz{\l}owski},
  {Ulaczyk}, {Rybicki}, {Iwanek}, {Wrona}, {Albrow}, {Chung}, {Gould}, {Han},
  {Hwang}, {Jung}, {Kim}, {Shin}, {Shvartzvald}, {Cha}, {Kim}, {Kim}, {Lee},
  {Lee}, {Lee}, {Park}, {Pogge}, {Bachelet}, {Christie}, {Hundertmark}, {Maoz},
  {McCormick}, {Natusch}, {Penny}, {Street}, {Tsapras}, {Beichman}, {Bryden},
  {Novati}, {Carey}, {Gaudi}, {Henderson}, {Johnson}, {Zhu}, {Bond}, {Abe},
  {Barry}, {Bennett}, {Bhattacharya}, {Donachie}, {Fujii}, {Fukui}, {Hirao},
  {Silva}, {Itow}, {Kirikawa}, {Kondo}, {Koshimoto}, {Alex Li}, {Matsubara},
  {Muraki}, {Miyazaki}, {Olmschenk}, {Ranc}, {Rattenbury}, {Satoh}, {Shoji},
  {Suzuki}, {Tanaka}, {Tristram}, {Yamawaki}, {Yonehara}, \& {MOA
  Collaboration}}]{OB190960}
{Yee}, J.~C., {Zang}, W., {Udalski}, A., {et~al.} 2021, \aj, 162, 180,
  \dodoi{10.3847/1538-3881/ac1582}

\bibitem[{{Yoo} {et~al.}(2004){Yoo}, {DePoy}, {Gal-Yam}, {Gaudi}, {Gould},
  {Han}, {Lipkin}, {Maoz}, {Ofek}, {Park}, {Pogge}, {Mu-Fun Collaboration},
  {Udalski}, {Soszy{\'n}ski}, {Wyrzykowski}, {Kubiak}, {Szyma{\'n}ski},
  {Pietrzy{\'n}ski}, {Szewczyk}, {{\.Z}ebru{\'n}}, \& {OGLE
  Collaboration}}]{Yoo2004}
{Yoo}, J., {DePoy}, D.~L., {Gal-Yam}, A., {et~al.} 2004, \apj, 603, 139,
  \dodoi{10.1086/381241}

\bibitem[{{Zang} {et~al.}(2021{\natexlab{a}}){Zang}, {Hwang}, {Udalski},
  {Wang}, {Zhu}, {Sumi}, {Yee}, {Gould}, {Mao}, {Zhang}, {Albrow}, {Chung},
  {Han}, {Jung}, {Ryu}, {Shin}, {Shvartzvald}, {Cha}, {Kim}, {Kim}, {Kim},
  {Lee}, {Lee}, {Lee}, {Park}, {Pogge}, {Mr{\'o}z}, {Skowron}, {Poleski},
  {Szyma{\'n}ski}, {Soszy{\'n}ski}, {Pietrukowicz}, {Koz{\l}owski}, {Ulaczyk},
  {Rybicki}, {Iwanek}, {Wrona}, {Gromadzki}, {Bond}, {Abe}, {Barry}, {Bennett},
  {Bhattacharya}, {Donachie}, {Fujii}, {Fukui}, {Hirao}, {Itow}, {Kirikawa},
  {Kondo}, {Koshimoto}, {Li}, {Matsubara}, {Muraki}, {Miyazaki}, {Olmschenk},
  {Ranc}, {Rattenbury}, {Satoh}, {Shoji}, {Ishitani Silva}, {Suzuki}, {Tanaka},
  {Tristram}, {Yamawaki}, {Yonehara}, {Beichman}, {Bryden}, {Calchi Novati},
  {Carey}, {Gaudi}, {Henderson}, {Johnson}, \& {Spitzer Team}}]{OB191053}
{Zang}, W., {Hwang}, K.-H., {Udalski}, A., {et~al.} 2021{\natexlab{a}}, \aj,
  162, 163, \dodoi{10.3847/1538-3881/ac12d4}

\bibitem[{{Zang} {et~al.}(2021{\natexlab{b}}){Zang}, {Han}, {Kondo}, {Yee},
  {Lee}, {Gould}, {Mao}, {de Almeida}, {Shvartzvald}, {Zhang}, {Albrow},
  {Chung}, {Hwang}, {Jung}, {Ryu}, {Shin}, {Cha}, {Kim}, {Kim}, {Kim}, {Lee},
  {Lee}, {Park}, {Pogge}, {Drummond}, {Tan}, {Nascimento J{\'u}nior}, {Maoz},
  {Penny}, {Zhu}, {Bond}, {Abe}, {Barry}, {Bennett}, {Bhattacharya},
  {Donachie}, {Fujii}, {Fukui}, {Hirao}, {Itow}, {Kirikawa}, {Koshimoto}, {Alex
  Li}, {Matsubara}, {Muraki}, {Miyazaki}, {Olmschenk}, {Ranc}, {Rattenbury},
  {Satoh}, {Shoji}, {Silva}, {Sumi}, {Suzuki}, {Tanaka}, {Tristram},
  {Yamawaki}, {Yonehara}, {Petric}, {Burdullis}, \& {Fouqu{\'e}}}]{KB200414}
{Zang}, W., {Han}, C., {Kondo}, I., {et~al.} 2021{\natexlab{b}}, Research in
  Astronomy and Astrophysics, 21, 239, \dodoi{10.1088/1674-4527/21/9/239}

\bibitem[{{Zang} {et~al.}(2022{\natexlab{a}}){Zang}, {Yang}, {Han}, {Lee},
  {Udalski}, {Gould}, {Mao}, {Zhang}, {Zhu}, {Albrow}, {Chung}, {Hwang},
  {Jung}, {Ryu}, {Shin}, {Shvartzvald}, {Yee}, {Cha}, {Kim}, {Kim}, {Kim},
  {Lee}, {Lee}, {Park}, {Pogge}, {Mr{\'o}z}, {Skowron}, {Poleski},
  {Szyma{\'n}ski}, {Soszy{\'n}ski}, {Pietrukowicz}, {Koz{\l}owski}, {Ulaczyk},
  {Rybicki}, {Iwanek}, {Wrona}, \& {Gromadzki}}]{2019_prime}
{Zang}, W., {Yang}, H., {Han}, C., {et~al.} 2022{\natexlab{a}}, \mnras, 515,
  928, \dodoi{10.1093/mnras/stac1883}

\bibitem[{{Zang} {et~al.}(2022{\natexlab{b}}){Zang}, {Shvartzvald}, {Udalski},
  {Yee}, {Lee}, {Sumi}, {Zhang}, {Yang}, {Mao}, {Novati}, {Gould}, {Zhu},
  {Beichman}, {Bryden}, {Carey}, {Gaudi}, {Henderson}, {Mr{\'o}z}, {Skowron},
  {Poleski}, {Szyma{\'n}ski}, {Soszy{\'n}ski}, {Pietrukowicz}, {Koz{\l}owski},
  {Ulaczyk}, {Rybicki}, {Iwanek}, {Wrona}, {Albrow}, {Chung}, {Han}, {Hwang},
  {Jung}, {Ryu}, {Shin}, {Cha}, {Kim}, {Kim}, {Kim}, {Lee}, {Lee}, {Park},
  {Pogge}, {Bond}, {Abe}, {Barry}, {Bennett}, {Bhattacharya}, {Donachie},
  {Fujii}, {Fukui}, {Hirao}, {Itow}, {Kirikawa}, {Kondo}, {Koshimoto}, {Li},
  {Matsubara}, {Muraki}, {Miyazaki}, {Ranc}, {Rattenbury}, {Satoh}, {Shoji},
  {Suzuki}, {Tanaka}, {Tristram}, {Yamawaki}, {Yonehara}, {Bachelet},
  {Hundertmark}, {Jaimes}, {Maoz}, {Penny}, {Street}, \& {Tsapras}}]{OB180799}
{Zang}, W., {Shvartzvald}, Y., {Udalski}, A., {et~al.} 2022{\natexlab{b}},
  \mnras, 514, 5952, \dodoi{10.1093/mnras/stac1631}

\bibitem[{{Zang} {et~al.}(2023){Zang}, {Jung}, {Yang}, {Zhang}, {Udalski},
  {Yee}, {Gould}, {Mao}, {Albrow}, {Chung}, {Han}, {Hwang}, {Ryu}, {Shin},
  {Shvartzvald}, {Cha}, {Kim}, {Kim}, {Kim}, {Lee}, {Lee}, {Lee}, {Park},
  {Pogge}, {KMTNet Collaboration}, {Mr{\'o}z}, {Skowron}, {Poleski},
  {Szyma{\'n}ski}, {Soszy{\'n}ski}, {Pietrukowicz}, {Koz{\l}owski}, {Ulaczyk},
  {Rybicki}, {Iwanek}, {Wrona}, {Gromadzki}, {OGLE Collaboration}, {Wang},
  {Zhang}, {Zhu}, \& {MAP Collaboration}}]{-4planet}
{Zang}, W., {Jung}, Y.~K., {Yang}, H., {et~al.} 2023, \aj, 165, 103,
  \dodoi{10.3847/1538-3881/acb34b}

\bibitem[{{Zang} {et~al.}(2025){Zang}, {Jung}, {Yee}, {Udalski}, {Wang}, {Zhu},
  {Sumi}, {Gould}, {Mao}, {Zhang}, {Albrow}, {Chung}, \& {Han}}]{OB160007}
{Zang}, W., {Jung}, Y.~K., {Yee}, J.~C., {et~al.} 2025, Science, in press

\end{thebibliography}

\begin{center}  
\renewcommand\arraystretch{1.2}
\begin{longtable*}{c c c c c c c}
    \caption{Information of unambiguous planetary events from the KMTNet 2017 subprime fields}
    \label{tab:sample_info}\\
    \hline
    \hline
    Event Name & KMTNet Name & $\log q$ & $s$ & Method & $\Delta\chi^2$ & Reference \\ 
    \hline
    KB171194 & KB171194 & $-4.582 \pm 0.058$ & $0.806 \pm 0.010$ & Discovery & & \cite{-4planet} \\
    \hline
    KB171003 & KB171003 & $-4.373 \pm 0.144$ & $0.910 \pm 0.005$ & Discovery  & 0.0 & \cite{-4planet} \\
     & & $-4.260 \pm 0.152$ & $0.889 \pm 0.004$ & & 0.2 & \\
    \hline
    OB171806 & KB171021 & $-4.352 \pm 0.171$ & $0.857 \pm 0.008$ & Discovery & 0.0 & \cite{-4planet} \\
    & & $-4.392 \pm 0.180$ & $0.861 \pm 0.007$ & & 0.2 & \\
    & & $-4.441 \pm 0.168$ & $1.181 \pm 0.011$ & & 8.3 & \\
    & & $-4.317 \pm 0.126$ & $1.190 \pm 0.012$ & & 8.4 & \\
    \hline
    OB171691 & KB170752 & $-4.013 \pm 0.152$ & $1.003 \pm 0.014$ & Recovery & 0.0 & \cite{OB171691} \\
    & & $-4.150 \pm 0.141$ & $1.058 \pm 0.011$ & & 0.4 & \\
    \hline
    KB170849 & KB170849 & $-3.998 \pm 0.053$ & $1.804 \pm 0.049$ & Discovery & & This work \\
    \hline
    KB171057 & KB171057 & $-3.925 \pm 0.101$ & $0.919 \pm 0.005$ & Discovery & & This work \\
    \hline
    OB170364 & KB171396 & $-3.341 \pm 0.079$ & $1.504 \pm 0.016$ & Recovery & & This work \\
    \hline
    KB172331 & KB172331 & $-2.891 \pm 0.044$ & $1.027 \pm 0.020$ & Recovery & & This work \\
    \hline
    KB171146 & KB171146 & $-2.699 \pm 0.076$ & $0.734 \pm 0.014$ & Recovery & 0.0 & \cite{KB171038} \\
    & & $-2.347 \pm 0.092$ & $1.148 \pm 0.014$ & & 9.5 & \\
    \hline
    KB172509 & KB172509 & $-2.360 \pm 0.053$ & $0.925 \pm 0.007$ & 
    Recovery & & \cite{Han3resonant} \\
    \hline
    KB171038 & KB171038 & $-2.276 \pm 0.025$ & $0.851 \pm 0.003$ & Recovery & & \cite{KB171038} \\
    \hline
    OB171099 & KB172336 & $-2.192 \pm 0.053$ & $1.137 \pm 0.014$ & Recovery & & \cite{Han3resonant} \\
    \hline
    OB171140 & KB171018 & $-2.142 \pm 0.039$ & $0.871 \pm 0.013$ & Recovery & 0.0 & \cite{OB171140} \\
     & & $-2.137 \pm 0.045$ & $0.870 \pm 0.014$ & & 1.0 & \\
    \hline
    OB171630 & KB171237 & $-2.114 \pm0 .009$ & $1.84\pm 0.02$ & Recovery & 0.0 & in prep \\
    & & $-2.119 \pm 0.008$ & $0.54 \pm 0.01$ & & 0.4 & \\
    \hline
    KB172197 & KB172197 & $-1.733 \pm 0.067$ & $0.723 \pm 0.017$ & Recovery & 0.0 & Han et al. in prep \\
     & & $-1.625 \pm 0.065$ & $1.540\pm 0.041$ & & 0.1 & \\
    \hline
    \hline
    \multicolumn{7}{c}{NOTE: For each planet, we only consider the models that have $\Delta\chi^2 < 10$ compared to the best-fit model.}\\
    \multicolumn{7}{c}{``Discovery'' represents that the planet was discovered using AnomlyFinder, and ``Recovery'' means} \\
    \multicolumn{7}{c}{that the planet was first discovered from by-eye searches and then recovered by AnomlyFinder.}
    
\end{longtable*}
\end{center}

\end{document}